# Comprehensive Analyses of the Strongly Carbon-chain Depleted Comets in Lowell Observatory's Narrowband Photometry Database

Allison N. Bair and David G. Schleicher
Lowell Observatory, 1400 W Mars Hill Rd, Flagstaff, AZ 86001, USA
Corresponding Author: bair@lowell.edu

ABSTRACT

We present measurements, analyses and results for the seventeen strongly carbon-chain depleted comets in Lowell Observatory's narrowband photometry database. The majority of comets exhibit a very similar, i.e. typical, composition in optical wavelengths, though the existence of anomalous comets with lower abundances of carbon-chain molecules relative to CN has long been known. M. F. A'Hearn et al. (1995) identified an entire class of these carbon-chain depleted comets, and the most recent full analysis of our database reveals there are varying degrees of carbon-chain depletion. Here we focus on the most depleted comets, the strongly carbon-chain depleted class, which is the largest non-typical compositional class to emerge from our taxonomic analyses. All comets in this class are strongly depleted in both $C_2$ and $C_3$ with respect to OH and CN, with depletions for these ratios being 3–27× below the mean abundance ratios for comets with typical composition. Several comets in this class additionally exhibit depletion in NH, with the largest depletions being 11× below mean typical values. A number of these comets exhibit asymmetry in production rates as a function of time and heliocentric distance, and one exhibits evidence of small secular changes. Almost all of the strongly depleted comets are Jupiter-family comets, indicating a Kuiper belt origin for this compositional class. Multiple lines of evidence from our full database indicate this composition is due to primordial conditions when and where these comets were formed and is not due to thermal heating after their arrival in the inner solar system.

## 1. INTRODUCTION

Comets are icy, minimally processed planetesimals left over from the formation of the Solar System. The ices in their nuclei store chemical remnants from the outer solar nebula where they formed and, as the least thermally processed objects easily accessible for detailed study, any anomalies in their chemical composition naturally lead to questions regarding the origin of these differences. Over the last few decades, investigations of ensembles of comets have revealed that the majority of comets exhibit a very similar chemical composition in optical wavelengths, but, as more objects have been measured, a few anomalous compositional groupings have emerged (e.g. M. F. A'Hearn et al. 1995, U. Fink 2009, A. L. Cochran et al. 2012). One of the most distinct anomalous groups identified is the depleted comets, first characterized by M. F. A'Hearn et al. (1995) in their analysis of narrowband photometry of 85 comets, 41 of which were considered well-determined. This depleted group of 12 objects was defined as having anomalously low production rates of $C_2$ (and usually also $C_3$), i.e. "carbon-chain" molecules, with respect to CN when compared to nearly all other, i.e. "typical" comets. The prototype of this depleted class, and the first comet clearly identified as having strong depletion in production rates of both $C_2$ and $C_3$ with respect to CN, is the Jupiter-family comet 21P/Giacobini-Zinner (see D. G. Schleicher et al.1987, A. L. Cochran & E. S. Barker 1987, J. E. Beaver et al. 1990, M. F. A'Hearn et al. 1995, A. L. Cochran et al. 2020, and D. G. Schleicher 2022, among others).

We have continued to expand the narrowband photometry database of M. F. A'Hearn et al. since the publication of their summary results in 1995, and our most recent full database analysis includes all observations from this program, obtained over 45 years from 1976 to 2021. As of mid-2021 the full database contained a total of 220 comets, with a large subset of 135 comets having well-determined production rates of five gas species (hereafter referred to as the restricted subset) that were used for compositional analyses. We have found that 26 of the comets in the restricted subset have depleted composition according to the guidelines originally defined in M. F. A'Hearn et al. (1995), but with more than triple the original restricted subset (135 now vs 41 originally), we were able to perform a more detailed analysis and have now defined three categories of carbon-chain depletion: strongly carbon-chain depleted, the largest of the non-typical classes, including 21P/G-Z and 11 additional comets in the restricted subset that are *strongly* depleted in *both* $C_2$-to-CN *and* $C_3$-to-CN; moderately carbon-chain depleted, including eight comets that are *somewhat* depleted in *both* $C_2$-to-CN *and* $C_3$-to-CN; and moderately $C_2$ depleted, including six comets that are *somewhat* depleted in $C_2$-to-CN *only*.

The strongly carbon-chain depleted class is one of the two most distinct taxonomic groups to arise from this recent compositional analysis, the other being the extremely CN-depleted class containing a single comet, 96P/Machholz 1 (see D. G. Schleicher 2008). For the carbon-chain (i.e. *both* $C_2$-to-CN *and* $C_3$-to-CN) depleted comets, distinguishing a boundary between the strongly and moderately depleted groups was a nuanced exercise; yet throughout our analyses, these two groupings continued to separate from one another and so we treat them as two separate taxonomical groups here and in our forthcoming full database summary paper. The primary characteristic that distinguishes the strongly carbon-chain depleted class is simply their stronger depletions in the carbon-chain species; these comets stand out as lower endmembers of what is, for the most part, a continuum of carbon-chain depletion. A second distinction is that over half of the strongly carbon-chain depleted comets are also moderately to strongly depleted in NH, while no members of the other carbon-chain depleted classes exhibit NH depletion. Finally, note that in addition to the 12 strongly carbon-chain depleted comets in our restricted subset, we have five more comets in the broader database (discussed in more detail in Section 3) that exhibit clear evidence of being strongly depleted in both $C_2$-to-CN and $C_3$-to-CN; we are including an analysis of *all 17* comets exhibiting strong carbon-chain depletion here.

While an overview and summary of the chemical and physical characteristics of *all* comets in our database will be included in our forthcoming full database paper, given the breadth of that project we simply will not have the room to include every data point for each individual comet observed since 1976. Detailed analyses have previously been published for three members of the strongly carbon-chain depleted class including the prototype of the depleted comets, 21P/G-Z (D. G. Schleicher et al. 1987, D. G. Schleicher 2022), the fragmented comet 73P/Schwassmann-Wachmann 3 (S-W 3; D. G. Schleicher & A. N. Bair 2011), and 43P/Wolf-Harrington (D. G. Schleicher et al. 1993). This publication serves as a resource for all 14 of the additional comet datasets that went into forming this distinct taxonomical class that will not be published elsewhere. In this publication, we utilize observations from our narrowband photometry database to further explore the strongly carbon-chain depleted class of comets, specifically how they relate to one another and to our database as a whole. Details involving the instrumentation, observations, data reductions, and an overview of the restricted subset used to derive our taxonomic classifications are in Section 2, while an overview of the database and details on the photometric observations of each strongly carbon-chain depleted comet are in Section 3. A discussion of our findings is presented in Section 4, and concluding remarks are in Section 5.

## 2. INSTRUMENTATION, OBSERVATIONS, REDUCTIONS, AND ABUNDANCE RATIOS

### 2.1 *Instrumentation and Observations*

In this paper we include detailed information on the strongly carbon-chain depleted comets, observed between 1977 and 2021, from the long-term photometric survey at Lowell Observatory; a summary list of the comets, their perihelion distances and dates, and the heliocentric distance ranges of our observations, are listed in Table 1. Observations were primarily completed at Lowell Observatory, and Perth Observatory was our primary southern location for supplementary observations, while the 88-inch (2.2-m) at Mauna Kea Observatory was used for a few observations for one comet; specific information on observing circumstances and the location and telescope used for each night of observations is included in Table 2. All the observations were obtained using conventional photoelectric photometers that were equipped with EMI 6256 photomultiplier tubes sensitive to UV and visible wavelengths, with either pulse-counting or DC amplifier systems, along with narrowband comet filters. Three different filter sets were used over the course of these observations: the original A'Hearn and Millis (AM) filters (M. F. A’Hearn et al. 1979), the International Halley Watch (IHW) filters (M .F. A’Hearn 1991), and the HB comet filters (T. L. Farnham et al. 2000) designed and produced for Comet Hale-Bopp’s apparition. The same gas emission bands are isolated by the different generations of filter sets (OH, NH, CN, $C_2$, and $C_3$), but the continuum locations have changed. Specifically, the UV and green continuum filters are located at 3675 Å and 5240 Å for the AM set, at 3650 Å and 4845 Å for the IHW set, and at 3448 Å and 5260 Å for the HB set; an intermediate blue continuum filter was added with the HB filters at 4450 Å. Note that the $C_3$ filter was not added to the original AM filter set until late 1977, and the OH and NH filters were not added to this set until late 1979. The IHW set had no NH filter, so we continued to use the one from the AM set until all filters were replaced by the HB filters in early 1997. Finally, OH and NH filters were not used for observations of Comets G-Z and S-W 3 obtained with the 24-inch (0.6-m) Planetary Patrol telescope at Perth Observatory due to the site's low elevation and resulting high atmospheric extinction in the UV.

[Table 1. Summary list of the strongly carbon-chain depleted comets; page 40]

[Table 2. Photometry observing circumstances and fluorescence efficiencies; page 41]

The number of measurement sets obtained for each comet varies greatly, with between 2 and 125 photometric sets obtained per comet (Table 1). For each set, emissions from each of the five gas filters, in addition to 2-3 continuum, i.e. dust, filters were typically measured, although occasionally fewer filters were used. Several integrations lasting fifteen seconds to a few minutes were obtained for each filter, depending on the observing conditions and the brightness of the comet, and were either interspersed with or followed by sky measurements typically 10-20 arcmin away from the comet. Photometer entrance apertures ranged in diameter from 10 to 205 arcsec, and the projected aperture radius ($\rho$) at the distance of the comets ranged from 1400 to 148,000 km. Table 2 includes the observational circumstances for each night, including the heliocentric and geocentric distances, $r_H$ and $\Delta$, respectively, and the phase angle ($\theta$) for each comet. The UT mid-time for each set, as well as the logarithm of the projected aperture radius at the distance of the comet (log $\rho$) are included in Table 3. Comet flux standard stars were measured nightly over a range of airmasses to determine atmospheric extinction and absolute instrumental calibrations; since standard stars were normally measured over an airmass range bracketing the comet, the errors associated with extinction and calibration are almost always small compared to other uncertainties such as the comet's signal-to-noise (M. F. A'Hearn et al. 1995)

[Table 3. Photometric fluxes and aperture abundances; page 43]

### 2.2 *Reductions*

We used our standard data reduction methods as detailed in M. F. A'Hearn et al. (1995). Specifics for how we obtained fluxes from each filter set are detailed in T. L. Farnham et al. (2000) for the HB filters, and in T. L. Farnham & D. G. Schleicher (2005) for revised procedures for the IHW and original AM filter sets. All of the data were re-reduced in a consistent manner for this new full database analysis, with the revised reductions yielding improved decontamination of the UV and green continuum filters by the $C_3$ and $C_2$ emission band wings, respectively, as compared to the original reductions published for some of these comets. The improvements in our reduction methods result in slightly different answers when compared to some of our previously published data, however *the compositional classifications for each comet remain the same*. The resulting emission band flux values for each gas species following continuum subtraction, along with the continuum fluxes, are listed in Table 3 for each observational set. The fluxes for each gas species were converted into column abundances by applying the appropriate fluorescence efficiencies (*L/N*). For $C_2$ and $C_3$, the fluorescence efficiencies were scaled as $r_H^{-2}$ and are given in M. F. A'Hearn et al. (1995), while the OH values vary with heliocentric velocity ($\dot{r}_H$) due to the Swings effect and are from D. G. Schleicher & M. F. A'Hearn (1988). For CN and NH, the fluorescence efficiencies vary due to the Swings effect with $\dot{r}_H$ as well as with heliocentric distance, and, beginning with our analyses of 73P/Schwassmann-Wachmann 3 (D. G. Schleicher & A. N. Bair 2011), we have additionally incorporated the fluorescence computations for CN from D. G. Schleicher (2010) and for NH from R. Meier et al. (1998; also see D. G. Schleicher & A. N. Bair 2011). The resulting nightly OH, CN, and NH *L/N* values are listed in Table 2.

To extrapolate total coma abundances from the column abundances, we applied a standard Haser model (L. Haser 1957), and the gas production rates were computed by dividing by the assumed daughter lifetimes. The Haser parent and daughter scalelengths and daughter lifetimes, with all values assumed to scale as $r_H^2$, are from M. F. A'Hearn et al. (1995); the final column abundances (log $M(\rho)$) for each observational set are listed in Table 3 while the derived production rates (log $Q$) are in Table 4. Finally, since OH has only one parent molecule, $H_2O$, the most abundant volatile in comets, we converted our OH Haser production rates to vectorial equivalent water production rates using the empirical relation derived by A. L. Cochran & D. G. Schleicher (1993) and discussed in D. G. Schleicher et al. (1998). The resulting mean derived water production value for each observational set is in the last column of Table 4.

[Table 4. Photometric production rates; page 46]

We continue to use the quantity $A(\theta)f\rho$, a proxy for dust production, for each continuum filter. This value, listed in Table 4, is the product of the dust albedo at the given phase angle, the filling factor for the aperture, and the projected aperture radius. There are no assumptions regarding the particle size distribution, and the value is independent of wavelength and aperture size if the dust grains are gray in color and have a canonical $1/\rho$ radial distribution (M. F. A'Hearn et al. 1984). Finally, since comets exhibit significant variations in scattering efficiency with phase angle, and because our viewing geometry can vary markedly both for individual comets throughout their apparition and from one comet to another, we apply a first-order phase angle adjustment to our $A(\theta)f\rho$ values using the Schleicher-Marcus composite phase curve and normalize to 0° phase, i.e. $A(0°)f\rho$, for use in certain purposes such as calculating the dust-to-gas ratio (see D. G. Schleicher & A. N. Bair 2011).

Since these objects have inherently low fluxes in the carbon-chain (and often NH) emission bands, many observations of these comets have high uncertainties, especially for $C_2$, $C_3$, and often NH. The uncertainties associated with each data point, included in Table 4, are based on photon statistics and reflect the 1-$\sigma$ values derived from the propagation of the observational uncertainties. If, after sky and continuum subtraction, a resulting emission band flux is negative, the tabulated value is listed as undefined since negative flux values are not possible.

### 2.3 *Abundance Ratios and the Restricted Subset of the Database*

The number of observations varies from comet to comet depending on a number of factors, including their inherent brightness and viewing geometry, and so for our compositional analyses we normalized the gas and dust production rates for each comet. This was done by calculating the production rate ratios (i.e. relative abundances) from the production rates ($Q$, listed in Table 4) for each minor species with respect to OH and to CN for each set of measurements, as defined in Section 2.1, for each comet. We then used the abundance ratios from each set to determine an unweighted mean value for each production rate ratio for each individual comet. In the cases where a measurement of NH, $C_2$, or $C_3$ went negative after continuum subtraction due to too low signal-to-noise, the ratio(s) from the measurement sets were included as a zero in the respective averages for the comets since an attempt at a measurement was made; we emphasize that there are no cases in which using this method resulted in a comet changing compositional class vs what it would have been had we omitted using that measurement entirely. Finally, we chose to do our averaging in linear space, although a reasonable argument could be made for averaging in logarithmic space, which gives slightly different numerical results.

Following a similar process as was used by M. F. A'Hearn et al. (1995), we developed a restricted subset of comets for use in determining our compositional taxonomy. Since our entire database at the time of these analyses contained 220 comets, including some with only one measurement set, high uncertainties, and/or measurements obtained at large heliocentric distances, a restricted subset was constructed in an effort to minimize the chances of poor or unrepresentative data being used in our compositional analyses. To be considered for inclusion in the restricted subset, a comet must have *at least three* measurement sets, *two* of which must include detections of *all five* of the gas species, and they must be obtained over at least *two nights* at heliocentric distances of *less than 3 au*. A few comets that nominally met these criteria were ultimately excluded from the restricted subset because either all detections of one of the gas species were from a single night, or because of high scatter and/or very large uncertainties. The resulting well-determined, restricted subset contains 135 comets – more than triple the subset of 41 comets used for analyses in A'Hearn et al. – whose mean production rate ratios were used to derive compositional classes through principal component analysis, cluster analysis, and visual assessments of ratio-ratio plots, discussed further in Section 4.1. Once we determined a classification system, we assigned the remaining, non-restricted comets to a compositional class when possible, again considering the number of detected species, the agreement of the measurements, and the uncertainties of the observations. Ultimately, we assigned compositional classes to 192 of the 220 comets; those that were not assigned to a class had too few observations or had measurements through too few filters, often combined with high uncertainties, that limited our confidence in their measurements. Several comets that were missing some filters entirely, having only been observed prior to 1980 or only from Perth, could not be assigned to compositional classes, though a few could still be classified when enough measurements of key gas species were available. This classification foundation allows us to assign classes to comets observed since the database was frozen in 2021 and has resulted in the discovery of one additional strongly carbon-chain depleted comet.

## 3. DATABASE OVERVIEW AND PHOTOMETRIC RESULTS

### 3.1 *The Data and an Overview of Compositional Classes*

As noted earlier, these strongly carbon-chain depleted comets are from Lowell Observatory's narrowband photometry database, a project that was started in 1976 and continues to the present. An analysis of the first 17 years of observations, which included 85 comets, was published in M. F. A'Hearn et al. (1995). In this original summary database publication, a restricted subset of 41 well-determined comets was used for taxonomical analyses, and 12 of those comets were determined to be depleted in $C_2$ with respect to both CN and OH, with most of the 12 additionally depleted in $C_3$. Most recently in mid-2021 we completed a re-analysis of all 45 years of photometry observations obtained since 1976, incorporating the observations included in the original M. F. A'Hearn et al. (1995) publication; at that time our full database contained 220 comets, 135 of which we consider well-determined and use for taxonomic analyses. These new analyses were completed using our updated filter calibrations and reduction methods described in Section 2.2, and results presented here supersede the values for comets that were published in A'Hearn et al.

Most of the 135 comets in the restricted subset of our database (see Section 2.3) have quite similar chemical compositions for the species that we measure and, following the definitions set in M. F. A'Hearn et al. (1995) we have placed these in the "typical" compositional class (102 comets; 76% of the of the restricted subset). The remaining comets fall into one of five other compositional categories, most of which include comets exhibiting carbon-chain depletion to varying degrees. One of the most distinct individual classes that emerged from our taxonomic analyses, and the one this publication is focused on, is the strongly carbon-chain depleted class. These 12 comets (9% of the restricted subset) are the largest non-typical compositional class and are *strongly* depleted in *both* carbon-chain species – $C_2$ *and* $C_3$ – with respect to both OH and CN, as compared to those in the typical compositional class, with some additionally depleted in NH with respect to OH and/or CN. An additional 8 comets (6%) in our restricted subset are *moderately* depleted in *both* carbon-chain species with respect to both OH and CN but show no depletion in NH (the moderately carbon-chain depleted class), while 6 comets are moderately depleted in $C_2$ *only* with respect to OH and CN and show no NH depletion (the moderately $C_2$ depleted class). This current classification system takes the single carbon-chain depleted class originally defined in A'Hearn et al. and subdivides it into the three categories of carbon-chain depletion just identified. This is possible due to our larger sample of carbon-chain depleted comets in the restricted subset (12 comets in A'Hearn et al. vs 26 now) and will be addressed further in our forthcoming analysis of the entire database. Finally, there are five additional comets that are not in the restricted subset that are strongly carbon-chain depleted, and we are including them in our analyses and discussions here. These objects were not in the restricted subset for various reasons, as noted in their individual summaries below, but their compositional ratios place them solidly into the strongly carbon-chain depleted class.

We focus this paper on just the strongly carbon-chain depleted comets as they are the largest – and one of the most distinct – non-typical compositional classes to emerge from our taxonomic analyses. All 17 of these objects are distinguished by having mean [log $Q(C_2)$-log $Q$(CN)], hereafter denoted as $C_2$-to-CN, production rate ratios that are 4.2× to 30× lower than the mean typical value for this ratio and mean [log $Q(C_2)$-log $Q$(OH)], hereafter $C_2$-to-OH, production rate ratios that are 4.9× to 27× lower than the mean typical value. They also *all* exhibit $C_3$ depletion, with mean [log $Q(C_3)$-log $Q$(CN)], hereafter $C_3$-to-CN, production rate ratios being 2.8× to 15× lower than the mean typical value, and mean [log $Q(C_3)$-log $Q$(OH)], hereafter $C_3$-to-OH, production rate ratios 2.5× to 19× lower than the mean typical value. NH is additionally somewhat to strongly depleted in 10 of these comets, with mean [log $Q$(NH)-log $Q$(CN)], hereafter NH-to-CN, production rate ratios ranging from 1.1× above to 11× below the mean typical value, and mean [log $Q$(NH)-log $Q$(OH)], hereafter NH-to-OH, ratios ranging from 0.8× above to 11× below the mean typical value. Finally, the mean [log $A(0°)f\rho$-log $Q$(OH)] using the phase-adjusted green continuum, hereafter dust-to-gas, ratios for the strongly depleted comets fall within the same range as the typical group; the strongly depleted objects differ somewhat, however, in that all of them fall within the upper half of the range for all dust-to-gas ratios. The mean production rate ratios for these individual comets, and the ranges of ratio values for each relevant compositional class, are listed in Table 5, with comets included in the restricted subset denoted by an asterisk.

[Table 5. Mean abundance ratios for the strongly carbon-chain depleted comets and ranges in composition for relevant compositional classes; page 49]

### 3.2 *Gas, Water, and Dust Production Rates, and Chemical Compositions of the Individual Comets*

Here, we detail the overall behavior for each strongly carbon-chain depleted comet. First, we present an overview of our observations for each comet and then, in the cases when enough data are available, we examine variations in gas and dust production rates with heliocentric distance ($r_H$ dependence). We have plotted the log value of each production rate for each comet (found in Table 4) as a function of the log of the heliocentric distance in Figures 1, 2, 3, 4, and 5 (note that the specific comets shown in each figure were chosen to minimize overlapping data points). We follow with an analysis of seasonal variations and of water production, noting the peak derived water production rate for each comet and the heliocentric distance, pre- or post-perihelion, at which it was observed. We report the median effective active area (i.e. the surface area required to be active) for each comet's nucleus based on water production rates (see Section 2.2) and, when the nucleus size for a comet is available, we calculate and report the median active surface fraction; these water-related values are listed in Table 6. This is followed by notes on each comet's composition as compared to the typical compositional class for the key mean production rate ratios of $C_2$-to-CN, $C_3$-to-CN, and NH-to-CN, and the phase-adjusted dust-to-gas ratio (also listed in Table 5). Finally, when possible, we do a brief comparison to results obtained by other researchers. The periodic comets are presented first, in order of their numerical designations, followed by the single long-period comet. As appropriate, we also provide alternative designations, including that of the discovery apparition and those associated with the apparitions that we obtained observations. The 12 comets that are in the restricted subset are denoted with an asterisk in the following subsections and in the tables.

[Table 6. Active areas and active fractions; page 50]

#### *21P/Giacobini-Zinner (G-Z); (1900 Y1) (1984e) (1985 XIII)**

We have observed the prototype of the depleted comets, 21P/Giacobini-Zinner (hereafter G-Z), during four apparitions – 1985, 1998, 2011, and 2018 – and note that it was first identified as having an unusual spectrum by N. T. Bobrovnikoff (1927). A thorough analysis and discussion of this comet was recently published in D. G. Schleicher (2022), so we only provide an overview here. A member of the Jupiter-family (JF), G-Z has an orbital period of 6.6 years and perihelion distances ($q$) between 1.01 and 1.03 au during our observations. We have 125 measurement sets for this comet, the most for any of our strongly depleted objects, ranging between 1.02 to 1.82 au pre-perihelion and 1.01 to 2.29 au post-perihelion (Figure 1). Its heliocentric distance dependence is very nonlinear and it reaches peak production rates about a month before perihelion, however, so slope computations for $r_H$-dependencies are used only for bulk comparisons between species for this comet (see Section 3.1 of D. G. Schleicher 2022). The slopes, combining all pre- and post-perihelion data, are similar for the carbon-bearing species (between -2.3 ± 0.4 and -3.0 ± 0.3 in log-log space), and somewhat steeper for OH and NH (-4.2 ± 0.4 and -4.5 ± 0.5), respectively.

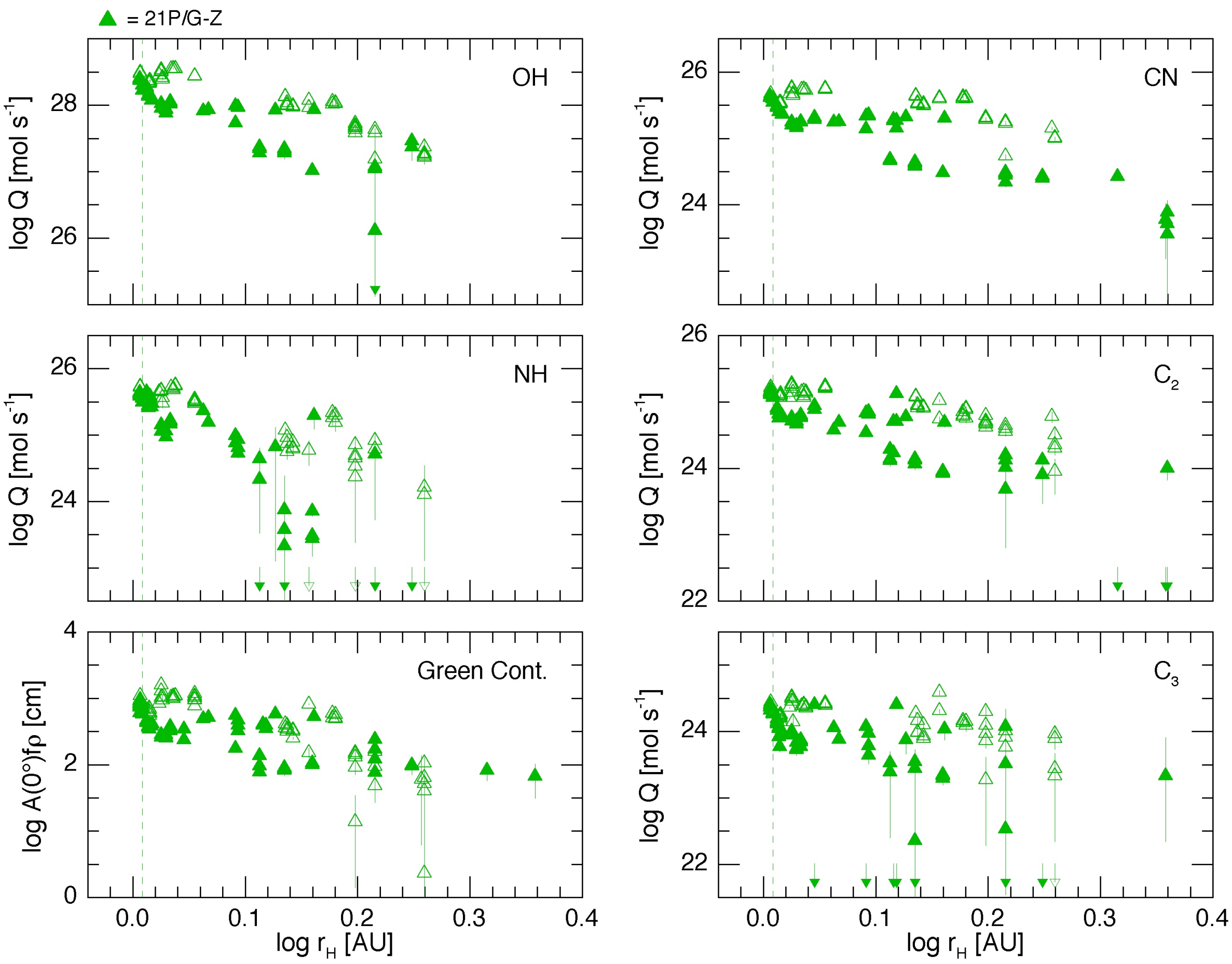

**Figure 1.** Logarithmic production rates for comet 21P/Giacobini-Zinner as a function of the log of the heliocentric distance. Open symbols represent data obtained before perihelion, while filled symbols are used for post-perihelion data. Downward arrows near the bottom of a panel indicate measurements that were attempted but went to zero after sky and continuum subtraction, and are listed as "und", or undefined, in Table 4. The vertical dashed line represents the perihelion distance and is color-coded to match the comet.

Dust is intermediate between the carbon-bearing species and OH at -3.5 ± 0.3, a trend similar to what is seen in many comets (and further discussed below in Section 4.4). It exhibits strong asymmetry associated with its peak production occurring 3-5 weeks before perihelion, with values that are 3× to 4× higher than at the same distance following perihelion. Our peak water production value, $4.7 \times 10^{28}$ mol $s^{-1}$, was obtained 25 days before perihelion during its 1985 apparition at 1.09 au. By three days before perihelion, its water production rate had dropped by a factor of about 1.5 down to $2.8 \times 10^{28}$ mol $s^{-1}$, consistent with results from L. A. McFadden et al. (1987) who found water production to peak approximately one month before perihelion, then decrease by roughly a factor of two by six days after perihelion. Using our water data, the median effective active area is 6.5 $km^2$, and, using an effective nucleus radius of 1.82 ± 0.05 km (J. Pittichová et al. 2008), its median active fraction is about 36%.

Its mean log production rate ratio of -0.52 for $C_2$-to-CN is at the upper border of the strongly carbon-chain depleted class (a value that is still 4.2× below the mean typical value!), while its log $C_3$-to-CN ratio of -1.29 falls in the midst of this group and is 5.9× below mean typical. The log NH-to-CN ratio of -0.20 is also depleted for G-Z, at 3.3× below the mean typical value. Its mean dust-to-gas ratio, based on log $A(0°)f\rho$– log $Q$(OH), is -25.3 cm s molecule$^{-1}$ and is near the lower end for comets in the strongly depleted class, but in the midrange for comets overall in our database. Our findings on G-Z's seasonal behavior and composition agree with results obtained by many others (e. g. A. L. Cochran & E. S. Barker 1987, L. A. McFadden et al. 1987, J. E. Beaver et al.1990, L.-M. Lara et el. 2003, M. R. Combi et al. 2011, A. L. Cochran et al. 2020, Y. Moulane et al. 2020), and Dello Russo et al. (2016) note that infrared measurements of $C_2H_2$ for this comet indicate it likely would fall into their hydrocarbon poor group.

*31P/Schwassmann-Wachmann 2 (S-W 2); (1929 B1) (1979k) (1981 VI) (1986h) (1987 XIX)**

Our observations of the JF comet 31P/Schwassmann-Wachmann 2 (hereafter S-W 2) are from three successive apparitions: 1981 (period = 6.5 yr, $q$ = 2.14 au), with two measurement sets; 1988 (period = 6.4 yr, $q$ = 2.07 au), with one set; and 1994 (period = 6.4 yr, $q$ = 2.07 au), with 8 sets; no data were obtained on subsequent passages due to an orbital perturbation following its 1994 apparition which increased its perihelion distance to 3.4 au. Our pre-perihelion observation range is 2.08 to 2.27 au, while post-perihelion it is 2.14 to 2.52 au. The signal-to-noise ratio for S-W 2 (blue circles in Figure 2) is overall quite low, and several species went undetected on some attempts. Nevertheless, our data suggest there is no offset between the pre- and post-perihelion production rates, nor between apparitions, but with the large amount of scatter it is difficult to say with certainty. Our peak water production rate of $1.0 \times 10^{28}$ mol s$^{-1}$ is from 71 days before perihelion at 2.20 au. Its median active area is 3.6 km$^2$ and, using a nucleus radius of 1.65 km (Y. R. Fernandez et al. 2013), the median active surface fraction is about 10%.

The mean log $C_2$-to-CN ratio for S-W 2 is among the lowest at -1.04 (14.2× below mean typical), while log $C_3$-to-CN is -1.33 (6.4× below mean typical) and it exhibits no NH depletion. Its log dust-to-gas ratio is -24.7 cm s molecule$^{-1}$, on the higher end for members in this class as well as for the overall restricted database. Finally, A. L. Cochran et al. (2012) found this comet to be $C_2$-to-CN depleted, but not depleted in $C_3$-to-CN, though one of their two $C_3$-to-CN measurements does fall just below their threshold for depletion for this ratio.

*43P/Wolf-Harrington; (1924 Y1) (1984g) (1984 XVII) (1990e) (1990 V)**

Results from our 1984 and 1990-1991 observations of the JF comet 43P/Wolf-Harrington were published in D. G. Schleicher et al. (1993); we now have two additional apparitions and updated reduction techniques, so while specific values originally reported have changed, the fundamental conclusions are confirmed. Wolf-Harrington (violet squares in Figure 3) had an orbital period of 6.5 years during our observations, and our 30 measurement sets come from four consecutive apparitions: 1984 ($q$ = 1.62 au), 1991 ($q$ = 1.61 au), 1997 ($q$ = 1.58 au), and 2004 ($q$ = 1.58 au); note that poor viewing geometry and an orbital change prevented observations during subsequent passages. Our 11 pre-perihelion observations are from the 1991 and 2004 apparitions, covering a heliocentric distance range of 1.62 to 1.95 au, while our 19 post-perihelion observations occurred during the 1984 and 1997 apparitions and cover a range of 1.58 to 1.87 au. Its coverage is only moderate, and there is a fair amount of scatter, especially before perihelion and for $C_2$ and $C_3$. We can, however, examine its bulk behavior post-perihelion for OH, CN, and dust, where OH has the steepest $r_H$-dependence with a slope in log-log space near -6, CN is somewhat shallower at about -5, and dust is near -2. Our measurements suggest its production rates are about 1.4× higher before perihelion for OH, CN, and $C_2$ but large uncertainties, especially for NH, make the true amount of offset unclear; it is also difficult to tell if Wolf-Harrington has shown any changes in production rates between apparitions. Its peak water production value is from our single data point during its 1984 apparition and seems anomalously high; we therefore use the next highest value of $5.1 \times 10^{27}$ mol s$^{-1}$, our closest data to perihelion during its 2004 apparition, obtained 33 days before reaching perihelion. The median active area is 3.0 km$^2$ and, using a nucleus radius of 2.1 km (G. Tancredi et al. 2006), the active fraction is about 5%.

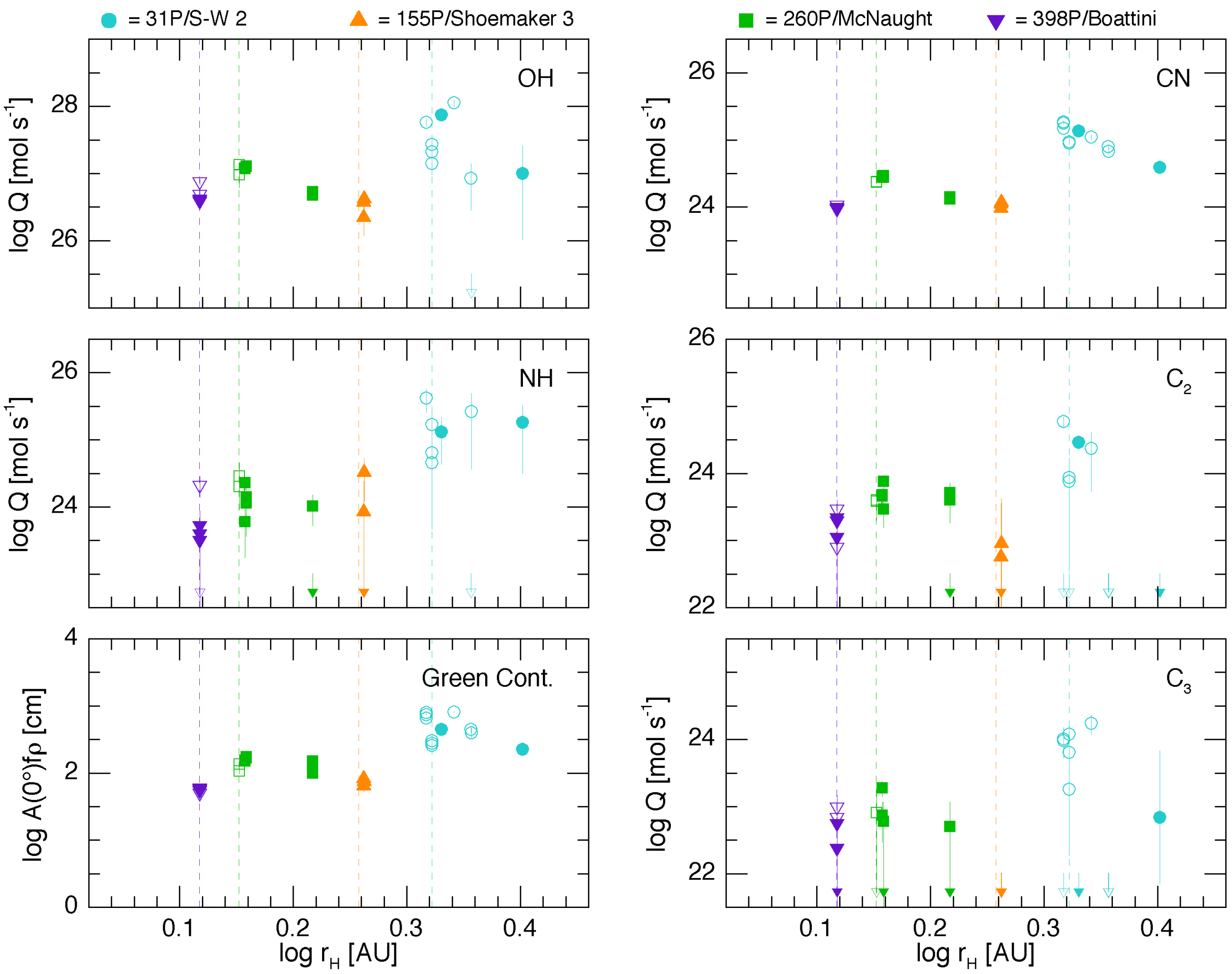

**Figure 2.** Logarithmic production rates for comets 31P/Schwassmann-Wachmann 2 (blue circles), 155P/Shoemaker 3 (orange triangles), 260P/McNaught (green squares), and 298P/Boattini (violet downward triangles). Further information same as in Figure 1.

Wolf-Harrington is among the most depleted comets in both its log $C_2$-to-CN and $C_3$-to-CN ratios, with values of -0.94 and -1.48, respectively, 11.3× and 9.2× below mean typical. It is also somewhat low in NH, with log NH-to-CN at -0.18 (3.2× below mean typical). Its log dust-to-gas ratio of -25.0 cm s molecule$^{-1}$ is within the midrange for the strongly depleted comets. Both U. Fink (2009) and A. L. Cochran et al. (2012) have also published production rates for Wolf-Harrington. Fink, who does not measure $C_3$, placed Wolf-Harrington among the "Giacobini-Zinner type" comets (low in $C_2$-to-$H_2O$ and $NH_2$-to-$H_2O$); while Cochran et al. have upper limits for both $C_2$ and $C_3$ that, when ratioed to CN, place Wolf-Harrington within their $C_2$-to-CN and $C_3$-to-CN depleted class.

*57P/duToit-Neujmin-Delporte (1941 O1)*

Our observations of the JF comet 57P/duToit-Neujmin-Delporte (d-N-D), which has an orbital period of 6.4 yr and a perihelion distance of 1.73 au, are from its 2021 apparition when we obtained three measurement sets over two nights following perihelion, when the comet was at 1.73 and 1.77 au. Only one set of measurements is complete with observations of all five gas species, however, since some NH and $C_3$ measurements went negative after continuum subtraction; for these reasons, and because it was observed after we "froze" our database in 2021, this comet is not included in our restricted subset. Our data of the detectable species, however, are self-consistent, with ratios firmly in the strongly depleted class (orange, downward triangles in Figure 4). Our peak water production of $1.3 \times 10^{28}$ mol s$^{-1}$ was obtained 18 days after d-N-D's closest approach to the Sun. Its effective active area of 12 km$^2$, combined with an effective nucleus radius of 0.96 km (Fernández et al. 2013), suggest it was active over more than 100% of its surface, consistent with reports of the nucleus fragmenting near the time of our observations, along with a several magnitude brightening (T. Lister et al. 2022), and is consistent with our not having observed it at a prior, better apparition in 2002.

Its $C_2$-to-CN and $C_3$-to-CN log production rate ratios of -0.75 and -1.32 are 7.2× and 6.4× lower than the mean typical values, respectively. d-N-D is additionally depleted in NH, with a NH-to-CN log ratio of -0.61 (8.4× below mean typical). Its log dust-to-gas ratio is -24.9 cm s molecule$^{-1}$ and within the midrange for the strongly depleted comets. Measurements obtained by E. Jehin et al. (2021) during the same time frame give a $C_2$-to-CN ratio of $-0.86 \pm 0.19$, in excellent agreement with our results.

*73P/Schwassmann-Wachmann 3 (S-W 3); (1930 J1) **

The individual measurements and detailed results for the fragmented JF comet 73P/Schwassmann-Wachmann 3 (S-W 3), which has a perihelion distance of 0.93 au and an orbital period of 5.4 yr, are available in D. G. Schleicher & A. N. Bair (2011); we provide a summary here. We obtained nine measurement sets over several nights after perihelion in 1995, with our observations beginning approximately 6 weeks after its initial outburst and subsequent fragmentation (J. Crovisier et al. 1995, Z. Sekanina 2005); note we did not measure OH and NH during the 1995 apparition. The comet had poor observing circumstances for its return in 2001, but during its close approach to Earth in 2006 we observed four components of its fragmented nucleus. We obtained 31 datasets for its main component, C, thought to be the same body observed in 1995 and with which we intercompare the two apparitions; the log $r_H$ plots can be found in Figure 1 of Schleicher & Bair (2011). Our observations span a heliocentric distance range that is long enough to examine its behavior as it approached and receded from the Sun, and inbound we have sufficient coverage from only 2006 where $r_H$-dependences for its production rates vary somewhat between the species, with NH steepest at -5.3 ± 0.4, followed by OH and $C_2$ (-3.5 ± 0.1 and -3.0 ± 0.2), and finally CN and $C_3$ (-2.3 ± 0.1 and -2.1 ± 0.4). The slope for dust unexpectedly follows a non-linear behavior, with a steep increase early that levels out and begins to decrease before perihelion, likely caused by a change in its dust properties over time. Outbound, only the 1995 apparition has adequate data, and the slopes are steeper (between -5 and -12) as would be expected for a post-fragmentation decrease in activity. Like most Jupiter-family comets, S-W 3 exhibits a seasonal effect, in this case with production rates after perihelion being 20%-40% lower than they were before. Finally, since the 1995 observations were obtained soon after its initial outburst, production rates were 8× (CN) to 25× (dust) higher than measured in 2006. Its relatively stable $r_H$-dependence in 2006 suggests most of its water production is coming from nucleus ice sublimation rather than icy grains, and our peak water production rate of 1.7 × $10^{28}$ mol $s^{-1}$ was measured 19 days before perihelion at a distance of 0.98 au (we have no OH measurements from 1995). Its median active area is 2.7 $km^2$ and, using a nucleus radius of 0.53 km for component C (I. Toth et al. 2008) the median active fraction is 76%.

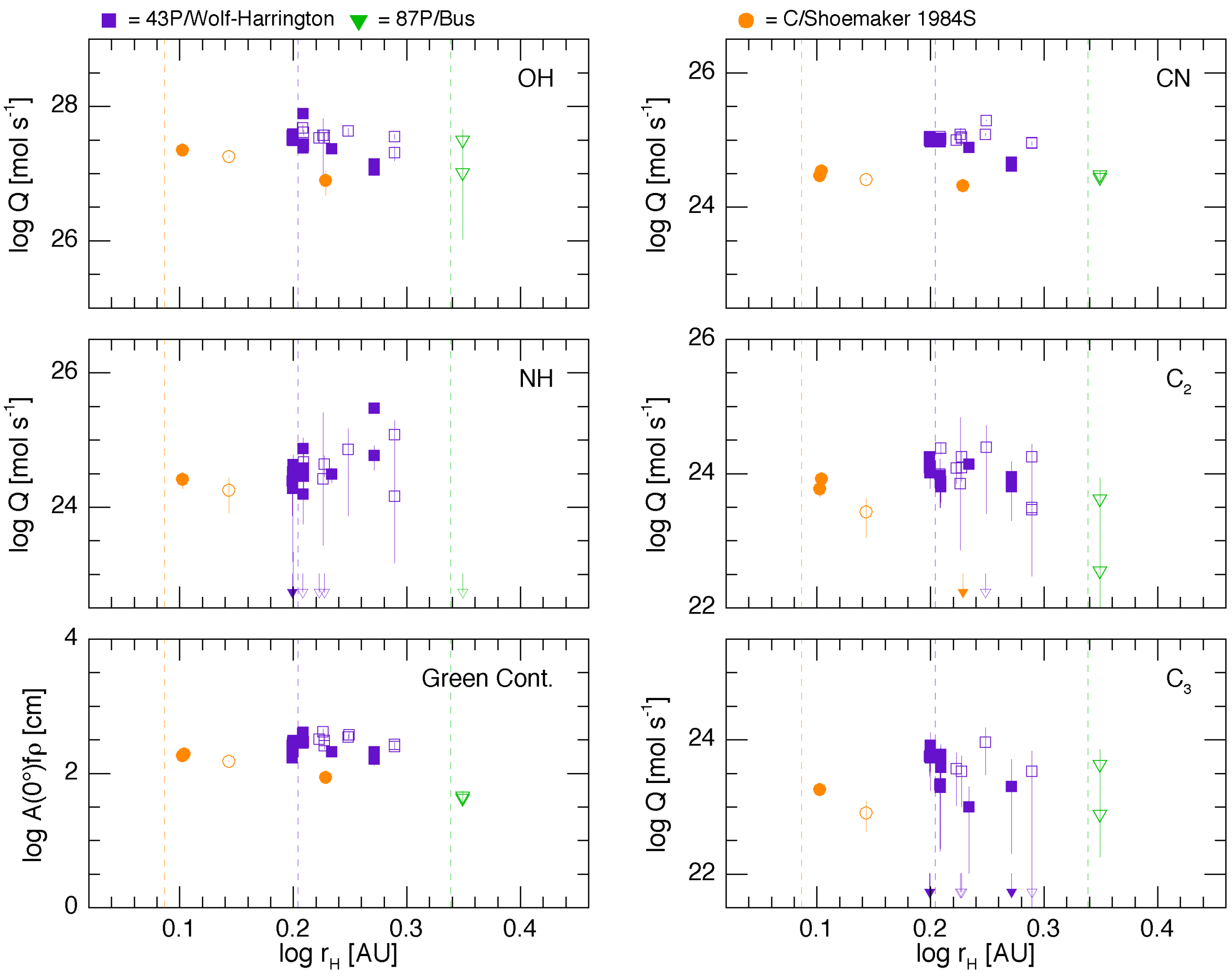

**Figure 3.** Logarithmic production rates for comets 43P/Wolf-Harrington (violet squares), 87P/Bus (green downward triangles), and C/Shoemaker (1984s) (orange circles). Further information same as in Figure 1.

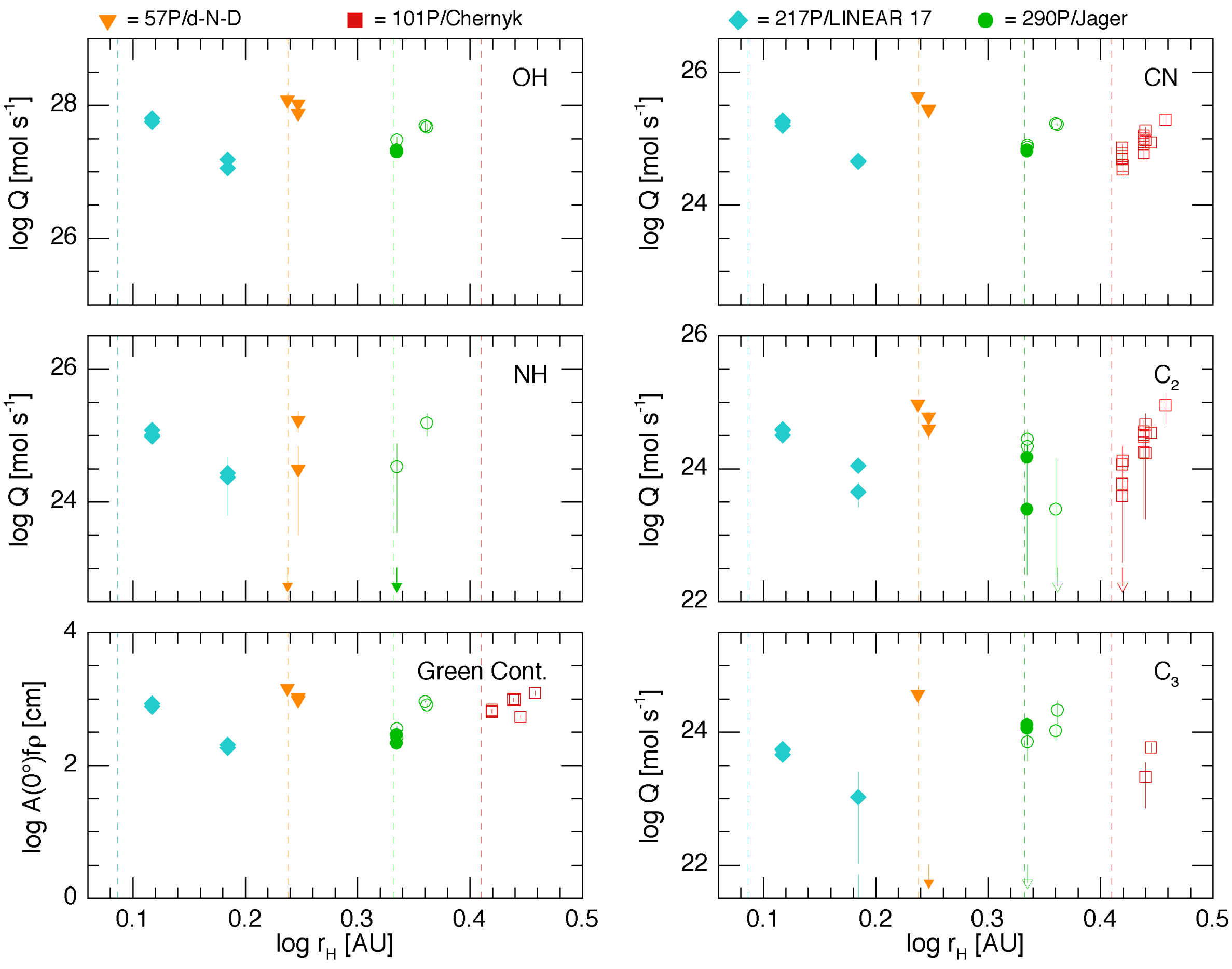

**Figure 4.** Logarithmic production rates for comets 57P/duToit-Neujmin-Delporte (orange downward triangles), 101P/Chernykh (red squares), 217P/LINEAR (blue diamonds), and 290P/Jager (green circles). Further information same as in Figure 1.

In addition to its main component, C, we successfully observed fragments B, G, and R in 2006. While we were able to get measurements for both B and G over several observing runs, both fragments experienced outbursts and we therefore made no attempt to determine their $r_H$-dependent fits; component R was quite faint and only detectable on a single night. See D. G. Schleicher & A. N. Bair (2011) for additional details about the gas (Section 3.1), water (Section 3.2), and dust (Section 3.4) production rates for these components.

Compositionally, combining 1995 measurements with those from the main component, C, in 2006, S-W 3 has a log $C_2$-to-CN ratio of -0.70, 6.5× below the mean typical value, and a log $C_3$-to-CN ratio of -1.53, 10.1× below mean typical; log NH-to-CN is somewhat low at -0.09, a factor of 2.6× lower than mean typical. Its log dust-to-gas ratio is -25.2 cm s molecule$^{-1}$, which is on the lower end for comets in this class. The composition of all four components measured in 2006 are consistent with one another, with all four fragments exhibiting strong carbon-chain depletion (D. G. Schleicher & A. N. Bair 2011, Figure 2 and Section 3.3). Its composition during our observations, which is dominated by freshly exposed material, is consistent with pre-fragmentation measurements obtained by U. Fink in 1990 (U. Fink 2009), *indicating the entire nucleus has a strongly depleted composition throughout*. Finally, our measurements agree with those made by other observers including I. Bertini et al. (2009), Croviser et al. 2009, and E. Jehin et al. (2022a, 2022b, 2022c), and infrared measurements by N. Dello Russo et al. (2007) show that S-W 3 components B and C by have similar, hydrocarbon poor composition, in addition to very low $NH_3$/HCN mixing ratios, indicating depletions in carbon-chain and NH parent species as well.

*87P/Bus (1981b) (1981 XI) (1981 E1)*

We have two sets of measurements, obtained on a single night, for Jupiter-family comet 87P/Bus. While not included in our restricted subset, the two CN measurements are nearly identical and both $C_2$ and $C_3$ are clearly depleted. Our observations are from 1981, when the comet had an orbital period of 6.5 yr, and were obtained at 2.23 au, 66 days before it reached its perihelion distance of 2.18 au (green, downward triangles in Figure 3). Its mean water production rate is $1.9 \times 10^{27}$ mol s$^{-1}$ with a derived median effective active area of 4.2 km$^2$. Its nucleus radius is 0.26 km (P. L. Lamy et al. 2011), which yields an active fraction of well over 100%, and we suspect our measurements were during an outburst given that it should have been too faint to successfully detect.

The log $C_2$-to-CN and $C_3$-to-CN production rate ratios for comet Bus are -1.12 and -1.07, respectively, which are 17× and 3.6× below the mean typical values for these ratios. We have no detectable NH measurements, so its NH abundance ratios are unknown. Its log dust-to-gas ratio is the lowest of the strongly carbon-chain depleted comets at -25.5 cm s molecule$^{-1}$.

*101P/Chernykh (1979l) (1978 IV) (1977 Q1)*

We obtained 14 measurement sets for the Jupiter-family comet 101P/Chernykh during its 1977 apparition when it had an orbital period of 15.9 yr and a perihelion distance of 2.57 au. Our data were acquired from 158 to 67 days before perihelion, between 2.87 and 2.63 au (red squares in Figure 4). Since this was before we acquired filters for OH and NH, we have no observations for these species, and while we have $C_2$ and CN measurements for all 14 sets, we only have two sets where $C_3$ data were obtained. Chernykh's log $C_2$-to-CN and $C_3$-to-CN production rate ratios are -0.61 and -1.35, respectively, which are 5.2× and 6.7× below the mean typical values. Though we have no OH observations, we can look at its dust-to-CN ratio as a proxy for comparison and find that Chernykh has the highest dust-to-CN ratio for the strongly carbon-chain depleted comets.

*114P/Wiseman-Skiff (1986 Y1)**

Our three sets of measurements for 114P/Wiseman-Skiff (W-S), a JF comet with an orbital period of 6.7 yr, were taken during its 2020 apparition, just after the comet reached perihelion at 1.58 au. Our data were taken on two nights within 10 days after perihelion (blue, downward triangles in Figure 5), and the peak water production during this time was $6.2 \times 10^{26}$ mol $s^{-1}$. Its water production corresponds to a median active area of 0.4 $km^2$ and, with a nucleus radius of 0.78 km (P. L. Lamy et al. 2009), its active fraction is quite small at just 0.6%.

Its log $C_2$-to-CN and $C_3$-to-CN production rate ratios are near the high end of the strongly depleted comets with values of -0.61 and -1.03, which are 5.3× and 3.2× below the mean typical values, respectively. Wiseman-Skiff has the lowest log NH-to-CN ratio of the strongly depleted comets at -0.72 (11× below mean typical), as well as a relatively low log dust-to-gas ratio of -25.3 cm s $molecule^{-1}$.

*123P/West-Hartley (1989 E3)**

The JF comet 123P/West-Hartley (West-Hartley) has an orbital period of 7.6 yr, and our 8 measurement sets are from its 2019 apparition when it reached a perihelion distance of 2.13 au. Our first two observations are from just 5 days before perihelion at 2.13 au, while our remaining six observations are from about one to four months after perihelion covering a heliocentric distance range of 2.13 to 2.30 au (orange squares in Figure 5). Our limited data suggest West-Hartley had somewhat higher production rates before perihelion for OH and CN by 1.3× and 1.2×, respectively, while NH, $C_2$, $C_3$, and dust show no definitive asymmetry. Our peak water production value of $1.3 \times 10^{27}$ mol $s^{-1}$ is from our single pre-perihelion night; its median active area is 1.4 $km^2$ and using the nucleus radius of 2.18 km from Y. R. Fernandez et al. (2013) its active fraction is about 2%.

Its mean log $C_2$-to-CN ratio is extremely low at -1.12, which is 17× below the mean typical value. West-Hartley is not nearly as depleted in log $C_3$-to-CN, but it is still low at -1.21, 5× below the mean typical value for this ratio, and it shows no depletion in NH. Its mean log dust-to-gas ratio is the highest of the strongly depleted comets (and among the highest in the restricted subset of the database) at -24.4 cm s $molecule^{-1}$.

*126P/IRAS (1983j) (1983 XIV) (1983 M1)**

126P/IRAS is a Halley-type comet with an unusually short orbital period of 13.2 years; its somewhat steeper inclination of 46° currently precludes it from being included in the Jupiter-family dynamical group. Our four sets of measurements are from its best, 1983 apparition, taken on three nights after it reached a perihelion distance of 1.70 au, and while the comet was at heliocentric distances spanning 1.71 to 1.99 au (violet diamonds in Figure 5). While its production rates for OH, CN, $C_3$, and dust are self-consistent, there is scatter for NH and $C_2$. The peak water production rate of $5.7 \times 10^{27}$ mol $s^{-1}$ is from our measurement closest to perihelion (+17 days and $r_H$ 1.71 au); its median effective active area is 4.5 $km^2$, and assuming a nucleus radius of 1.57 km (O. Groussin et al. 2004), approximately 14% of its surface is active.

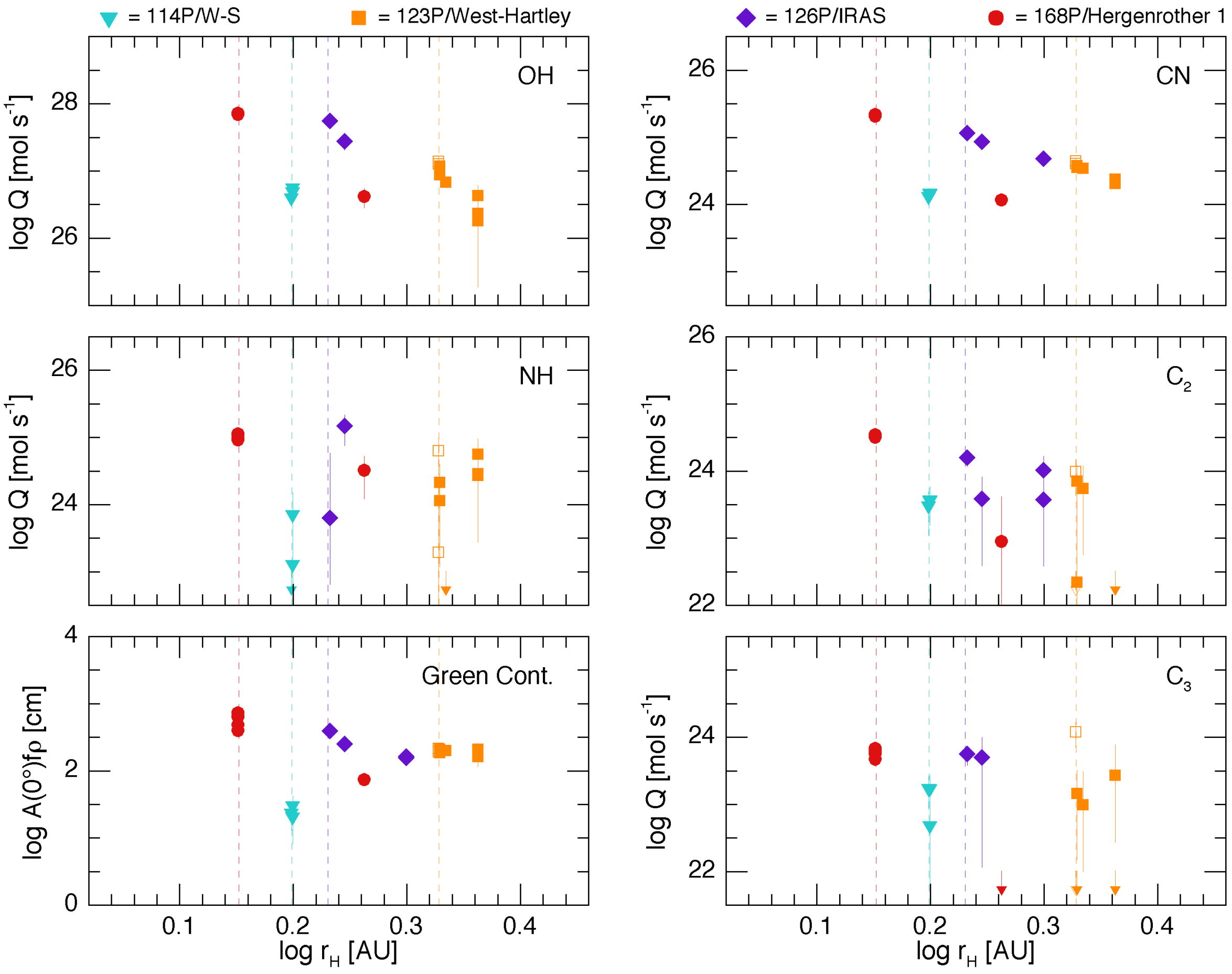

**Figure 5.** Logarithmic production rates for comets 114P/Wiseman-Skiff (blue downward triangles), 123P/West-Hartley (orange squares), 126P/IRAS (violet diamonds), and 168P/Hergenrother (red circles). Further information same as in Figure 1.

The mean log $C_2$-to-CN ratio for IRAS is among the lowest at -1.06, nearly 15× lower than the mean typical value, while its log $C_3$-to-CN ratio is -1.27 and 5.6× lower than mean typical. IRAS exhibits no NH depletion, and its mean log dust-to-gas ratio is among the lower values in this class at -25.1 cm s molecule$^{-1}$. A. L. Cochran et al. (1989, 2012) also observed IRAS during its 1983 apparition, and their spectra span a 3-month interval that almost entirely overlaps with ours. While they did detect a signal for CN, their observations of $C_2$ and $C_3$ yielded only upper limits that fall above the cut-off values for their depleted comet classes; given the circumstances, the authors did not designate a class for this comet.

*155P/Shoemaker 3 (1986 A1) (2002 R2)*

Comet 155P/Shoemaker 3 is represented by orange triangles in Figure 2. This JF comet has an orbital period of 17.1 years, and our observations are from its 2003 apparition when it reached a perihelion distance of 1.81 au. Our three data sets were obtained over two nights soon after perihelion at 1.83 au. Its production rates of gas and dust are among the lowest measured for the depleted comets, and all 3 of our attempts to measure $C_3$ were unsuccessful. Additionally, 1 out of 3 attempts to measure both NH and $C_2$ were unsuccessful. The uncertainties associated with our two remaining $C_2$ points are quite high, at over 350%. As such, although nominally meeting requirements for inclusion in the restricted subset for analyses, this comet was excluded. Its three CN measurements are self-consistent, however, and its relative $C_2$ production rates indicate it belongs in the carbon-chain depleted class, and its counts of $C_3$ relative to $C_2$ before continuum subtraction also suggest strong depletion in $C_3$. Its peak water production rate of $4.2 \times 10^{26}$ mol $s^{-1}$ is from our second night of data, 23 days after perihelion, and its median active area is only 0.5 $km^2$; there is no radius measurement for this comet's nucleus that we are aware of.

This comet has the lowest log $C_2$-to-CN ratio for the classified comets in our database at -1.37, a number 30× lower than the mean typical value, while it shows no NH depletion. The log dust-to-gas ratio is one of the highest in this class – and among the highest in the restricted database – at -24.6 cm s $molecule^{-1}$.

*168P/Hergenrother 1 (1998 W2)*

Our observations of 168P/Hergenrother 1, a JF comet with an orbital period of 6.9 years, are from its best, 2012 apparition. We have five measurement sets from when the comet was at 1.42 au, just 8 days after reaching perihelion. As is evident in Figure 5 (red circles), our observations, though only from a single night, are self-consistent for all measured species and the comet is clearly strongly carbon-chain depleted. Our OH data yield a water production rate of $8.1 \times 10^{27}$ mol $s^{-1}$ and the derived median effective active area is 5.1 $km^2$. Reports from others indicate this comet had three outbursts, starting approximately 3 weeks before our observations began and then continuing for several weeks after, which resulted in nucleus fragmentation and 6 detected companions (G. Sostero et al. 2012; E. A. Mueller et al. 2012; C. Hergenrother et al. 2012). As such, our water values are certainly elevated above its baseline production rates. We could not locate any nucleus size measurements for Hergenrother 1 following its 2012 fragmentation, though if we use the value of 0.48 $km^2$ obtained by Y. R. Fernandez et al. (2013) from 2007, our water production rates yield an active fraction of about 175%. Note this comet was not recovered during its 2019 apparition, suggesting a major change or disintegration occurred after it was last observed in 2012 (Q. Ye et al. 2019).

The mean log $C_2$-to-CN ratio for Hergenrother 1 is -0.80, a value 8.2× below mean typical, while its mean $C_3$-to-CN ratio is 11× below the mean typical value at -1.55. It is additionally somewhat depleted in NH, with a mean log NH-to-CN ratio of -0.31 (4.3× below mean typical). Its log mean dust-to-gas ratio is -25.1 cm s $molecule^{-1}$. Compositional measurements from D. M. Pierce & A. L. Cochran (2021), taken within days of our observations, are in excellent agreement and show strong depletion in $C_2$ and $C_3$ as well as depletion in $NH_2$.

*217P/LINEAR (2001 MD7) (2009 F3) (LINEAR 17)**

A JF comet with an orbital period of 7.8 years, 217P/LINEAR is represented by blue diamonds in Figure 4. Our five sets of measurements, taken during two nights following perihelion at heliocentric distances of 1.31 and 1.53 au, are from its 2009 apparition when it reached a perihelion distance of 1.22 au. The data closest to perihelion, at 1.31 au, are in close agreement with one another and there is a steep drop-off in production rates by the time of our next observations at 1.53 au, where there is a larger discrepancy between the individual $C_2$ and $C_3$ data and uncertainties are higher (note that one $C_3$ point is so low that it is off of the scale in Figure 4). Our coverage in distance is only moderate, and the $r_H$-dependent slopes are extremely steep for OH and NH (nearly -10 and -9 in log-log space, respectively), and slightly less so for CN (near -8). Its $C_3$ and $C_2$ slopes appear similarly steep but scatter prevents us from extracting a reasonable value, and dust is unusually steep during this short time frame, near -9. These steep $r_H$-dependencies for the gas and dust are likely the result of our first measurements occurring within days of a documented outburst (Y. Sarugaku et al. 2010) and our last measurements being more than a month after the dust grains produced by the outburst had presumably mostly cleared the aperture, explaining the steep drop-off between measurements. Our peak water production of $7.5 \times 10^{27}$ mol $s^{-1}$ is from our first night of observations, 36 days after perihelion at an $r_H$ of 1.31 au, and the median effective active area is 3.4 $km^2$. There is no nucleus measurement for LINEAR that we are aware of.

The log $C_2$-to-CN ratio for LINEAR is -0.71, 6.6× below mean typical, and its log $C_3$-to-CN ratio is among the lowest in our database at -1.64, a value 13.3× lower than mean typical. Its NH-to-CN ratio of -0.23 is 3.5× below mean typical, and its log dust-to-gas ratio is -24.9 cm s $molecule^{-1}$.

*260P/McNaught (2005 K3)**

Our observations of 260P/McNaught, a JF comet with an orbital period of 7.0 yr, are from its 2019 apparition when it reached a perihelion distance of 1.42 au. We have two pre-perihelion measurement sets at 1.42 au, about a week before the comet reached perihelion, and 7 post-perihelion sets from 1.44 au to 1.65 au (green squares in Figure 2). Its moderate coverage in heliocentric distance after perihelion allows us to compare differences in its $r_H$-dependence for OH, CN, and dust ($C_2$ and $C_3$ have too much scatter for a meaningful comparison). The slope for OH is nearly -7 in log-log space, while CN is somewhat shallower near -6; dust is shallowest with a value near -2. Examining the production rate measurements surrounding perihelion, $C_2$, CN, and dust are somewhat higher after perihelion than they were before, while it is difficult to discern if there are any pre-post asymmetries for the remaining species. By the time McNaught reached 1.65 au after perihelion, production rates for OH and CN had dropped by factors 2 or more; any trends for the other species are difficult to discern due to scatter. Our peak water production rate of $1.5 \times 10^{27}$ mol $s^{-1}$ is from nine days before perihelion, and the comet's effective active area is 0.9 $km^2$. Assuming a nucleus radius of 1.54 km (Y. R. Fernandez et al. 2013), McNaught is active over about 3% of its surface.

Its log $C_2$-to-CN ratio is 6.8× below mean typical at -0.72, and McNaught has the lowest log $C_3$-to-CN ratio of all the comets measured in our database at -1.68 – a factor of 14.6× lower than the mean typical value. For log NH-to-CN, this comet is 3.5× below mean typical at -0.23, and its log dust-to-gas ratio is -24.8 cm s molecule$^{-1}$. Spectra obtained by D. M. Pierce and A. L. Cochran (2021) also show McNaught to be $C_2$-to-CN depleted, though their NH-to-CN and $C_3$-to-CN data points are within their typical ranges.

*290P/Jager (1998 U3) (2013 N1)**

Our six measurement sets for the Jupiter-family comet 290P/Jager are from its successive 1999 and 2014 apparitions. This comet has a 15.1 yr orbital period with a perihelion distance of 2.15 au; four of our datasets are from before perihelion, ranging 2.30 to 2.16 au, while the other two were obtained after perihelion at 2.16 au. The 2014 data were taken at approximately the same heliocentric distance of 2.16 au on each side of perihelion, and its pre-perihelion production rates are around 1.2× to 1.5× higher for OH, CN, and $C_2$ (green circles in Figure 4), while $C_3$ and dust appear symmetric. Our peak water production measurement of $4.5 \times 10^{27}$ mol s$^{-1}$ was taken 88 days before perihelion during its 1999 apparition, and its median effective active area is 4.8 km$^2$. There is no radius measurement for Jager that we are aware of.

Its log $C_2$-to-CN ratio is 8.4× below the mean typical value at -0.81, while it has the highest log $C_3$-to-CN ratio among the strongly depleted comets at -0.96, a value 2.8× lower than mean typical. Its log NH-to-CN ratio is very low with a value of -0.56, which 7.6× below mean typical. For log dust-to-gas, Jager is in the mid-range for all comets at -24.8 cm s molecule$^{-1}$.

*398P/Boattini (2009 Q4)**

Our observations for 398P/Boattini, a JF comet now having an orbital period of 5.5 yr and perihelion distance of 1.31 au, are from two nights surrounding perihelion in late 2020 and early 2021, following a slow decrease in perihelia distances. We obtained a total of 5 measurement sets at 1.31 au: two from before perihelion and 3 from after (violet downward triangles Figure 2). Due to uncertainties and scatter, it is not clear if there is any asymmetry in the production rates about perihelion. Our derived peak water production rate of $9.1 \times 10^{26}$ mol s$^{-1}$ occurred 10 days before perihelion, and it has a median effective active surface area of just 0.3 km$^2$, the smallest in the strongly depleted class as well as one of the smallest in our entire database, but its fractional active area remains unknown.

Its mean log $C_2$-to-CN production rate ratio of -0.75 is 7.2× lower than the mean typical value, while its log $C_3$-to-CN value of -1.32 is 6.4× lower than mean typical. Boattini is additionally depleted in NH-to-CN, with a log ratio of -0.19 which is 3.2× below mean typical; its mean log dust-to-gas ratio is -24.9 cm s molecule$^{-1}$. Upper limits for the $C_2$-to-CN ratio obtained by E. Jehin et al. (2020) corroborate this comet's strong $C_2$-to-CN depletion.

*C/Shoemaker (1984s) (1985 II) (1984 U2)**

C/Shoemaker (1984s) is an old long-period comet with a relatively short orbit of 270 years and a low inclination of 14°; our four sets of measurements were obtained during its 1985 apparition when it reached a perihelion distance of 1.22 au. Our observations are from before perihelion at 1.39 au and after perihelion from 1.27 to 1.69 au. The production rates (orange circles in Figure 3) are overall quite low, with one of our attempted $C_2$ measurements going undetected, and our data show no obvious asymmetry. The peak water production rate is $2.7 \times 10^{27}$ mol $s^{-1}$, measured 23 days after perihelion at 1.27 au, and the median active area is 1.3 $km^2$.

The log $C_2$-to-CN ratio for Shoemaker (1984s) is -0.87, 9.5× below mean typical, while log $C_3$-to-CN is -1.33, 6.5× below mean typical. It is additionally depleted in NH, with log NH-to-CN at -0.11, 2.7× below mean typical. Its log mean dust-to-gas ratio is -25.0 cm s molecule$^{-1}$. A. L. Cochran et al. (2012) obtained upper limits for this comet and also categorized it as depleted in both $C_2$-to-CN and $C_3$-to-CN.

## 4. DISCUSSION OF RESULTS

A few of the better-studied strongly carbon-chain depleted comets have dedicated publications, and this study aims to put them into context with the additional objects that exhibit similar compositions but do not have sufficient data to merit their own focused publications. All but one of these 17 objects are in short-period orbits and, like many periodic comets, most of these are inherently faint due to their generally smaller active source regions – a result of frequent passes into the inner Solar System. This becomes apparent when looking at the number of observations and the range of heliocentric distance covered for each of these objects, as even several comets that were observed over multiple apparitions have relatively few measurements. Just one of the comets is *not* in a short-period orbit: an old long-period (OLP) comet with an orbital period of 264 years that is in a low-inclination orbit and has similar characteristics to the short-period objects. Almost all of these comets have observations near perihelion, when they were likely brightest, while about one-quarter of them have adequate heliocentric distance coverage to identify trends with production rates as they approached and receded from the Sun. About half of these comets were observed both before and after perihelion, with most exhibiting seasonality in production rates, being higher on one side of perihelion than the other. The peak water production values reached by these strongly depleted comets are generally low, but still within the range of values we have measured for the comets with typical composition, while for mean dust-to-gas ratios they fall within the upper portion of the range covered by the typical comets. *All of them are strongly depleted in both their mean $C_2$-to-CN and $C_3$-to-CN abundance ratios, as well as in their mean ratios of $C_2$-to-OH and $C_3$-to-OH.* About half additionally exhibit moderate to strong NH depletion, while the other half are in the middle to lower range of the mean NH abundance ratios measured for all comets in the restricted subset of our database.

### 4.1 *Gas Composition and Dust-to-Gas Ratios*

As the endmembers of the carbon-chain depleted comets, all 17 of these objects exhibit strong depletions in *both $C_2$ and $C_3$* with respect to CN and to OH (when measured), as compared to comets in the other compositional classes in our database (see Table 5). Their individual mean $C_2$-to-CN (followed by $C_2$-to-OH in parentheses) abundance ratios range from 4.2× to 17× (4.9× to 27×) *below* the mean value for the ratio for all typical comets in the restricted subset of our database, while the average $C_2$-to-CN depletion for the strongly depleted class as a whole is 8.2× (9.1×) below the mean typical value. For $C_3$-to-CN ($C_3$-to-OH), their mean abundance ratios range from 2.8× to 14× (2.5× to 19×) *below* the mean typical value for this ratio, while the average depletion is 5.9× (6.2×) below the mean value for the typical comets. This strong depletion is evident in Figure 6, where we show four key taxonomy plots for the mean abundance ratios of all comets in the well-determined restricted subset of our database (see Section 2.3), along with the additional strongly depleted comets from our broader database. The strongly carbon-chain depleted class is represented by dark blue triangles; those meeting the requirements for the restricted subset are filled, while the additional comets from our broader database are skeletal symbols. Many of the strongly carbon-chain depleted comets additionally have low NH-to-OH and NH-to-CN abundance ratios, with about half of them being low enough to exhibit moderate to strong NH depletion. Finally, as is evident in Figure 6, none of the carbon-chain depleted comets exhibit depletion in CN with respect to OH.

In some taxonomy plots in Figure 6, the strongly carbon-chain depleted class stands out quite clearly, while in other instances its separation from the adjoining moderately carbon-chain depleted class, especially along the borders, is more nuanced. The distinctions between compositional groupings are clearest in the plots of abundance ratios with respect to CN (upper and lower right panels). In the $C_3$-to-CN vs $C_2$-to-CN plot (lower right), all of the carbon-chain depleted classes are distinct, but it is also evident that carbon-chain depletion is a continuum with no clear boundary between those with moderate (light blue) and strong (dark blue) depletion when looking at just these ratios. When folding NH into the picture, however, the strongly carbon-chain depleted class as a distinct taxonomical subset becomes more apparent, as is seen in the NH-to-CN vs $C_2$-to-CN plot (upper right). Turning to the plots of abundance ratios with respect to OH (upper and lower left panels), the separation of strongly and moderately carbon-chain depleted comets is not always immediately as evident, with some moderately depleted members having $C_2$-to-OH and/or $C_3$-to-OH abundance ratios as low as some in the strongly depleted class. If one looks at the levels of depletion in reference to the diagonal along which the typical comets fall, however, the reasoning for which class was ultimately assigned to each comet becomes more evident. The carbon-chain depleted classes have minimal overlap in the $C_2$-to-OH vs CN-to-OH plot (upper left), where three moderately depleted comets (light blue) have $C_2$-to-OH abundance ratios similar to some in the strongly depleted class (dark blue). These particular strongly depleted comets additionally exhibit mild to moderate NH depletion, but the nearby moderately depleted comets have NH abundances in the typical range; this is one reason why these specific comets were distinguished through various analyses as belonging in the strongly carbon-chain depleted class.

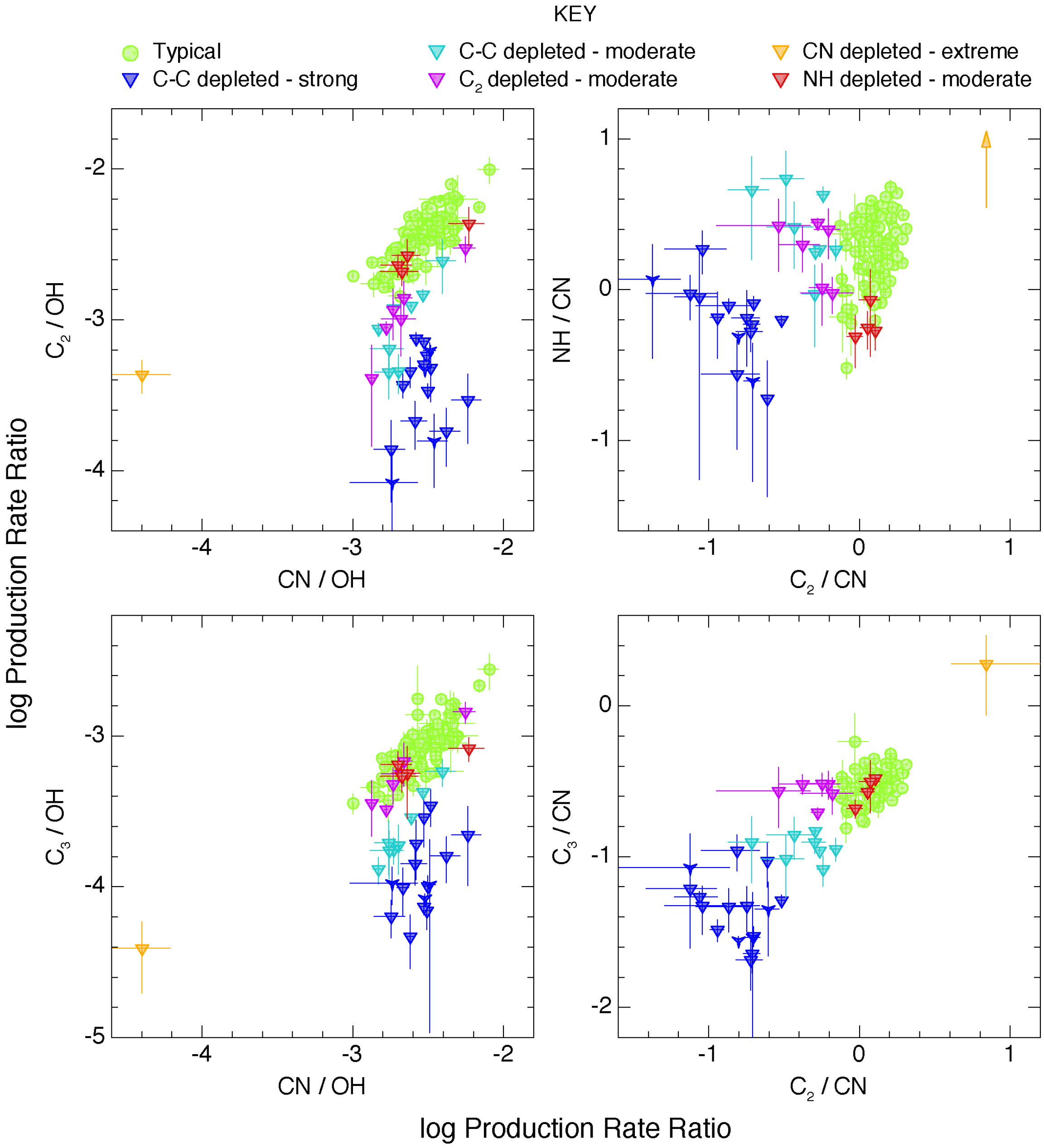

**Figure 6.** Key logarithmic abundance ratio plots from the restricted subset of our database, with the five additional strongly carbon-chain comets from our full database included and represented by skeletal symbols. Logarithmic mean production rate ratios of minor species with respect to OH and CN are presented and coded for our six compositional classes.

We emphasize that the selected ratio-ratio plots in Figure 6 are just part of the ensemble of analyses we completed and that we used multiple analyses to determine which comets were placed into which compositional categories. In addition to visually analyzing the ratio-ratio plots, we employed principal component analysis, which examines the total variance in the mean production rate ratios, and cluster analysis which looks at the degrees of similarity among the objects, to provide further insights. In Figure 6, it is evident that carbon-chain depletion is a continuum and there is no distinct division with the boundary lines that separate the depleted classes being somewhat arbitrary. The strongly and moderately depleted groups separated out from one another in multiple analyses, however, and we drew this delineation to distinguish the more strongly depleted members of this important compositional class. Additional plots will be shown and discussed further in our upcoming full database paper. *Finally, we note that nucleus sizes, the mean heliocentric distance of the observations, and the perihelion distance for a given comet have no correlation with these compositional classifications.*

Other researchers have also constructed compositional classification systems for comets, and the most useful ones for direct comparison here are the spectroscopic surveys in optical wavelengths published by A. L. Cochran et al. (1992, 2012) and U. Fink (U. Fink & M. D. Hicks 1996, U. Fink 2009), as these studies are quite extensive and measure several of the same or comparable species as we do, specifically $C_2$, CN, OH or its parent $H_2O$, NH or $NH_2$, and/or $C_3$. The parameters used for data reduction in each study, such as fluorescence efficiencies and Haser scalelengths, differ between the surveys so direct comparisons of their absolute values are not advisable; the relative offsets of abundance ratios are largely independent of these variables, however, and can be quite telling. Since these authors have not subdivided their depleted comets into moderately and strongly depleted classes, we make the following comparisons using all of our carbon-chain depleted comets, not just the strongly depleted ones. Of the 135 comets in our restricted subset with well-determined production rates, 54 are also in the restricted subsets of A. L. Cochran et al. (2012; 59 comets in their restricted subset) and/or U. Fink (2009; 50 comets in their restricted subset) including four from each of our strongly and moderately carbon-chain depleted classes. All eight of these comets that we identify as being carbon-chain depleted, whether strongly or moderately so, are identified as being depleted within the classification systems in the other surveys. In the case of the comets that we additionally identify as being NH depleted, there is also agreement among the surveys when NH and/or $NH_2$ data are available. The remaining 46 comets in common between these three restricted subsets are in our typical compositional class, and there is general agreement among the surveys on their classifications as well. In the few instances where we disagree with others, the comets are near borders where others have classified them as depleted and we have not.

Finally, looking at dust, we see that all comets exhibiting any level of carbon-chain depletion are in approximately the upper half of the dust-to-gas range of all 135 comets, as well as in the upper half of the range for all Jupiter-family comets within the restricted subset of our database. This agrees with findings from U. Fink (2009), who noted their low $C_2$ comets had noticeably higher dust-to-gas ratios than those with typical compositions. As with all of our observed comets, and especially those within the Jupiter-family, there is a strong correlation of dust-to-gas ratios with perihelion distance, with the objects having smaller perihelion distances exhibiting lower dust-to-gas ratios as compared to those that do not get as close to the Sun; this correlation with dust-to-gas ratios and heliocentric distance was also noted by M. A'Hearn et al. (1995) and Fink (2009).

Turning to the color of the dust, as a whole the dust grains exhibit very little color, with nearly all having a reddening per 1000Å around 10% which is comparable to most other JF comets in our database.

### 4.2 *Gas Composition with Heliocentric Distance*

When considering the distance from the Sun at which each comet in our full database was observed, *there is no correlation between heliocentric distance and carbon-chain depletion*. This holds true when examining the abundance ratios over a range of heliocentric distances for an individual comet, as well as when looking at the *mean* abundance ratios for each comet in regard to the mean heliocentric distance of its observations. This is important to emphasize since within our database, independent of taxonomical classification, we do see a clear systematic trend of decreasing $C_2$-to-CN with increasing distance from the Sun that is minimal or absent for the other abundance ratios. This is due to the differences in slopes with heliocentric distance for the species that we measure, with $C_2$ almost always having the steepest $r_H$-dependence of the carbon-bearing species. We know that most, if not all, of the larger change in $C_2$ production rates with heliocentric distance is an artifact of the basic 2-generation Haser model and its associated parameters used to compute the gas production rates. This model is intended for a simple parent/daughter relation, but $C_2$ has at least 3 parents ($C_2H_6$, $C_2H_2$, and $C_3$) as well as possible grandparents, and as a result the model does not successfully reproduce the spatial profiles for the parent molecules. A trend in $C_2$ abundance ratios with heliocentric distance was also noted by R. L. Newburn & H. Spinrad (1989) and R. Schulz et al. (1998), but neither U. Fink (2009) nor A. L. Cochran et al. (2012) noted an obvious trend in $C_2$ in their large databases; the choice of $C_2$ scalelengths is the likely source of differences seen between these studies. For a thorough discussion of Haser scalelengths and how a variety of commonly used scalelengths vary with heliocentric distance, see U. Fink & M. R. Combi's (2004) analysis where they found that they could bring the results of several investigators into reasonable agreement by using a common set of scalelengths.

Even with this trend involving the $C_2$-to-CN abundance ratio, there is a separation between the moderately and strongly carbon-chain depleted comets, and even more so between the typical and strongly carbon-chain depleted classes. Two of the strongly carbon-chain depleted comets, G-Z and S-W 3, have an adequate range of observations to examine their *abundance ratios* with heliocentric distance (Figure 7). Near perihelion, where we expect the highest $C_2$-to-CN ratios given the trend just discussed, both of these comets have values well below the boundary between the typical and carbon-chain depleted compositional classes, which occurs at approximately -0.13. Since carbon-chain depletion is a continuum, and several factors fold into which class we ultimately assigned to a given comet, there are a few individual measurements for both of these comets that overlap the range of the moderately carbon-chain depleted class. For S-W 3 (right panel), all log $C_2$-to-CN values are clearly strongly depleted and below -0.30 (a factor of 1.5× below the boundary between typical and depleted comets, and more than 2.5× below the mean typical value), and all but four of the 125 data points for G-Z are also below this same threshold. We note that for the individual measurement sets for the remaining 15 strongly carbon-chain depleted comets, every individual log $C_2$-to-CN value is at least a factor of 2.5× lower than the mean typical value for this abundance ratio.

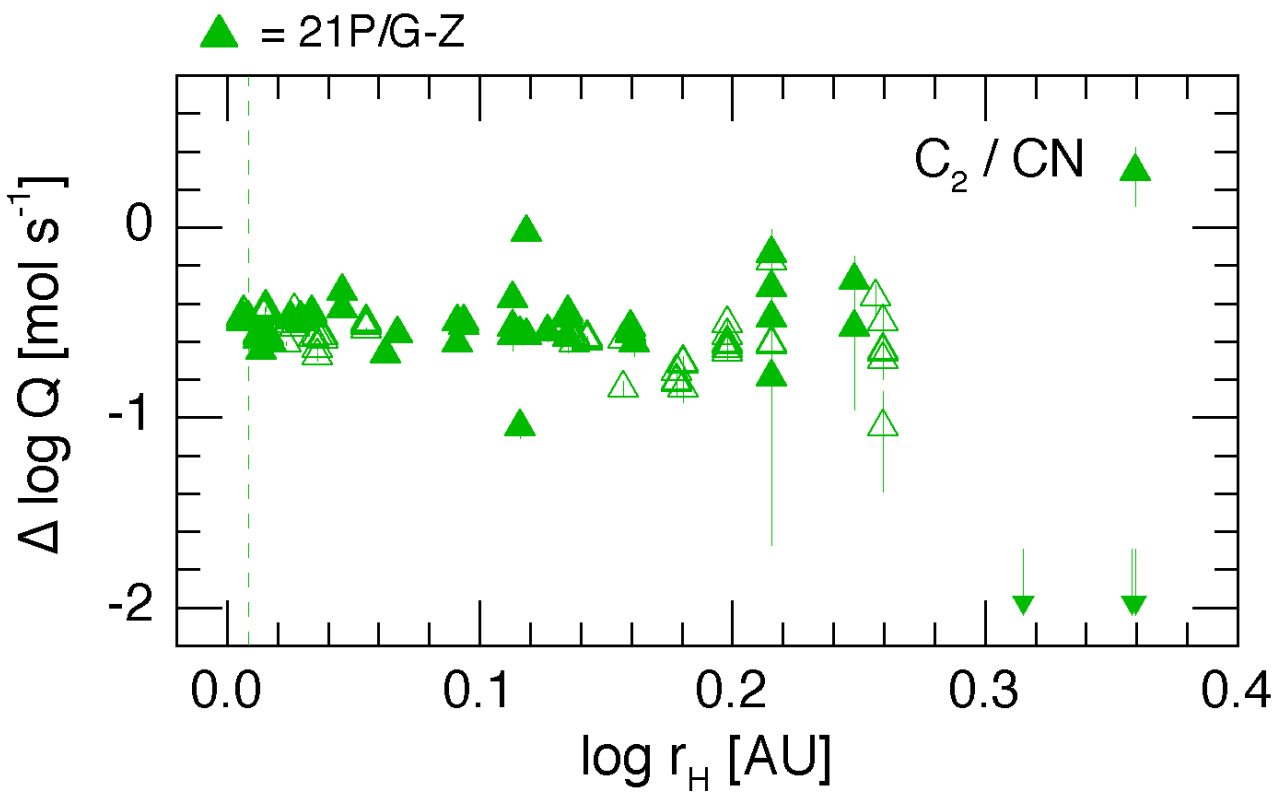

**Figure 7.** $C_2$-to-CN logarithmic production rate ratios vs heliocentric distance ($r_H$) for comets 21P/G-Z (top) and 73P/S-W 3 (1995 measurements of the fragmenting main nucleus + Component C from 2006; bottom). There is a trend of increasing C2-to-CN production rates with decreasing heliocentric throughout our database, independent of compositional class. Nearly all abundance ratio sets for G-Z, and all for S-W 3 – even at smaller heliocentric distances – are clearly well below the nominal boundary of approximately -0.13, above which we consider an abundance ratio to be typical.

The separation between compositional classes with heliocentric distance can be seen in Figure 8 where in the top panel we plot the *mean* $C_2$-to-CN values for every comet in our restricted subset against the mean heliocentric distance of its observations. We show $C_3$-to-CN for comparison in the lower panel, as we don't see a trend with this abundance ratio. The offset between typical and depleted comets in both plots is consistent, whatever the mean heliocentric distance of the observations. In conclusion, we emphasize that though there are some trends with heliocentric distance in our data, *none of our strongly carbon-chain depleted comets exhibit behavior that would potentially move them into the typical compositional class or even the moderately carbon-chain depleted class, as these heliocentric trends are small when compared to the amount of depletion that these objects exhibit.*

## 4.3 *Dynamical Origins*

Dynamically, nearly all of these strongly depleted comets are periodic objects that presumably come from the scattered disc region of the Kuiper belt. Sixteen of the 17 comets are on short-period orbits of less than 20 years, with 15 of these belonging to the Jupiter-family (JF) dynamical group (Tisserand Invariants with respect to Jupiter ($T_J$) > 2.0), as illustrated in Figure 9 where we plot mean production rate ratios of the comets in the restricted database as a function of $T_J$. Just one strongly carbon-chain depleted comet belongs to the Halley-type group ($T_J$ < 2.0) – comet 126P/IRAS with a relatively short orbital period for its dynamical group (13.2 yr) but a higher inclination of 46°, which nominally precludes it from being a JF comet with its resulting $T_J$ of 1.96. However, prior to its relatively close approach to Saturn in 1950, the value of $T_J$ was 2.01, and orbital integrations indicate that it has oscillated over the $T_J$ boundary separating the JF and Halley-type periodic comets in the past. The remaining object is the old long period comet C/Shoemaker (1984s) (1984 U2) which is on a 270 yr, low inclination (14°) orbit that reaches aphelion within the scattered disc at 82 au, possibly indicating a scattered disc origin for this comet as well. See A'Hearn et al. (1995), p. 224-225 for details on how the dynamical groups were determined.

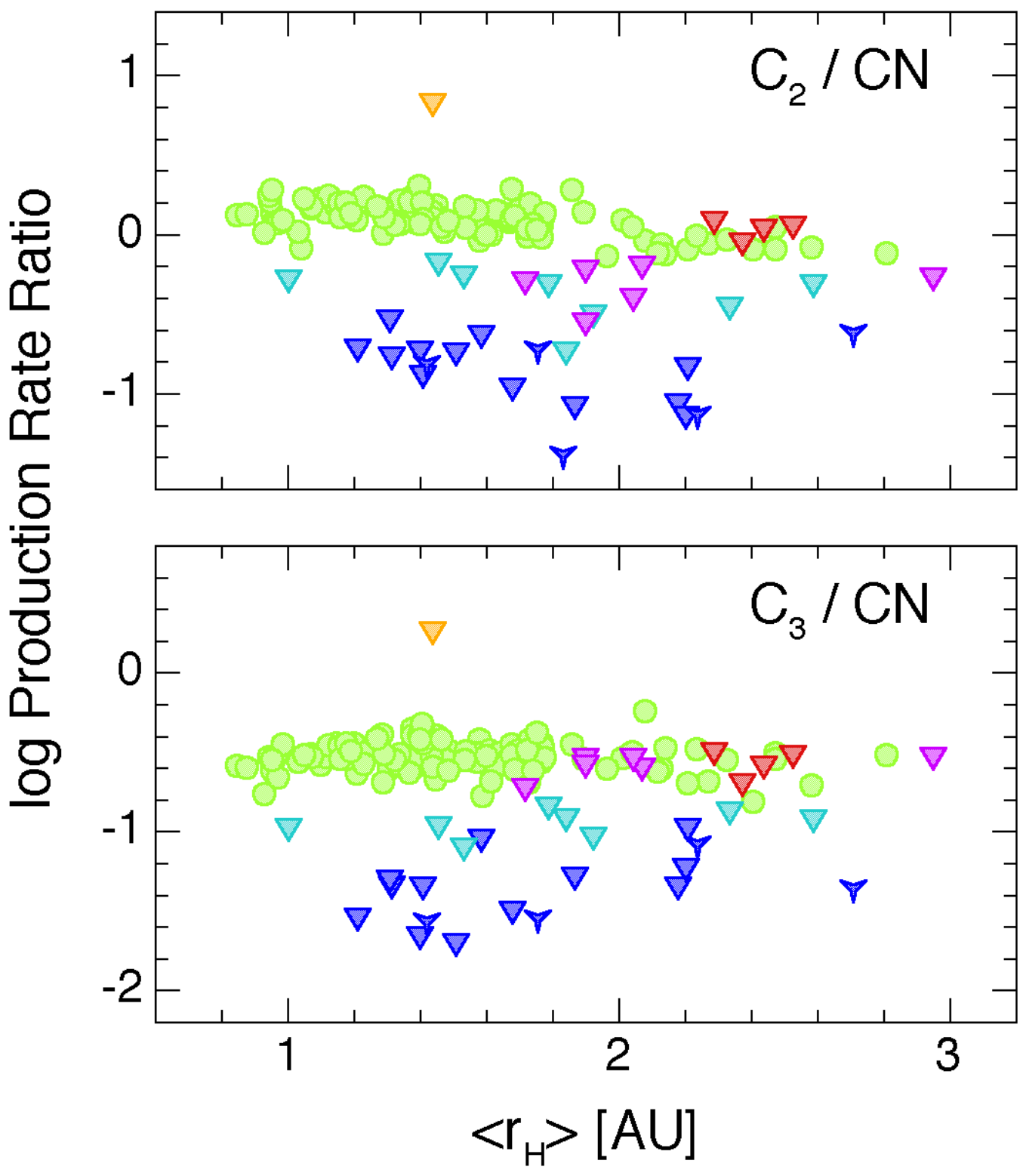

**Figure 8.** Logarithmic mean production rate ratios of $C_2$-to-CN and $C_3$-to-CN vs mean heliocentric distance ($r_H$). Colors and symbols are the same as in Figure 6. For C2-to-CN, between 1 and 3 au there's 0.3 change in the slope of the fit (log space) for comets with typical carbon-chain values, which is equal to almost a factor of 2. For the strongly depleted class, the progression is also present and is just slightly steeper with a 0.4 change in the fit, or a factor of 2.5. This progression is the result of Haser model parameters as discussed in the text and is not seen for $C_3$-to-CN. For both ratios, the offset between typical and strongly carbon-chain depleted compositional classes is clear at all heliocentric distances.

To put this in context with the other comets we have observed, we compare the dynamical origins of comets in the strongly carbon-chain depleted class to the objects in our restricted subset, and then to our entire database. Within the well-determined *restricted subset* of 135 comets (see Section 2.3), we find that 10 of the 47 JF comets (21%) are strongly carbon-chain depleted, but just one each of the 12 Halley-type (8%) and 24 old long period (4%) comets exhibit strong carbon-chain depletion, while none of the 31 young long period or 21 dynamically new objects do. If we expand our statistics on strong depletion to include all comets from our *entire database* with assigned compositional classes, our total sample of objects increases by over 40% to 193 comets and yields similar percentages of strong carbon-chain depletion by dynamical group – 28% of JF comets, 7% of Halley-type, 3% of old long period, and still none that are young or dynamically new. Finally, if we look at all comets exhibiting some degree of carbon-chain depletion (both strongly and moderately depleted, as discussed above in Section 4.1) within our restricted subset, we find that 38% of the JF comets are either strongly or moderately depleted in both carbon-chain molecules, but still just 8% and 4% of the Halley-type and old long period comets, respectively, exhibit any degree of carbon-chain depletion. A single young long period comet is moderately carbon chain depleted (5% of the sample) and none of the dynamically new objects are. Just over half of the strongly carbon-chain depleted comets additionally have some degree of depletion in NH, while none of the moderately depleted comets do.

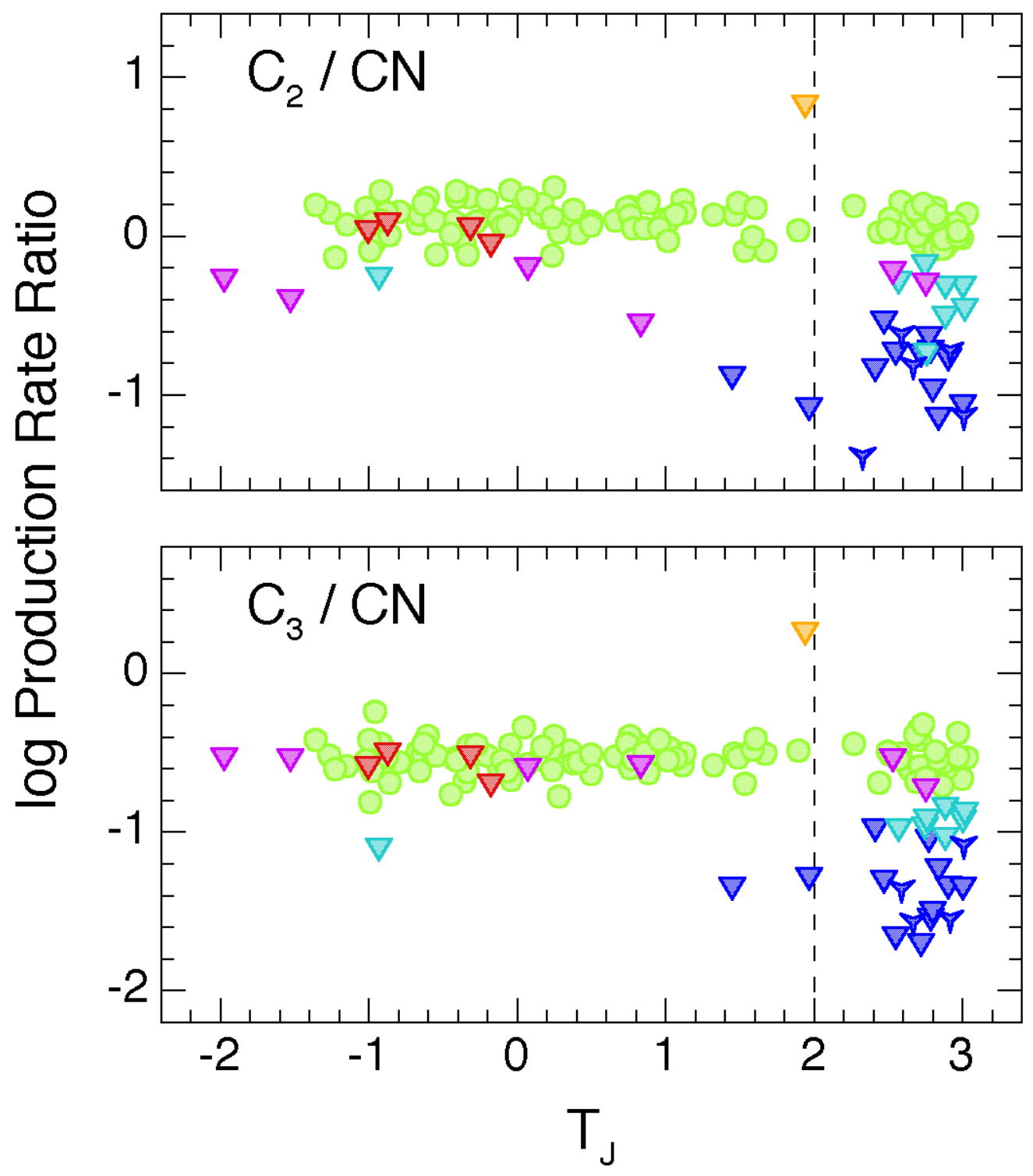

**Figure 9.** Logarithmic mean production rate ratios of $C_2$-to-CN and $C_3$-to-CN as a function of the Tisserand invariant ($T_J$) with respect to Jupiter. Colors and symbols are the same as in Figure 6. The association of carbon-chain depleted comets with the Jupiter-family ($T_J > 2$) dynamical class is very strong. Nearly 90% of comets believed to originate from the Oort cloud have typical composition (green) and most carbon-chain deplete comets (light and dark blue) are Jupiter-family, presumed to have originated in the Kuiper Belt.

It cannot be a coincidence that nearly 40% of the JF comets exhibit some degree of carbon-chain depletion, while members of the other dynamical groups have carbon-chain abundances almost entirely within the typical range. The evidence in our database imply carbon-chain depletion is a continuum (see the taxonomy plots in Figure 6), owing to the conditions in the early solar nebula when and where the comets were formed, and that there is no distinct division between the strongly and moderately depleted classes based on their carbon-chain abundance ratios alone. Current models indicate the Oort cloud comets, which are almost entirely typical in composition, formed closer to the Sun than did the comets that are found in the Kuiper belt, and specifically the scattered disc that is the likely source region for JF comets (N. A. Kaib & K. Volk 2024, and references therein). We continue to assume that the heliocentric distances, and thus temperatures, of condensation out of the nebula into cometesimals was a key factor for carbon-chain composition, and we have tentatively concluded that carbon-chain depletion increases with greater distances and lower temperatures, with a potential additional temperature threshold that would enable the formation of the NH depleted comets (e. g. M. F. A'Hearn et al. 1995 and D. G. Schleicher 2022).

That this composition is primordial and *not* evolutionary comes from a few different lines of evidence, the first being the small sample of fragmenting or disintegrating comets that have compositional information available. The strongest evidence comes from observations of the fragmenting comet 73P/S-W 3, which was observed to have a major outburst and subsequent fragmentation during its 1995 apparition (J. Crovisier et al. 1995, Z. Sekanina 2005), with several components documented during its poor apparition in 2000-2001 (I. Toth et al. 2005). Throughout its excellent apparition in 2006 (which brought it within 0.08 au of Earth) we obtained compositional measurements on four of its components, some of which were still actively outbursting and fragmenting, exposing fresh interior surfaces that dominated the observed activity (see Section 3.2 above and D. G. Schleicher & A. N. Bair 2011). *The relative abundances for the species we measure were consistent across each component, and every component that we observed in 2006 exhibited strong carbon-chain depletion, in agreement with the compositional measurements we obtained immediately following its initial outburst in 1995 and consistent with pre-fragmentation measurements obtained by U. Fink in 1990 (U. Fink & M. D. Hicks 1996).* Another line of evidence comes from our observations of the moderately carbon-chain depleted comet, C/LINEAR (1999 S4), which was shedding materials during the time of our observations; this comet also did not change its chemical composition, even with the constant exposure of fresh material from its disintegrating nucleus (T. L. Farnham et al. 2001).

Conversely, if we consider the short-period comets in our database that have highly thermally evolved surfaces (i.e. have few, discrete source regions and very small active fractions) we find that thermal processing does not result in carbon-chain depletion. Nine of the eleven comets with active fractions under one percent in our database, including 2P/Encke, 28P/Neujmin 1, 49P/Arend-Rigaux, and 209P/LINEAR, among others, have typical composition, while just two, 114P/Wiseman-Skiff and 260P/McNaught, are carbon-chain depleted. That we have highly evolved comets from multiple compositional classes shows that the observed range of active fractions is a natural progression of the depletion of surface volatiles, and it does not cause an alteration of their chemical composition by thermal processes. Within our entire database, we see no connection between the degree of carbon-chain depletion and aging, with no correlation between carbon-chain composition and effective active area, active fraction or perihelion distance.

### 4.4 *Gas and Dust Production Rates with Heliocentric Distance, Seasonal Asymmetries, and Secular Trends*

When looking at changes in overall gas and dust production rates with heliocentric distance, seasonal asymmetries, and secular trends, we see no differences between objects in the strongly carbon-chain depleted class and those that are in the other compositional classes in our full database. There are, however, behaviors that these comets exhibit that are characteristic of comets as a whole, and some that are distinct to the periodic comets. Just five of the strongly carbon-chain depleted comets have sufficient to moderate heliocentric distance ranges that make it feasible to examine trends in production rates for at least some species as they approach and/or recede from the Sun, while nine have measurements at comparable distances on both sides of perihelion that allow us to examine seasonal activity. We have observed five over multiple apparitions, with two observed at similar locations in their orbits allowing for a comparison of activity over time.

First, examining slopes of production rates with distance from the Sun, five strongly carbon-chain depleted comets have measurements over sufficient ranges in heliocentric distance to discern if trends in behavior are present (Section 3.2 and Figures 1-5). Similar to most comets from all compositional classes of our full database, the three carbon-bearing species for these comets behave similarly to one another with relatively shallow $r_H$-dependences for their production rates, and with $C_2$ being somewhat steeper when we can measure its slope. Likewise, in our full database OH and NH often exhibit similar behaviors, with steeper $r_H$-dependences than what is seen for the carbon-bearing species, which is also the case for these comets. The slope for dust is usually shallower than all of the gases, and while this is the case for three of these five comets (43P/Wolf-Harrington, 73P/S-W 3, and 260P/McNaught) the dust slope is intermediate between the carbon-bearing species and OH for 21P/G-Z and as steep as OH in 217P/LINEAR. The two strongly depleted comets with longer timelines of observations, G-Z and S-W 3, have overall shallower slopes, while the $r_H$-dependencies for the other three comets are especially steep, likely reflecting their much smaller range of distance covered. We emphasize that these broad trends of relative steepness between the species are seen in our entire database and are independent of compositional class and dynamical origin.

There are some differences in production rates with heliocentric distance *based on dynamical age* that emerge throughout our entire database, however, with the periodic comets almost always having steeper $r_H$-dependences for all measured species than what is seen in the younger dynamical groups. Additionally, most JF comets in our full database exhibit seasonal behaviors in activity. Of the nine strongly depleted comets with measurements at comparable distances on both sides of perihelion, six exhibit clear seasonal behaviors for at least some of the species, with five comets having higher production rates before perihelion and one higher after, while the remaining three have enough scatter to make a comparison difficult. None of these results are surprising, as these short-period comets generally have smaller and more isolated, and possibly fewer, active surface regions on their nuclei, a result of having spent a relatively longer amount of time in the inner solar system with numerous passes closer to the Sun. Their thermally evolved nuclei therefore usually exhibit seasonal effects, as they are quite sensitive to the changing illumination from the Sun throughout their orbits. They additionally often have distinct turn-on times – and then brighten and fade somewhat quickly as their active areas receive and then lose sunlight, resulting in the observed steeper $r_H$-dependencies for all species and their frequent asymmetries in production rates about perihelion. A direct outcome of this is that periodic comets are usually observed over a shorter range of heliocentric distances, and that their strong seasonal behaviors, and generally smaller active surface regions, additionally result in fewer periodic comets having reliable measured $r_H$-dependencies.

Only a few comets from our entire database exhibit secular changes in production rates over the course of multiple orbits, and when we do see changes that appear to be associated with a diminishing supply of volatiles, they are usually quite small. This is consistent with the estimate from D. Nesvorný et al. (2017) that periodic comets remain visible for a few hundred passages, implying source regions are generally long-lived; even with several decades of observations for some of these comets, we don't expect to see large secular changes. We have observed five strongly depleted comets over multiple apparitions, but only two, Wolf-Harrington and G-Z, have appropriate measurements at the same location along their orbits to compare production rates between apparitions. For Wolf-Harrington (4 apparitions between 1984 and 2004), the production rates are consistent over our 20 years of observations, with no evidence of secular changes. For G-Z, however, there appears to be a small secular decrease over 33 years, from 1985 and 2018 (5 apparitions) that is discussed in detail in Schleicher (2022). Trends involving production rates with heliocentric distance and secular activity are especially difficult to determine for the strongly depleted comets given the inherent faintness of most. Even when we can discern trends, it is often only for OH and CN due to the depletion – and resulting low signal-to-noise – for $C_2$, $C_3$, and often also for NH. As such, we have only measured these characteristics in a small number of these comets. Seasonal differences in production rates about perihelion are more frequently detected, since these objects, similar to most comets in short-period orbits, usually have distinct active regions and are most often at or near peak brightness during their closest approach to the Sun.

### 4.5 *Outbursts and Fragmentation, Water Production, Effective Active Areas, and Active Fractions*

Outburst events, where comets suddenly brighten due to a rapid increase in activity, are usually short-lived and modern surveys are revealing they are more common than previously known (e. g. M. Kelley et al. 2024, T. Lister et al. 2022). The fragmentation of comet nuclei, though less common, has been documented in several dozen comets (H. Beohnhardt 2004, Y. R. Fernandez 2009) and while a primary component often survives a fragmentation event, the smaller components rarely remain observable for subsequent apparitions. For the comets we have measured in our database, there appears to be no strong correlation between outbursts and either chemical composition or dynamical group, and of the 13 JF comets in our database with documented outbursts, nearly equal proportions are from the typical and carbon-chain depleted classes. For the JF comets we have observed that have had documented fragmentation events, however, there is a larger proportion of carbon-chain depleted comets than typical, with five being carbon-chain depleted (four strongly and one moderately), and just one is that is typical. When we expand to include the non-JF comets in our database, however, we note that we have observed several additional comets noted to have fragmented, most of which have typical composition.

About half of our strongly carbon-chain depleted comets have been documented to experience outbursts and/or fragmentation events, which, depending on the timing, could influence our derived production rates, median effective active areas, and active fractions. The five with documented outburst activity are 43P/West-Hartley (M. S. P. Kelley et al. 2019), 155P/Shoemaker 3 (D. Bodewits et al. 2019), 217P/LINEAR (Y. Sarugaku et al. 2010), 260P/McNaught (M. S. P. Kelley et al. 2019), and 398P/Boattini (T. L. Farnham et al. 2021); of these, only LINEAR was known to be in outburst near the time of our observations. The four in this class known to have experienced fragmentation events are 57P/d-N-D (Y. R. Fernandez 2009), 73P/S-W 3 (Z. Sekanina 2007), 101P/Chernykh (J. Luu & D. Jewitt 1991, H. Boehnhardt 2004) and 168P/Hergenrother 1 (G. Sostero et al. 2012). Of these, all but Chernykh were reported to be in this mode during the time of at least one of our observations. For d-N-D and Hergenrother 1, the nucleus components were still close together during our observations and we did not observe multiple fragments; S-W 3 remains the only comet for which we have been able to discern and observe multiple components of its nucleus.

The distributions of mean water production, active area and active fraction for the strongly depleted comets, listed in Table 6, are comparable to those seen for all periodic comets regardless of their compositional class. Additionally, we see no trends in the mean water related variables with either perihelion distance or the mean heliocentric distance of observations for the strongly depleted comets; this is likely due to the varying evolutionary stages of these comets, as JF comets are continually being moved into the inner solar system. Comet G-Z has the highest mean water production rate of the strongly carbon-chain depleted class, and its effective active area and active fraction are the largest of the comets not documented to be fragmenting or outbursting. The three comets known to be splitting during at least some of our observations have the highest mean water production rates following G-Z; they are d-N-D, S-W3, and Hergenrother 1. Their effective active areas are among the highest of the strongly depleted comets, and, when nucleus sizes are known, they unsurprisingly have high active fractions ranging from 76% to >100% (Table 6). Note that we have no water production rates for Chernykh, which was documented to split in 1991 (J. Luu and D. Jewitt 1991), since our observations are from 1977 and before we had an OH filter. If we use CN as a proxy measurement, however, its effective active area is the highest of this group at 13 $km^2$ (d-N-D, the next highest, is 12 $km^2$ using this same proxy) and its active fraction is over 100%, suggesting Chernykh may have been fragmenting or in outburst during this earlier apparition as well. At the other end of the range, almost a third of these comets – 114P/Wiseman-Skiff, 43P/West-Hartley, 155P/Shoemaker 3, 260P/McNaught and 398P/Boattini – exhibit very low water production rates and have correspondingly low effective active areas and active fractions, when available, at ≤3%. (Table 6). Their water-related values are among the lowest for the carbon-chain depleted class, for the Jupiter family comets, and for the entire database.

For comets known to have outbursts (but *not* fragmentation), for the most part we see no discernible differences in either water production rates, effective active areas, or active fractions (when available) as compared to objects with no reported outbursts. The one exception is LINEAR, the only one known to be in outburst at the time of our observations, which, after G-Z, has the highest water-related values of the non-fragmenting comets. We note that one additional strongly carbon-chain depleted comet, Bus, has water production rates that yield an active fraction well over 100%. Our observations were obtained on just a single night in 1981, and while we are unaware of any reports of fragmentation or outburst activity, we suspect our measurements were during an outburst since it should have been too faint to successfully detect otherwise.

Finally, while the strongly carbon-chain depleted comets do have relatively high dust-to-gas ratios as compared to most typical Jupiter-family comets, we see no correlation in dust-to-gas ratios with outbursts, either within the strongly depleted classes or within the full restricted database. This agrees with findings from Kelley et al. (2024), where the authors found no obvious correlation between a comet's probability to outburst and its dust-to-gas ratio. We note, however, that of the six JF comets we have measured with documented fragmentation events, the five that are carbon-chain depleted (four strongly and one moderately) have relatively high dust-to-gas ratios while the single typical comet has a very low dust-to-gas ratio. This suggests that within the JF dynamical group these dustier, carbon-chain depleted comets may be more prone to fragmentation events than the typical comets, though a larger sample size is needed to say with any certainty.

In conclusion, the strongly carbon-chain depleted comets have water production rates, effective active areas, and active fractions that span a range of values that entirely overlap with the typical comets, both JF and otherwise. Additionally, their seasonal behaviors are no different than what we see for most periodic comets. This provides strong evidence that surface evolution is not indicative of composition, and more specifically of carbon-chain depletion. We additionally find no obvious correlation with dust-to-gas ratios and outbursts. We do, however, see a higher proportion of fragmenting JF comets that have depleted composition, but there are a number of dynamically new and young long period comets with typical composition that have had documented fragmentation and/or disintegration events.

## 5. SUMMARY AND CONCLUSIONS

The Lowell narrowband photometry database has been continually updated since the M. F. A'Hearn et al. (1995) publication that first introduced the carbon-chain depleted compositional class, and our recent full analysis, which incorporates the A'Hearn et al. database, contains 220 comets and reveals that there are varying degrees of carbon-chain depletion. We used the mean values from a restricted subset of 135 well-determined comets to determine taxonomic classifications through statistical analyses and visual assessments of various production rate ratio plots. Using this restricted subset, we ultimately subdivided the carbon-chain depleted comets into three compositional classes, with the strongly depleted class, containing comets strongly depleted in *both* $C_2$ *and* $C_3$, being the most distinct. We placed comets from the full database into compositional classes where possible and have identified 17 comets that we consider to be strongly carbon-chain depleted. This is the largest non-typical compositional class in our database, and it consistently stood out throughout our various analyses– not only separating from comets with typical composition but also from those exhibiting more moderate levels of carbon-chain depletion.

For carbon-chain depletion (i.e. depleted in *both* $C_2$ *and* $C_3$ with respect to CN and OH), there is no distinct division to distinguish between the moderately and strongly depleted classes; instead, there is a continuum regarding the degree of depletion. Many of the strongly depleted comets additionally exhibit depletion in NH, however, and this ultimately determined the placement of some of the comets near the boundary between these two classes, as our various analyses considered production rate ratios of all gas species. This strongly depleted class is distinct in that all of the comets have mean abundance ratios of $C_2$ and $C_3$ with respect to both CN and OH that are significantly below the mean ratios for the typical comets, ranging from about 3× to 30× lower than mean typical values (see Section 3.1 and Table 5).

We see no correlations in carbon-chain depletion with the heliocentric distances of observations, the perihelion distance, the size of a comet's nucleus, or whether a nucleus has experienced outbursts; fragmentation events appear to be more common in carbon-chain depleted comets within the Jupiter family dynamical class, however, but our small sample size precludes any strong conclusions. We do see a correlation between dust-to-gas ratios and carbon-chain depletion, with all carbon-chain depleted comets (both strongly and moderately) within the upper ~half of measured dust-to-gas ratios for the JF comets as well as for all dynamical classes. There are trends with heliocentric distance in $C_2$ production rates throughout our database that were measured in many comets *regardless of compositional class* which we attribute to our chosen Haser scalelengths used to derive production rates; these trends are quite small when compared to the degree of depletion exhibited by these strongly carbon-chain depleted objects.

Several key pieces of evidence from our database indicate carbon-chain depletion is due to primordial conditions when and where these comets formed and is *not* due to evolution from thermal heating after these objects arrived in the inner solar system. The strongest evidence comes from our observations of the fragmented comet 73P/Schwassmann-Wachmann 3 in 2006, where our compositional measurements indicated strong carbon-chain depletion across four components of the nucleus. These measurements of freshly exposed materials confirm that the interior of this comet is strongly depleted, and are consistent with pre-fragmentation measurements from 1990 by U. Fink (2009) and our own measurements obtained during its fragmentation in 1995 (D. G. Schleicher & A. N. Bair 2011), eliminating evolution as the cause of depletion. A second line of evidence comes from our observations of several highly thermally evolved short-period comets, i.e. those on Jupiter-family or Halley-type orbits that have few, discrete source regions, very small active fractions, and generally smaller perihelion distances. Most of these highly evolved comets are typical in composition, and the proportions of thermally processed typical and carbon-chain depleted comets are similar to the relative amounts of typical and carbon-chain depleted short-period comets in our database, indicating there is no relation between thermal evolution and carbon-chain depletion. Finally, for dust production, the more thermally evolved comets in our full database trend towards having lower dust-to-gas ratios, while all of our carbon-chain depleted comets have dust-to-gas ratios within the mid- to high-range of values we have measured, and one would expect them to be lower if thermal processing were the cause of depletion. *Taken together, these lines of evidence are a strong indication that comets exhibiting carbon-chain depletion have this composition due to the primordial temperature and pressure conditions in the early solar nebula when and where these comets formed, rather than to more recent thermal evolution that occurred subsequent to their arrival in the inner solar system.*

There is a strong connection between carbon-chain depletion and the Jupiter-family dynamical class. This was one of the most important findings in A'Hearn et al.'s (1995) analysis, and our findings here strongly reinforce this. We find that nearly all of the 17 carbon-chain depleted comets are from the Jupiter-family, presumed to originate in the scattered disc of the Kuiper belt, and that the two currently in non-JF orbits have parameters that suggest a likely scattered disc origin as well. Nearly 40% of the JF comets in our database exhibit some degree of carbon-chain depletion, and we hypothesize that there was a temperature threshold in the outer solar system when and where these objects were coalescing that resulted in the continuum of carbon-chain depletion we observe, with a possible additional threshold even farther from the Sun that caused the NH depletion observed in some of the strongly carbon-chain depleted comets.

We emphasize the importance of long-term, large-scale studies of comets to provide insights on the ranges of chemical compositions and overall behaviors of comets across dynamical groups, which help determine what is "normal" and what is anomalous. Many of the strongly carbon-chain depleted comets are relatively faint, with resulting short timelines of observations, despite our efforts to further characterize them. Future observations of these comets, when viable, will contribute to our understanding of this interesting compositional group, especially when observations can extend the timelines of observations and/or provide more insights on their seasonal behaviors. Of the 17 strongly carbon-chain depleted comets, the first to have a reasonable apparition is 260P/McNaught in the second half of 2026, followed by 21P/G-Z in 2031, 398P/Boattini in 2032, and 73P/S-W 3 in 2033. Unfortunately, only one of the five unrestricted objects, 57P/duToit-Neujmin-Delporte in 2034, has a viable apparition within the next two decades. Finally, both 126P/IRAS in 2036 and 114P/Wiseman-Skiff in 2039 have good apparitions to conclude the decade. The use of different observational techniques and instruments will provide insights into what additional parent and daughter molecules are (or are not) depleted in these comets. Advancements in the near-IR have contributed information on the likely parent molecules of some species we measure, such as $H_2O$, HCN, $NH_3$, $C_2H_2$ and $C_2H_6$ (e.g. Dello Russo et al. 2016), though brightness limits for these techniques mean few objects have been measured beyond their closest approaches to Earth and relatively few periodic comets have been measured thus far. Future observations with facilities such as JWST can provide information on the composition of cometary ices, and specifically relative amounts of CO and $CO_2$, exploring whether or not carbon-chain depletion is related to these fundamental volatiles.

## ACKNOWLEDGEMENTS

We thank R. Millis, M. A'Hearn, T. Farnham, D. Osip, and D. Thompson for acquiring some of these reported observations. We also thank Lowell Observatory for making facilities available over such a long interval of time, thus permitting many of these studies. We additionally thank M. Knight for feedback on early drafts of this manuscript, along with two anonymous referees. This research was originally supported by grants from NASA's Planetary Astronomy Program and most recently by the Solar System Observations Program (Grant Number 80NSSC18K0856). We also gratefully acknowledge support from the Marcus Cometary Research Fund at Lowell Observatory.

Facilities: Lowell Observatory 72 inch (1.8 m) Perkins Telescope; Lowell Observatory 42 inch (1.1 m) John S. Hall Telescope; Lowell Observatory 31 inch (0.8 m) telescope (LO:0.8m); Perth Observatory 24 inch (0.6 m) Planetary Patrol Telescope; Mauna Kea Observatory 88 inch (2.2 m) Telescope.

Table 1. Summary list of the strongly carbon-chain depleted comets.

| Comet | Dynamical Group [a] | $T_{Perihelion}$ | | | Period (year) | Perihelion (au) | $r_H$ pre-$q$ (au) min | $r_H$ pre-$q$ (au) max | $r_H$ post-$q$ (au) min | $r_H$ post-$q$ (au) max | # of Data Sets |
|---|---|---|---|---|---|---|---|---|---|---|---|
| 21P/Giacobini-Zinner* | JF | 1985 | Sep | 6 | 6.6 | 1.028 | 1.055 | 1.515 | 1.029 | 1.450 | 27 |
| | | 1998 | Nov | 22 | 6.6 | 1.034 | 1.034 | 1.435 | 1.060 | 2.289 | 36 |
| | | 2012 | Feb | 12 | 6.9 | 1.030 | 1.806 | 1.806 | ... | ... | 1 |
| | | 2018 | Sep | 10 | 6.5 | 1.013 | 1.015 | 1.818 | 1.014 | 2.066 | 61 |
| 31P/Schwassmann-Wachmann 2* | JF | 1981 | Mar | 17 | 6.5 | 2.135 | 2.196 | 2.196 | 2.140 | 2.140 | 2 |
| | | 1987 | Aug | 30 | 6.4 | 2.071 | ... | ... | 2.524 | 2.524 | 1 |
| | | 1994 | Jan | 23 | 6.4 | 2.070 | 2.076 | 2.275 | ... | ... | 8 |
| 43P/Wolf-Harrington* | JF | 1984 | Sep | 23 | 6.5 | 1.616 | ... | ... | 1.617 | 1.617 | 1 |
| | | 1991 | Apr | 5 | 6.5 | 1.608 | 1.684 | 1.946 | ... | ... | 6 |
| | | 1997 | Sep | 30 | 6.5 | 1.582 | ... | ... | 1.582 | 1.867 | 18 |
| | | 2004 | Mar | 18 | 6.5 | 1.579 | 1.615 | 1.688 | ... | ... | 5 |
| 57P/duToit-Neujmin-Delporte | JF | 2021 | Oct | 17 | 6.4 | 1.720 | ... | ... | 1.729 | 1.767 | 3 |
| 73P/Schwassmann-Wachmann 3* | JF | 1995 | Sep | 24 | 5.3 | 0.933 | ... | ... | 1.009 | 1.256 | 9 |
| | | 2006 | Jun | 7 | 5.4 | 0.939 | 0.982 | 1.652 | 1.725 | 1.725 | 31 |
| 87P/Bus | JF | 1981 | Jun | 11 | 6.5 | 2.183 | 2.234 | 2.234 | ... | ... | 2 |
| 101P/Chernyk | JF | 1978 | Feb | 14 | 15.9 | 2.567 | 2.626 | 2.871 | ... | ... | 14 |
| 114P/Wiseman-Skiff* | JF | 2020 | Jan | 14 | 6.7 | 1.579 | ... | ... | 1.579 | 1.583 | 3 |
| 123P/West-Hartley* | JF | 2019 | Feb | 5 | 7.6 | 2.127 | 2.127 | 2.127 | 2.133 | 2.304 | 8 |
| 126P/IRAS* | HT | 1983 | Aug | 24 | 13.2 | 1.697 | ... | ... | 1.707 | 1.994 | 4 |
| 155P/Shoemaker 3 | JF | 2002 | Dec | 15 | 17.1 | 1.814 | ... | ... | 1.829 | 1.831 | 3 |
| 168P/Hergenrother 1 | JF | 2012 | Oct | 2 | 6.9 | 1.415 | ... | ... | 1.417 | 1.417 | 5 |
| 217P/LINEAR* | JF | 2009 | Sep | 9 | 7.8 | 1.224 | ... | ... | 1.310 | 1.529 | 5 |
| 260P/McNaught* | JF | 2019 | Sep | 10 | 7.0 | 1.417 | 1.421 | 1.421 | 1.437 | 1.648 | 9 |
| 290P/Jager* | JF | 1999 | Mar | 10 | 14.9 | 2.134 | 2.291 | 2.301 | ... | ... | 2 |
| | | 2014 | Mar | 12 | 15.2 | 2.156 | 2.162 | 2.162 | 2.160 | 2.160 | 4 |
| 398P/Boattini* | JF | 2020 | Dec | 27 | 5.5 | 1.306 | 1.311 | 1.311 | 1.311 | 1.311 | 5 |
| C/Shoemaker 1984s (1984 U2)* | OLP | 1985 | Jan | 4 | 270 | 1.215 | 1.391 | 1.391 | 1.266 | 1.693 | 4 |

[a] JF = Jupiter-family, HT = Halley-type, OLP = Old long period

*Comets denoted with an asterisk are included in the restricted subset of 135 comets used for taxonomic analyses

Table 2. Photometry observing circumstances and fluorescence efficiencies for the strongly carbon-chain depleted comets. [a]

| UT Date | ΔT | $r_H$ | Δ | Phase | Phase | $\dot{r}_H$ | log $L/N$ (erg s$^{-1}$ molecule$^{-1}$) | | | Teles. [c] |
|---|---|---|---|---|---|---|---|---|---|---|
| | (day) | (au) | (au) | (°) | Adj. [b] | (km s$^{-1}$) | OH | NH | CN | |
| 31P/Schwassmann-Wachmann 2 | | | | | | | | | | |
| 1981 Jan 5.3 | −71.1 | 2.196 | 1.227 | 5.7 | 0.097 | −2.9 | −15.377 | −13.686 | −13.056 | L72 |
| 1981 Apr 5.2 | +18.7 | 2.140 | 1.931 | 27.9 | 0.366 | +0.8 | −15.433 | −13.827 | −13.220 | L72 |
| 1988 Mar 18.3 | +200.9 | 2.524 | 1.581 | 9.2 | 0.151 | +6.5 | −15.331 | −13.801 | −13.155 | L72 |
| 1993 Sep 18.5 | −127.0 | 2.275 | 2.358 | 25.0 | 0.341 | −5.1 | −15.474 | −13.788 | −13.161 | L42 |
| 1993 Sep 19.4 | −126.0 | 2.272 | 2.344 | 25.1 | 0.342 | −5.1 | −15.472 | −13.788 | −13.161 | L42 |
| 1993 Dec 8.4 | −46.1 | 2.100 | 1.342 | 21.6 | 0.308 | −2.2 | −15.425 | −13.818 | −13.158 | L31 |
| 1993 Dec 9.3 | −45.1 | 2.099 | 1.333 | 21.3 | 0.305 | −2.1 | −15.427 | −13.821 | −13.162 | L31 |
| 1994 Jan 4.3 | −19.2 | 2.076 | 1.139 | 11.2 | 0.180 | −0.9 | −15.450 | −13.839 | −13.199 | L42 |
| 43P/Wolf-Harrington | | | | | | | | | | |
| 1984 Sep 29.5 | +6.8 | 1.617 | 1.723 | 34.7 | 0.414 | +0.7 | −15.192 | −13.592 | −12.979 | L72 |
| 1990 Dec 12.1 | −113.6 | 1.946 | 1.643 | 30.4 | 0.385 | −8.8 | −15.295 | −13.650 | −13.031 | L72 |
| 1991 Jan 18.2 | −76.6 | 1.775 | 1.836 | 31.6 | 0.394 | −7.0 | −15.283 | −13.570 | −12.924 | L42 |
| 1991 Jan 19.2 | −75.6 | 1.771 | 1.841 | 31.5 | 0.393 | −6.9 | −15.281 | −13.569 | −12.924 | L42 |
| 1991 Feb 13.2 | −50.6 | 1.684 | 1.962 | 30.2 | 0.384 | −5.0 | −15.212 | −13.550 | −12.896 | L42 |
| 1997 Sep 30.4 | +0.7 | 1.582 | 1.636 | 36.2 | 0.422 | +0.1 | −15.225 | −13.592 | −12.975 | L31 |
| 1997 Oct 4.5 | +4.7 | 1.583 | 1.611 | 36.5 | 0.424 | +0.6 | −15.218 | −13.577 | −12.967 | L31 |
| 1997 Oct 9.5 | +9.8 | 1.585 | 1.580 | 36.8 | 0.425 | +1.1 | −15.203 | −13.562 | −12.943 | L31 |
| 1997 Nov 1.5 | +32.8 | 1.617 | 1.447 | 37.3 | 0.428 | +3.6 | −15.071 | −13.496 | −12.818 | L31 |
| 1997 Nov 2.5 | +33.7 | 1.619 | 1.441 | 37.3 | 0.428 | +3.7 | −15.067 | −13.493 | −12.815 | L31 |
| 1997 Dec 4.4 | +65.7 | 1.714 | 1.283 | 34.8 | 0.415 | +6.5 | −15.036 | −13.489 | −12.807 | L42 |
| 1998 Jan 8.5 | +100.7 | 1.867 | 1.160 | 26.9 | 0.358 | +8.5 | −15.102 | −13.562 | −12.883 | L31 |
| 2004 Jan 18.2 | −59.6 | 1.688 | 1.631 | 34.4 | 0.412 | −6.0 | −15.224 | −13.535 | −12.886 | L42 |
| 2004 Jan 23.2 | −54.6 | 1.671 | 1.657 | 34.4 | 0.412 | −5.6 | −15.210 | −13.532 | −12.883 | L42 |
| 2004 Feb 11.2 | −35.6 | 1.619 | 1.755 | 33.7 | 0.408 | −3.8 | −15.182 | −13.553 | −12.883 | L42 |
| 2004 Feb 13.2 | −33.6 | 1.615 | 1.766 | 33.6 | 0.407 | −3.6 | −15.181 | −13.556 | −12.886 | L42 |
| 57P/duToit-Neujmin-Delporte | | | | | | | | | | |
| 2021 Nov 5.1 | +18.7 | 1.729 | 1.971 | 30.2 | 0.384 | +1.6 | −15.254 | −13.618 | −12.996 | L42 |
| 2021 Nov 30.1 | +43.7 | 1.767 | 2.178 | 26.4 | 0.354 | +3.6 | −15.148 | −13.569 | −12.900 | L42 |
| 87P/Bus | | | | | | | | | | |
| 1981 Apr 5.4 | −66.9 | 2.234 | 1.276 | 9.8 | 0.160 | −2.6 | −15.431 | −13.854 | −13.195 | L72 |
| 101P/Chernykh | | | | | | | | | | |
| 1977 Oct 9.4 | −157.9 | 2.871 | 1.892 | 5.7 | 0.0973 | −6.1 | −15.646 | −13.959 | −13.367 | L31 |
| 1977 Oct 5.2 | −132.1 | 2.785 | 1.808 | 5.4 | 0.092 | −5.4 | −15.611 | −13.943 | −13.339 | L42 |
| 1977 Oct 15.3 | −122.1 | 2.755 | 1.822 | 9.0 | 0.148 | −5.0 | −15.599 | −13.947 | −13.331 | L42 |
| 1977 Oct 19.3 | −118.0 | 2.744 | 1.835 | 10.4 | 0.169 | −4.9 | −15.595 | −13.947 | −13.327 | L42 |
| 1977 Dec 8.2 | −68.2 | 2.628 | 2.222 | 21.4 | 0.306 | −3.0 | −15.567 | −13.975 | −13.319 | L72 |
| 1977 Dec 9.2 | −67.2 | 2.626 | 2.233 | 21.5 | 0.307 | −3.0 | −15.567 | −13.971 | −13.319 | L72 |
| 114P/Wiseman-Skiff | | | | | | | | | | |
| 2020 Jan 14.1 | +0.1 | 1.579 | 0.862 | 33.5 | 0.407 | −0.0 | −15.223 | −13.592 | −12.975 | L42 |
| 2020 Jan 24.2 | +10.1 | 1.583 | 0.931 | 35.4 | 0.418 | +1.1 | −15.202 | −13.561 | −12.943 | L42 |
| 123P/West-Hartley | | | | | | | | | | |
| 2019 Jan 31.3 | −4.8 | 2.127 | 1.277 | 17.4 | 0.261 | −0.2 | −15.481 | −13.851 | −13.231 | L42 |
| 2019 Feb 26.3 | +21.2 | 2.133 | 1.200 | 11.8 | 0.189 | +1.1 | −15.461 | −13.815 | −13.204 | L42 |
| 2019 Mar 25.2 | +48.1 | 2.160 | 1.271 | 15.6 | 0.239 | +2.4 | −15.397 | −13.780 | −13.145 | L42 |

Table 2—Continued

| UT Date | ΔT (day) | $r_H$ (au) | Δ (au) | Phase (°) | Phase Adj.[b] | $\dot{r}_H$ (km $s^{-1}$) | log $L/N$ (erg $s^{-1}$ $molecule^{-1}$) OH | NH | CN | Teles.[c] |
|---|---|---|---|---|---|---|---|---|---|---|
| 2019 May 30.2 | +114.1 | 2.304 | 1.946 | 25.9 | 0.349 | +5.0 | −15.317 | −13.747 | −13.095 | L42 |
| 126P/IRAS | | | | | | | | | | |
| 1983 Sep 10.3 | +17.6 | 1.707 | 0.792 | 20.7 | 0.298 | +2.1 | −15.171 | −13.590 | −12.951 | L72 |
| 1983 Oct 5.4 | +42.7 | 1.759 | 0.834 | 18.0 | 0.268 | +5.0 | −15.041 | −13.530 | −12.851 | L31 |
| 1983 Nov 29.2 | +97.5 | 1.994 | 1.455 | 28.1 | 0.368 | +9.4 | −15.088 | −13.627 | −12.943 | L72 |
| 155P/Shoemaker 3 | | | | | | | | | | |
| 2003 Jan 5.4 | +21.5 | 1.829 | 0.954 | 19.6 | 0.286 | +2.4 | −15.253 | −13.640 | −12.996 | L42 |
| 2003 Jan 7.3 | +23.5 | 1.831 | 0.947 | 18.8 | 0.277 | +2.6 | −15.240 | −13.635 | −12.987 | L42 |
| 168P/Hergenrother 1 | | | | | | | | | | |
| 2012 Oct 9.2 | +7.2 | 1.417 | 0.445 | 16.5 | 0.250 | +1.1 | −15.106 | −13.466 | −12.848 | L42 |
| 217P/LINEAR | | | | | | | | | | |
| 2009 Oct 15.5 | +36.5 | 1.310 | 0.611 | 46.8 | 0.463 | +7.8 | −14.801 | −13.270 | −12.561 | L42 |
| 2009 Nov 21.3 | +73.3 | 1.529 | 0.651 | 26.2 | 0.352 | +12.3 | −14.697 | −13.461 | −12.742 | L42 |
| 260P/McNaught | | | | | | | | | | |
| 2019 Aug 31.3 | −9.7 | 1.421 | 0.631 | 38.8 | 0.435 | −1.5 | −15.103 | −13.516 | −12.854 | L42 |
| 2019 Oct 1.3 | +21.3 | 1.437 | 0.563 | 31.5 | 0.393 | +3.2 | −14.991 | −13.413 | −12.730 | L42 |
| 2019 Oct 3.2 | +23.2 | 1.441 | 0.562 | 30.9 | 0.389 | +3.5 | −14.975 | −13.405 | −12.719 | L42 |
| 2019 Nov 25.3 | +76.3 | 1.648 | 0.737 | 19.9 | 0.289 | +9.3 | −14.971 | −13.471 | −12.770 | L42 |
| 290P/Jager | | | | | | | | | | |
| 1998 Dec 8.2 | −91.9 | 2.301 | 1.399 | 12.6 | 0.200 | −6.0 | −15.493 | −13.780 | −13.167 | L31 |
| 1998 Dec 11.3 | −88.8 | 2.291 | 1.374 | 11.6 | 0.186 | −5.8 | −15.487 | −13.780 | −13.164 | L31 |
| 2014 Feb 24.2 | −16.3 | 2.162 | 1.501 | 23.6 | 0.328 | −1.2 | −15.476 | −13.873 | −13.224 | L42 |
| 2014 Mar 25.2 | +12.7 | 2.160 | 1.791 | 27.3 | 0.361 | +0.9 | −15.479 | −13.833 | −13.224 | L42 |
| 398P/Boattini | | | | | | | | | | |
| 2020 Dec 16.1 | −10.6 | 1.311 | 0.378 | 25.7 | 0.347 | −1.8 | −15.025 | −13.442 | −12.775 | L42 |
| 2021 Jan 6.2 | +10.4 | 1.311 | 0.389 | 27.7 | 0.365 | +1.8 | −15.002 | −13.377 | −12.735 | L42 |
| C/Shoemaker 1984s (1984 U2) | | | | | | | | | | |
| 1984 Nov 20.3 | −44.6 | 1.391 | 0.435 | 18.6 | 0.275 | −12.5 | −14.785 | −13.423 | −12.757 | L72 |
| 1985 Jan 27.3 | +23.4 | 1.266 | 0.444 | 42.3 | 0.448 | +7.5 | −14.730 | −13.242 | −12.530 | MK |
| 1985 Jan 28.3 | +24.4 | 1.271 | 0.448 | 42.1 | 0.447 | +7.7 | −14.733 | −13.246 | −12.533 | MK |
| 1985 Mar 24.3 | +79.4 | 1.693 | 0.945 | 30.2 | 0.384 | +16.9 | −14.735 | −13.600 | −12.845 | L72 |

[a]All parameters are given for the midpoint of each night's observations. Data for comets 21P/Giacobini-Zinner and 73P/Schwassmann-Wachmann 3 are found in Schleicher (2022) and Schleicher & Bair (2011), respectively.

[b]Adjustment to 0° solar phase angle to $\log(A(\theta)f\rho)$ values based on assumed phase function (see text).

[c]Telescope ID: L72 = Lowell 72-inch (1.8-m); L42 = Lowell 42-inch (1.1-m); L31 = Lowell 31-inch (0.8-m); MK = Mauna Kea 88-inch

Table 3. Photometric fluxes and aperture abundances for the strongly carbon-chain depleted comets.

| UT Date | Aperture | | log Emission Band Flux[a] (erg cm$^{-2}$ s$^{-1}$) | | | | | log Continuum Flux[a] (erg cm$^{-2}$ s$^{-1}$ Å$^{-1}$) | | | log $M(\rho)$[a] (molecule) | | | | |
|---|---|---|---|---|---|---|---|---|---|---|---|---|---|---|---|
| | Size (arcsec) | log $\rho$ (km) | OH | NH | CN | $C_3$ | $C_2$ | UV | Blue | Green | OH | NH | CN | $C_3$ | $C_2$ |
| 31P/Schwassmann-Wachmann 2 | | | | | | | | | | | | | | | |
| 1981 Jan 5.3 | 14.1 | 3.80 | −12.13 | ... | −12.69 | −12.57 | −13.43 | −14.32 | ... | −13.94 | 30.91 | ... | 28.11 | 27.74 | 27.23 |
| 1981 Apr 5.2 | 28.5 | 4.30 | −11.84 | −13.24 | −12.18 | *und* | −12.84 | −14.64 | ... | −14.34 | 31.61 | 28.61 | 29.06 | *und* | 28.18 |
| 1988 Mar 18.3 | 40.1 | 4.36 | −12.44 | −12.90 | −12.47 | −13.56 | *und* | −14.66 | ... | −14.31 | 30.74 | 28.75 | 28.53 | 27.09 | *und* |
| 1993 Sep 18.5 | 73.7 | 4.80 | *und* | −12.28 | −11.86 | *und* | *und* | −14.40 | ... | −14.08 | *und* | 29.71 | 29.50 | *und* | *und* |
| 1993 Sep 19.4 | 104.1 | 4.95 | −12.01 | *und* | −11.58 | *und* | *und* | −14.02 | ... | −13.87 | 31.65 | *und* | 29.77 | *und* | *und* |
| 1993 Dec 8.3 | 114.7 | 4.75 | −11.24 | −12.06 | −11.27 | −11.58 | −12.33 | −14.11 | ... | −13.69 | 31.89 | 29.46 | 29.59 | 28.77 | 28.37 |
| 1993 Dec 8.4 | 114.7 | 4.75 | −11.35 | −12.62 | −11.29 | −11.84 | *und* | −14.06 | ... | −13.72 | 31.78 | 28.90 | 29.57 | 28.50 | *und* |
| 1993 Dec 9.3 | 114.7 | 4.74 | −11.52 | −12.47 | −11.28 | −12.39 | −12.39 | −13.98 | ... | −13.65 | 31.61 | 29.05 | 29.58 | 27.95 | 28.30 |
| 1994 Jan 4.3 | 51.7 | 4.33 | −11.45 | −12.23 | −11.52 | −11.93 | *und* | −13.82 | ... | −13.40 | 31.56 | 29.17 | 29.24 | 28.27 | *und* |
| 1994 Jan 4.3 | 51.7 | 4.33 | ... | ... | −11.60 | −11.95 | −11.98 | −13.71 | ... | −13.45 | ... | ... | 29.16 | 28.24 | 28.56 |
| 1994 Jan 4.3 | 51.7 | 4.33 | ... | ... | −11.49 | *und* | *und* | −13.83 | ... | −13.37 | ... | ... | 29.27 | *und* | *und* |
| 43P/Wolf-Harrington | | | | | | | | | | | | | | | |
| 1984 Sep 29.5 | 40.1 | 4.40 | −11.18 | −12.83 | −11.66 | −12.65 | −12.80 | −14.30 | ... | −13.97 | 31.93 | 28.68 | 29.24 | 27.69 | 27.88 |
| 1990 Dec 12.1 | 28.5 | 4.23 | ... | ... | −12.09 | ... | −12.89 | −14.68 | ... | −14.43 | ... | ... | 28.82 | ... | 27.92 |
| 1990 Dec 12.1 | 28.5 | 4.23 | −12.19 | −13.95 | −12.08 | *und* | −13.65 | −14.68 | ... | −14.41 | 30.98 | 27.58 | 28.83 | *und* | 27.15 |
| 1990 Dec 12.1 | 40.1 | 4.38 | −11.70 | −12.78 | −11.86 | −12.60 | −13.45 | −14.63 | ... | −14.26 | 31.47 | 28.75 | 29.05 | 27.86 | 27.36 |
| 1991 Jan 18.2 | 49.8 | 4.52 | ... | ... | −11.26 | ... | −12.27 | −14.24 | ... | −14.00 | ... | ... | 29.64 | ... | 28.55 |
| 1991 Jan 19.1 | 49.8 | 4.52 | −11.44 | −12.72 | −11.46 | −12.04 | *und* | −14.69 | ... | −14.03 | 31.82 | 28.83 | 29.44 | 28.43 | *und* |
| 1991 Feb 13.2 | 70.2 | 4.70 | −11.20 | −12.90 | −11.22 | *und* | −12.55 | *und* | ... | −13.77 | 32.05 | 28.69 | 29.71 | *und* | 28.29 |
| 1997 Sep 30.4 | 114.7 | 4.83 | −10.86 | −12.58 | −11.05 | −11.85 | −11.75 | −14.11 | −13.75 | −13.88 | 32.24 | 28.89 | 29.80 | 28.42 | 28.87 |
| 1997 Sep 30.4 | 114.7 | 4.83 | ... | ... | −11.00 | ... | ... | ... | −13.75 | −13.76 | ... | ... | 29.85 | ... | ... |
| 1997 Sep 30.5 | 81.4 | 4.68 | −11.12 | *und* | −11.23 | *und* | −12.11 | −13.91 | −13.97 | −13.83 | 31.99 | *und* | 29.63 | *und* | 28.52 |
| 1997 Oct 4.5 | 114.7 | 4.83 | −10.89 | −12.43 | −10.98 | −11.83 | −11.92 | −14.08 | −13.76 | −13.68 | 32.20 | 29.01 | 29.85 | 28.43 | 28.69 |
| 1997 Oct 4.5 | 114.7 | 4.83 | ... | ... | −11.00 | ... | ... | ... | −13.76 | −13.66 | ... | ... | 29.83 | ... | ... |
| 1997 Oct 4.5 | 81.4 | 4.68 | −11.02 | −12.90 | −11.19 | *und* | −12.17 | −14.15 | −13.78 | −13.79 | 32.06 | 28.54 | 29.64 | *und* | 28.44 |
| 1997 Oct 4.5 | 81.4 | 4.68 | ... | ... | −11.17 | ... | ... | ... | −13.78 | −13.78 | ... | ... | 29.66 | ... | ... |
| 1997 Oct 9.4 | 114.7 | 4.82 | −10.89 | −12.30 | −11.00 | −11.66 | −11.88 | −14.76 | −13.82 | −13.77 | 32.16 | 29.11 | 29.80 | 28.59 | 28.71 |
| 1997 Oct 9.4 | 114.7 | 4.82 | ... | ... | −10.98 | ... | ... | ... | −13.82 | −13.75 | ... | ... | 29.81 | ... | ... |
| 1997 Oct 9.5 | 81.4 | 4.67 | −11.07 | −14.82 | −11.22 | −11.94 | −12.11 | −14.73 | −13.84 | −13.81 | 31.98 | 26.59 | 29.57 | 28.30 | 28.49 |
| 1997 Oct 9.5 | 81.4 | 4.67 | ... | ... | −11.16 | ... | ... | ... | −13.84 | −13.74 | ... | ... | 29.63 | ... | ... |
| 1997 Nov 1.4 | 114.7 | 4.78 | −10.85 | −12.66 | −10.85 | −11.77 | −12.02 | −13.91 | −13.65 | −13.62 | 31.99 | 28.61 | 29.74 | 28.42 | 28.52 |
| 1997 Nov 1.5 | 81.4 | 4.63 | −11.00 | −12.55 | −11.07 | −12.35 | −12.24 | −14.18 | −13.81 | −13.72 | 31.84 | 28.71 | 29.52 | 27.84 | 28.29 |
| 1997 Nov 2.5 | 114.7 | 4.78 | −10.79 | −12.26 | −10.82 | −11.88 | −12.18 | −14.27 | −13.68 | −13.60 | 32.05 | 29.00 | 29.76 | 28.31 | 28.35 |
| 1997 Nov 2.5 | 81.4 | 4.63 | −11.03 | −12.62 | −11.03 | −12.04 | −12.30 | −14.47 | −13.80 | −13.75 | 31.80 | 28.64 | 29.55 | 28.14 | 28.23 |
| 1997 Dec 4.4 | 104.1 | 4.69 | −10.88 | −12.42 | −10.98 | −12.53 | −11.93 | −13.88 | −13.82 | −13.78 | 31.82 | 28.73 | 29.49 | 27.60 | 28.55 |
| 1998 Jan 8.4 | 81.4 | 4.53 | −11.35 | −11.71 | −11.49 | −12.33 | −12.35 | −13.91 | −14.21 | −13.97 | 31.33 | 29.43 | 28.97 | 27.79 | 28.11 |
| 1998 Jan 8.5 | 57.6 | 4.38 | −11.68 | −12.66 | −11.66 | *und* | −12.74 | −14.51 | −14.05 | −14.01 | 31.00 | 28.48 | 28.80 | *und* | 27.73 |
| 2004 Jan 18.2 | 92.7 | 4.74 | −10.99 | −12.44 | −11.05 | *und* | −11.94 | −14.19 | −13.78 | −13.83 | 32.11 | 28.97 | 29.71 | *und* | 28.74 |
| 2004 Jan 18.2 | 59.4 | 4.55 | −11.32 | *und* | −11.33 | −12.30 | −12.37 | −14.30 | −13.93 | −13.94 | 31.78 | *und* | 29.44 | 28.02 | 28.31 |
| 2004 Jan 23.2 | 59.4 | 4.55 | −11.30 | *und* | −11.34 | −12.27 | −12.37 | −14.42 | −13.98 | −13.92 | 31.80 | *und* | 29.43 | 28.07 | 28.31 |
| 2004 Feb 11.2 | 59.4 | 4.58 | −11.19 | −12.72 | −11.34 | −12.07 | −12.04 | −14.36 | −13.95 | −13.95 | 31.93 | 28.77 | 29.49 | 28.29 | 28.66 |
| 2004 Feb 13.2 | 46.3 | 4.47 | −11.29 | *und* | −11.46 | −12.61 | −12.59 | −14.34 | −13.95 | −13.95 | 31.84 | *und* | 29.37 | 27.75 | 28.12 |
| 57P/duToit-Neujmin-Delporte | | | | | | | | | | | | | | | |

Table 3—Continued

| UT Date | Aperture Size (arcsec) | Aperture log ρ (km) | log Emission Band Flux[a] (erg cm$^{-2}$ s$^{-1}$) OH | NH | CN | $C_3$ | $C_2$ | log Continuum Flux[a] (erg cm$^{-2}$ s$^{-1}$ Å$^{-1}$) UV | Blue | Green | log $M(\rho)$[a] (molecule) OH | NH | CN | $C_3$ | $C_2$ |
|---|---|---|---|---|---|---|---|---|---|---|---|---|---|---|---|
| 2021 Nov 5.1 | 48.6 | 4.54 | −10.98 | *und* | −11.01 | −11.46 | −11.69 | −13.92 | −13.47 | −13.43 | 32.31 | *und* | 30.02 | 29.06 | 29.17 |
| 2021 Nov 30.1 | 97.2 | 4.89 | −10.53 | −12.67 | −10.73 | *und* | −11.52 | −13.54 | −13.40 | −13.34 | 32.74 | 29.03 | 30.29 | *und* | 29.45 |
| 2021 Nov 30.1 | 97.2 | 4.89 | −10.67 | −11.93 | −10.74 | *und* | −11.70 | −13.52 | −13.35 | −13.31 | 32.60 | 29.76 | 30.28 | *und* | 29.27 |
| 87P/Bus | | | | | | | | | | | | | | | |
| 1981 Apr 5.3 | 40.1 | 4.27 | −11.94 | ... | −12.52 | −12.56 | −13.43 | −15.29 | ... | −14.86 | 31.16 | ... | 28.34 | 27.80 | 27.28 |
| 1981 Apr 5.4 | 40.1 | 4.27 | −12.43 | *und* | −12.56 | −13.30 | −14.51 | ... | ... | −14.82 | 30.66 | *und* | 28.30 | 27.06 | 26.20 |
| 101P/Chernykh | | | | | | | | | | | | | | | |
| 1977 Oct 9.4 | 77.9 | 4.73 | ... | ... | −11.63 | ... | −12.06 | ... | ... | −13.44 | ... | ... | 29.74 | ... | 29.21 |
| 1977 Oct 5.2 | 242.0 | 5.20 | ... | ... | −11.22 | −12.00 | −11.70 | ... | ... | −13.25 | ... | ... | 30.09 | 28.86 | 29.50 |
| 1977 Oct 15.2 | 73.7 | 4.69 | ... | ... | −11.91 | −12.89 | −12.75 | ... | ... | −13.56 | ... | ... | 29.39 | 27.96 | 28.44 |
| 1977 Oct 15.3 | 73.7 | 4.69 | ... | ... | −11.77 | ... | −12.33 | ... | ... | −13.57 | ... | ... | 29.53 | ... | 28.87 |
| 1977 Oct 19.3 | 48.0 | 4.50 | ... | ... | −12.40 | ... | −12.80 | ... | ... | −13.76 | ... | ... | 28.90 | ... | 28.40 |
| 1977 Oct 19.3 | 48.0 | 4.50 | ... | ... | −12.18 | ... | −12.77 | ... | ... | −13.77 | ... | ... | 29.12 | ... | 28.43 |
| 1977 Oct 19.3 | 48.0 | 4.50 | ... | ... | −12.14 | ... | −12.72 | ... | ... | −13.76 | ... | ... | 29.17 | ... | 28.48 |
| 1977 Oct 19.3 | 48.0 | 4.50 | ... | ... | −12.27 | ... | −13.04 | ... | ... | −13.76 | ... | ... | 29.03 | ... | 28.16 |
| 1977 Dec 8.1 | 34.0 | 4.44 | ... | ... | −12.89 | ... | *und* | ... | ... | −14.24 | ... | ... | 28.58 | ... | *und* |
| 1977 Dec 8.2 | 34.0 | 4.44 | ... | ... | −12.83 | ... | −13.37 | ... | ... | −14.28 | ... | ... | 28.63 | ... | 27.96 |
| 1977 Dec 9.1 | 34.0 | 4.44 | ... | ... | −12.73 | ... | *und* | ... | ... | −14.27 | ... | ... | 28.73 | ... | *und* |
| 1977 Dec 9.1 | 34.0 | 4.44 | ... | ... | −12.56 | ... | −13.91 | ... | ... | −14.28 | ... | ... | 28.90 | ... | 27.43 |
| 1977 Dec 9.2 | 34.0 | 4.44 | ... | ... | −12.69 | ... | −13.43 | ... | ... | −14.28 | ... | ... | 28.77 | ... | 27.90 |
| 1977 Dec 9.2 | 34.0 | 4.44 | ... | ... | −12.65 | ... | −13.72 | ... | ... | −14.27 | ... | ... | 28.81 | ... | 27.61 |
| 114P/Wiseman-Skiff | | | | | | | | | | | | | | | |
| 2020 Jan 14.1 | 97.2 | 4.48 | −11.75 | *und* | −11.83 | −12.05 | −12.43 | *und* | −14.43 | −14.51 | 30.79 | *und* | 28.47 | 27.67 | 27.63 |
| 2020 Jan 24.2 | 97.2 | 4.52 | −11.60 | −13.09 | −11.76 | −12.65 | −12.36 | −14.75 | −14.49 | −14.61 | 30.99 | 27.86 | 28.57 | 27.14 | 27.77 |
| 2020 Jan 24.2 | 97.2 | 4.52 | −11.66 | −13.82 | −11.78 | −12.10 | −12.37 | *und* | −14.45 | −14.45 | 30.93 | 27.13 | 28.55 | 27.69 | 27.76 |
| 123P/West-Hartley | | | | | | | | | | | | | | | |
| 2019 Jan 31.3 | 97.2 | 4.65 | −11.69 | −12.63 | −11.80 | −11.63 | −12.39 | −15.03 | −13.97 | −13.89 | 31.45 | 28.88 | 29.09 | 28.68 | 28.28 |
| 2019 Jan 31.3 | 97.2 | 4.65 | −11.73 | −14.14 | −11.76 | *und* | *und* | −14.26 | −13.88 | −13.84 | 31.41 | 27.37 | 29.13 | *und* | *und* |
| 2019 Feb 26.3 | 97.2 | 4.63 | −11.87 | −13.32 | −11.83 | *und* | −12.52 | −14.14 | −13.73 | −13.79 | 31.20 | 28.10 | 28.98 | *und* | 28.09 |
| 2019 Feb 26.3 | 97.2 | 4.63 | −11.73 | −13.05 | −11.80 | −12.53 | −14.03 | −14.19 | −13.79 | −13.74 | 31.34 | 28.37 | 29.02 | 27.74 | 26.58 |
| 2019 Mar 25.2 | 97.2 | 4.65 | −11.92 | *und* | −11.79 | −12.73 | −12.66 | −14.26 | −13.85 | −13.85 | 31.13 | *und* | 29.01 | 27.60 | 28.02 |
| 2019 May 30.2 | 97.2 | 4.84 | −12.16 | −12.98 | −12.10 | −12.55 | *und* | *und* | −14.16 | −14.21 | 31.19 | 28.79 | 29.03 | 28.21 | *und* |
| 2019 May 30.2 | 97.2 | 4.84 | −12.42 | −12.69 | −12.06 | *und* | *und* | −14.64 | −14.33 | −14.19 | 30.92 | 29.09 | 29.07 | *und* | *und* |
| 2019 May 30.3 | 97.2 | 4.84 | −12.53 | −13.01 | −12.03 | *und* | *und* | −14.35 | −14.46 | −14.29 | 30.81 | 28.77 | 29.09 | *und* | *und* |
| 126P/IRAS | | | | | | | | | | | | | | | |
| 1983 Sep 10.3 | 28.5 | 3.91 | −11.46 | −14.08 | −11.72 | −12.19 | −12.65 | −14.06 | ... | −13.74 | 30.96 | 26.76 | 28.48 | 27.52 | 27.41 |
| 1983 Oct 5.4 | 114.7 | 4.54 | −10.66 | −11.65 | −10.82 | −11.59 | −12.35 | −13.85 | ... | −13.35 | 31.67 | 29.17 | 29.32 | 28.19 | 27.78 |
| 1983 Nov 29.2 | 28.5 | 4.18 | ... | ... | −12.27 | ... | −13.59 | −14.79 | ... | −14.59 | ... | ... | 28.45 | ... | 27.13 |
| 1983 Nov 29.2 | 28.5 | 4.18 | ... | ... | ... | ... | −13.15 | ... | ... | −14.61 | ... | ... | ... | ... | 27.57 |
| 155P/Shoemaker 3 | | | | | | | | | | | | | | | |
| 2003 Jan 5.4 | 74.2 | 4.41 | −12.09 | *und* | −12.24 | *und* | *und* | −14.68 | −14.26 | −14.14 | 30.57 | *und* | 28.16 | *und* | *und* |
| 2003 Jan 5.4 | 74.2 | 4.41 | −12.32 | −13.36 | −12.19 | *und* | −13.54 | −15.07 | −14.23 | −14.25 | 30.34 | 27.69 | 28.22 | *und* | 26.74 |
| 2003 Jan 7.3 | 74.2 | 4.41 | −12.03 | −12.77 | −12.14 | *und* | −13.35 | −14.69 | −14.23 | −14.16 | 30.62 | 28.27 | 28.24 | *und* | 26.92 |
| 168P/Hergenrother 1 | | | | | | | | | | | | | | | |

Table 3—Continued

| UT Date | Aperture | | log Emission Band Flux[a] (erg cm$^{-2}$ s$^{-1}$) | | | | | log Continuum Flux[a] (erg cm$^{-2}$ s$^{-1}$ Å$^{-1}$) | | | log $M(\rho)$[a] (molecule) | | | | |
|---|---|---|---|---|---|---|---|---|---|---|---|---|---|---|---|
| | Size (arcsec) | log $\rho$ (km) | OH | NH | CN | $C_3$ | $C_2$ | UV | Blue | Green | OH | NH | CN | $C_3$ | $C_2$ |
| 2012 Oct 9.2 | 97.2 | 4.20 | −10.21 | −11.63 | −10.31 | −11.09 | −11.10 | −12.97 | −12.56 | −12.54 | 31.64 | 28.58 | 29.28 | 27.96 | 28.29 |
| 2012 Oct 9.2 | 204.5 | 4.52 | −9.71 | −11.17 | −9.82 | −10.94 | −10.62 | −12.81 | −12.43 | −12.42 | 32.14 | 29.04 | 29.78 | 28.11 | 28.77 |
| 2012 Oct 9.2 | 62.4 | 4.00 | −10.53 | −11.93 | −10.61 | −11.36 | −11.43 | −13.11 | −12.69 | −12.67 | 31.32 | 28.28 | 28.98 | 27.69 | 27.97 |
| 2012 Oct 9.2 | 155.9 | 4.40 | −9.90 | −11.34 | −9.98 | −10.87 | −10.77 | −12.87 | −12.47 | −12.46 | 31.95 | 28.87 | 29.61 | 28.18 | 28.62 |
| 2012 Oct 9.2 | 97.2 | 4.20 | −10.21 | −11.60 | −10.31 | −11.11 | −11.10 | −12.97 | −12.56 | −12.53 | 31.64 | 28.61 | 29.29 | 27.93 | 28.30 |
| 217P/LINEAR | | | | | | | | | | | | | | | |
| 2009 Oct 15.4 | 97.2 | 4.33 | −9.98 | −11.39 | −10.13 | −11.23 | −11.00 | −13.07 | −12.70 | −12.69 | 31.84 | 28.90 | 29.45 | 28.02 | 28.60 |
| 2009 Oct 15.5 | 155.9 | 4.54 | −9.67 | −11.15 | −9.83 | −11.09 | −10.73 | −12.91 | −12.55 | −12.54 | 32.15 | 29.14 | 29.76 | 28.17 | 28.87 |
| 2009 Oct 15.5 | 62.4 | 4.14 | −10.33 | −11.81 | −10.47 | −11.50 | −11.38 | −13.29 | −12.91 | −12.88 | 31.49 | 28.48 | 29.11 | 27.75 | 28.22 |
| 2009 Nov 21.2 | 77.8 | 4.26 | −10.73 | −12.49 | −11.12 | −14.36 | −11.91 | −13.88 | −13.52 | −13.46 | 31.04 | 28.05 | 28.70 | 25.09 | 27.88 |
| 2009 Nov 21.3 | 97.2 | 4.36 | −10.71 | −12.39 | −10.98 | −12.11 | −12.16 | −14.00 | −13.47 | −13.41 | 31.07 | 28.15 | 28.83 | 27.34 | 27.64 |
| 260P/McNaught | | | | | | | | | | | | | | | |
| 2019 Aug 31.3 | 97.2 | 4.35 | −11.14 | −12.30 | −11.34 | *und* | −12.12 | −13.83 | −13.67 | −13.55 | 31.01 | 28.26 | 28.56 | *und* | 27.58 |
| 2019 Aug 31.3 | 97.2 | 4.35 | −11.00 | −12.46 | −11.33 | −12.14 | −12.10 | −14.00 | −13.68 | −13.65 | 31.16 | 28.10 | 28.57 | 27.21 | 27.60 |
| 2019 Oct 1.3 | 97.2 | 4.30 | −10.91 | −12.29 | −11.11 | −11.73 | −12.03 | −13.85 | −13.47 | −13.42 | 31.03 | 28.08 | 28.57 | 27.53 | 27.58 |
| 2019 Oct 1.3 | 97.2 | 4.30 | −10.92 | −12.87 | −11.11 | −12.14 | −12.01 | −13.79 | −13.45 | −13.41 | 31.02 | 27.50 | 28.57 | 27.12 | 27.60 |
| 2019 Oct 3.2 | 77.8 | 4.20 | −11.03 | −12.74 | −11.23 | *und* | −11.96 | −13.81 | −13.45 | −13.46 | 30.89 | 27.61 | 28.44 | *und* | 27.65 |
| 2019 Oct 3.2 | 97.2 | 4.30 | −10.88 | −12.49 | −11.11 | −12.24 | −12.22 | −13.70 | −13.40 | −13.38 | 31.05 | 27.86 | 28.55 | 27.03 | 27.39 |
| 2019 Nov 25.2 | 97.2 | 4.41 | −11.36 | −12.82 | −11.61 | −12.54 | −12.33 | −14.03 | −13.55 | −13.55 | 30.79 | 27.84 | 28.35 | 27.08 | 27.64 |
| 2019 Nov 25.3 | 77.8 | 4.32 | −11.55 | *und* | −11.71 | *und* | *und* | −14.08 | −13.74 | −13.69 | 30.61 | *und* | 28.24 | *und* | *und* |
| 2019 Nov 25.3 | 97.2 | 4.41 | −11.41 | *und* | −11.60 | *und* | −12.22 | −13.96 | −13.64 | −13.73 | 30.74 | *und* | 28.36 | *und* | 27.74 |
| 290P/Jager | | | | | | | | | | | | | | | |
| 1998 Dec 8.2 | 120.7 | 4.79 | −11.08 | −12.08 | −11.05 | −11.39 | *und* | −13.98 | −13.25 | −13.21 | 32.15 | 29.44 | 29.86 | 29.07 | *und* |
| 1998 Dec 11.3 | 46.3 | 4.36 | −11.72 | ... | −11.67 | −12.15 | −13.61 | −13.94 | −13.57 | −13.54 | 31.49 | ... | 29.22 | 28.30 | 27.18 |
| 2014 Feb 24.2 | 48.6 | 4.42 | −11.87 | *und* | −12.04 | −12.26 | −12.46 | −14.53 | −14.11 | −14.07 | 31.41 | *und* | 28.99 | 28.21 | 28.36 |
| 2014 Feb 24.3 | 97.2 | 4.72 | −11.57 | −12.95 | −11.55 | *und* | −12.10 | −14.09 | −13.90 | −13.90 | 31.71 | 28.72 | 29.47 | *und* | 28.72 |
| 2014 Mar 25.2 | 97.2 | 4.80 | −11.60 | *und* | −11.68 | −11.82 | −12.30 | −14.74 | −14.09 | −14.09 | 31.83 | *und* | 29.50 | 28.80 | 28.67 |
| 2014 Mar 25.2 | 97.2 | 4.80 | −11.58 | *und* | −11.66 | −11.78 | −13.09 | −14.27 | −14.24 | −13.96 | 31.85 | *und* | 29.52 | 28.85 | 27.89 |
| 398P/Boattini | | | | | | | | | | | | | | | |
| 2020 Dec 16.1 | 97.2 | 4.12 | −11.22 | *und* | −11.45 | −11.76 | −12.59 | −13.98 | −13.55 | −13.53 | 30.41 | *und* | 27.93 | 27.07 | 26.60 |
| 2020 Dec 16.1 | 97.2 | 4.12 | −11.03 | −12.24 | −11.45 | −11.92 | −12.02 | −13.97 | −13.61 | −13.60 | 30.60 | 27.81 | 27.93 | 26.92 | 27.17 |
| 2021 Jan 6.2 | 97.2 | 4.14 | −11.30 | −13.01 | −11.45 | −12.02 | −12.20 | −14.03 | −13.60 | −13.57 | 30.33 | 27.00 | 27.92 | 26.85 | 27.01 |
| 2021 Jan 6.2 | 97.2 | 4.14 | −11.28 | −12.90 | −11.48 | *und* | −12.15 | −13.92 | −13.63 | −13.59 | 30.35 | 27.10 | 27.88 | *und* | 27.06 |
| 2021 Jan 6.2 | 97.2 | 4.14 | −11.29 | −12.78 | −11.47 | −12.39 | −12.44 | −14.05 | −13.55 | −13.55 | 30.34 | 27.23 | 27.90 | 26.47 | 26.77 |
| C/Shoemaker 1984s (1984 U2) | | | | | | | | | | | | | | | |
| 1984 Nov 20.3 | 40.1 | 3.80 | −11.11 | −13.00 | −11.72 | −12.42 | −12.78 | −13.89 | ... | −13.52 | 30.40 | 27.15 | 27.76 | 26.59 | 26.58 |
| 1985 Jan 27.3 | 28.5 | 3.66 | −11.16 | −12.87 | −11.63 | −12.17 | −12.55 | −14.07 | ... | −13.70 | 30.32 | 27.12 | 27.64 | 26.78 | 26.74 |
| 1985 Jan 28.3 | 20.0 | 3.51 | ... | ... | −11.83 | ... | −12.68 | −14.07 | ... | −13.82 | ... | ... | 27.46 | ... | 26.63 |
| 1985 Mar 24.3 | 28.5 | 3.99 | −11.88 | ... | −12.38 | ... | *und* | −14.78 | ... | −14.54 | 30.25 | ... | 27.87 | ... | *und* |

[a] "*und*" stands for "undefined". For the gases, this means that the emission flux was measured but was less than zero following sky and continuum removal. For the continuum, this means the continuum flux was measured but following sky subtraction it was less than zero. Data for comets 21P/Giacobini-Zinner and 73P/Schwassmann-Wachmann 3 are found in Schleicher (2022) and Schleicher & Bair (2011), respectively.

Table 4. Photometric production rates for the strongly carbon-chain depleted comets.

| UT Date | ΔT | log $r_H$ | log $\rho$ | log $Q^{a,b}$ (molecules $s^{-1}$) | | | | | log $A(\theta)f\rho^{a,b}$ (cm) | | | log $Q$ [a] |
|---|---|---|---|---|---|---|---|---|---|---|---|---|
| | (day) | (au) | (km) | OH | NH | CN | $C_3$ | $C_2$ | UV | Blue | Green | $H_2O$ |
| 31P/Schwassmann-Wachmann 2 | | | | | | | | | | | | |
| 1981 Jan 5.3 | −71.1 | 0.342 | 3.80 | 28.05 .05 | ... | 25.04 .04 | 24.24 .12 | 24.37 .25 | 2.62 .03 | ... | 2.81 .01 | 28.02 |
| 1981 Apr 5.2 | +18.7 | 0.330 | 4.30 | 27.87 .07 | 25.12 .22 | 25.13 .02 | *und* | 24.46 .09 | 2.17 .08 | ... | 2.28 .02 | 27.84 |
| 1988 Mar 18.3 | +200.9 | 0.402 | 4.36 | 27.00 .41 | 25.26 .26 | 24.59 .05 | 22.84 .99 | *und* | 2.09 .13 | ... | 2.20 .03 | 26.93 |
| 1993 Sep 18.5 | −127.0 | 0.357 | 4.80 | *und* | 25.42 .27 | 24.83 .05 | *und* | *und* | 2.17 .23 | ... | 2.26 .07 | *und* |
| 1993 Sep 19.4 | −126.0 | 0.356 | 4.95 | 26.93 .22 | *und* | 24.90 .03 | *und* | *und* | 2.40 .12 | ... | 2.30 .04 | 26.88 |
| 1993 Dec 8.3 | −46.1 | 0.322 | 4.75 | 27.43 .07 | 25.23 .18 | 24.97 .02 | 24.08 .13 | 23.94 .29 | 1.95 .12 | ... | 2.14 .03 | 27.40 |
| 1993 Dec 8.4 | −46.1 | 0.322 | 4.75 | 27.32 .07 | 24.66 .38 | 24.95 .01 | 23.81 .20 | *und* | 2.00 .10 | ... | 2.10 .03 | 27.29 |
| 1993 Dec 9.3 | −45.1 | 0.322 | 4.74 | 27.15 .12 | 24.81 .33 | 24.97 .01 | 23.26 .52 | 23.88 .29 | 2.08 .10 | ... | 2.17 .03 | 27.13 |
| 1994 Jan 4.3 | −19.2 | 0.317 | 4.33 | 27.76 .05 | 25.62 .14 | 25.25 .02 | 24.01 .17 | *und* | 2.50 .04 | ... | 2.69 .01 | 27.74 |
| 1994 Jan 4.3 | −19.2 | 0.317 | 4.33 | ... | ... | 25.17 .02 | 23.98 .16 | 24.77 .08 | 2.61 .03 | ... | 2.64 .01 | ... |
| 1994 Jan 4.3 | −19.2 | 0.317 | 4.33 | ... | ... | 25.27 .02 | *und* | *und* | 2.49 .05 | ... | 2.72 .01 | ... |
| 43P/Wolf-Harrington | | | | | | | | | | | | |
| 1984 Sep 29.5 | +6.8 | 0.209 | 4.40 | 27.89 .04 | 24.87 .16 | 25.02 .01 | 23.34 .27 | 23.86 .20 | 2.10 .05 | ... | 2.20 .01 | 27.92 |
| 1990 Dec 12.1 | −113.6 | 0.289 | 4.23 | ... | ... | 24.95 .02 | ... | 24.25 .19 | 2.00 .11 | ... | 2.02 .03 | ... |
| 1990 Dec 12.1 | −113.6 | 0.289 | 4.23 | 27.31 .09 | 24.16 .88 | 24.96 .02 | *und* | 23.49 .60 | 2.01 .10 | ... | 2.04 .03 | 27.30 |
| 1990 Dec 12.1 | −113.6 | 0.289 | 4.38 | 27.55 .04 | 25.08 .21 | 24.95 .02 | 23.53 .30 | 23.46 .59 | 1.90 .12 | ... | 2.04 .03 | 27.54 |
| 1991 Jan 18.2 | −76.6 | 0.249 | 4.52 | ... | ... | 25.29 .01 | ... | 24.39 .33 | 2.17 .38 | ... | 2.18 .04 | ... |
| 1991 Jan 19.1 | −75.6 | 0.248 | 4.52 | 27.63 .05 | 24.86 .31 | 25.08 .02 | 23.96 .22 | *und* | 1.72 .22 | ... | 2.14 .04 | 27.64 |
| 1991 Feb 13.2 | −50.6 | 0.226 | 4.70 | 27.57 .25 | 24.42 .99 | 25.08 .10 | *und* | 23.85 .99 | *und* | ... | 2.24 .04 | 27.59 |
| 1997 Sep 30.4 | +0.7 | 0.199 | 4.83 | 27.56 .03 | 24.39 .23 | 24.99 .01 | 23.75 .17 | 24.25 .08 | 1.87 .13 | 1.90 .05 | 1.80 .06 | 27.59 |
| 1997 Sep 30.4 | +0.7 | 0.199 | 4.83 | ... | ... | 25.04 .01 | ... | ... | ... | *und* | 1.91 .04 | ... |
| 1997 Sep 30.5 | +0.8 | 0.199 | 4.68 | 27.50 .02 | *und* | 25.00 .01 | *und* | 24.08 .11 | 2.23 .06 | 1.84 .05 | 2.00 .04 | 27.54 |
| 1997 Oct 4.5 | +4.7 | 0.199 | 4.83 | 27.52 .02 | 24.52 .17 | 25.05 .01 | 23.76 .15 | 24.07 .09 | 1.91 .12 | 1.89 .04 | 1.99 .04 | 27.55 |
| 1997 Oct 4.5 | +4.7 | 0.199 | 4.83 | ... | ... | 25.02 .01 | ... | ... | ... | *und* | 2.00 .03 | ... |
| 1997 Oct 4.5 | +4.8 | 0.199 | 4.68 | 27.59 .02 | 24.28 .28 | 25.02 .01 | *und* | 24.01 .15 | 1.98 .11 | 2.02 .04 | 2.04 .04 | 27.62 |
| 1997 Oct 4.5 | +4.8 | 0.199 | 4.68 | ... | ... | 25.04 .01 | ... | ... | ... | *und* | 2.03 .04 | ... |
| 1997 Oct 9.4 | +9.7 | 0.200 | 4.82 | 27.49 .03 | 24.63 .15 | 25.00 .01 | 23.92 .19 | 24.11 .08 | 1.21 .37 | 1.83 .04 | 1.90 .04 | 27.53 |
| 1997 Oct 9.4 | +9.7 | 0.200 | 4.82 | ... | ... | 25.02 .01 | ... | ... | ... | *und* | 1.90 .03 | ... |
| 1997 Oct 9.5 | +9.8 | 0.200 | 4.67 | 27.52 .02 | 22.34 .99 | 24.97 .01 | 23.73 .22 | 24.07 .10 | 1.39 .27 | 1.96 .04 | 2.01 .04 | 27.55 |
| 1997 Oct 9.5 | +9.8 | 0.200 | 4.67 | ... | ... | 25.03 .01 | ... | ... | ... | *und* | 2.06 .03 | ... |
| 1997 Nov 1.4 | +32.7 | 0.209 | 4.78 | 27.38 .03 | 24.19 .21 | 25.00 .00 | 23.77 .11 | 23.97 .10 | 2.04 .07 | 1.97 .03 | 2.03 .03 | 27.41 |
| 1997 Nov 1.5 | +32.8 | 0.209 | 4.63 | 27.44 .02 | 24.53 .14 | 24.97 .01 | 23.29 .27 | 23.94 .13 | 1.92 .09 | 1.97 .03 | 2.07 .03 | 27.47 |
| 1997 Nov 2.5 | +33.7 | 0.209 | 4.78 | 27.44 .02 | 24.58 .12 | 25.02 .01 | 23.65 .19 | 23.80 .15 | 1.68 .15 | 1.94 .03 | 2.04 .03 | 27.47 |
| 1997 Nov 2.5 | +33.8 | 0.209 | 4.63 | 27.41 .02 | 24.46 .16 | 25.01 .01 | 23.59 .20 | 23.88 .14 | 1.63 .16 | 1.97 .03 | 2.04 .03 | 27.44 |
| 1997 Dec 4.4 | +65.7 | 0.234 | 4.69 | 27.37 .02 | 24.49 .10 | 24.89 .00 | 23.00 .30 | 24.14 .06 | 2.12 .05 | 1.84 .03 | 1.91 .03 | 27.38 |
| 1998 Jan 8.4 | +100.7 | 0.271 | 4.53 | 27.14 .05 | 25.47 .06 | 24.61 .02 | 23.30 .41 | 23.95 .23 | 2.22 .11 | 1.59 .15 | 1.85 .07 | 27.14 |
| 1998 Jan 8.5 | +100.7 | 0.271 | 4.38 | 27.05 .06 | 24.77 .15 | 24.67 .01 | *und* | 23.80 .23 | 1.77 .14 | 1.90 .04 | 1.96 .03 | 27.05 |
| 2004 Jan 18.2 | −59.6 | 0.227 | 4.74 | 27.57 .02 | 24.64 .13 | 25.04 .01 | *und* | 24.25 .08 | 1.95 .11 | 2.03 .04 | 2.00 .04 | 27.60 |
| 2004 Jan 18.2 | −59.5 | 0.227 | 4.55 | 27.53 .05 | *und* | 25.02 .01 | 23.53 .23 | 24.09 .14 | 2.03 .10 | 2.07 .04 | 2.08 .04 | 27.55 |
| 2004 Jan 23.2 | −54.6 | 0.223 | 4.55 | 27.53 .04 | *und* | 25.00 .01 | 23.57 .24 | 24.08 .15 | 1.90 .14 | 2.01 .05 | 2.09 .04 | 27.55 |
| 2004 Feb 11.2 | −35.6 | 0.209 | 4.58 | 27.61 .04 | 24.67 .16 | 25.01 .01 | 23.78 .15 | 24.38 .08 | 1.96 .11 | 2.04 .05 | 2.07 .04 | 27.64 |
| 2004 Feb 13.2 | −33.6 | 0.208 | 4.47 | 27.68 .05 | *und* | 25.05 .01 | 23.33 .34 | 23.99 .23 | 2.09 .09 | 2.16 .04 | 2.18 .04 | 27.71 |
| 57P/duToit-Neujmin-Delporte | | | | | | | | | | | | |
| 2021 Nov 5.1 | +18.7 | 0.238 | 4.54 | 28.08 .09 | *und* | 25.63 .01 | 24.57 .10 | 24.97 .06 | 2.59 .11 | 2.72 .03 | 2.78 .02 | 28.10 |

Table 4—Continued

| UT Date | ΔT | log $r_H$ | log $\rho$ | log $Q^{a,b}$ (molecules $s^{-1}$) | | | | | log $A(\theta)f\rho^{a,b}$ (cm) | | | log $Q^{a}$ |
|---|---|---|---|---|---|---|---|---|---|---|---|---|
| | (day) | (au) | (km) | OH | NH | CN | $C_3$ | $C_2$ | UV | Blue | Green | $H_2O$ |
| 2021 Nov 30.1 | +43.7 | 0.247 | 4.89 | 28.02 .03 | 24.49 .35 | 25.44 .01 | *und* | 24.78 .07 | 2.74 .06 | 2.55 .04 | 2.63 .03 | 28.03 |
| 2021 Nov 30.1 | +43.7 | 0.247 | 4.89 | 27.88 .07 | 25.23 .13 | 25.43 .01 | *und* | 24.60 .12 | 2.76 .08 | 2.60 .04 | 2.67 .03 | 27.89 |
| 87P/Bus | | | | | | | | | | | | |
| 1981 Apr 5.3 | −66.9 | 0.349 | 4.27 | 27.50 .16 | ... | 24.48 .06 | 23.63 .23 | 23.62 .32 | 1.23 .28 | ... | 1.47 .08 | 27.46 |
| 1981 Apr 5.4 | −66.9 | 0.349 | 4.27 | 27.01 .35 | *und* | 24.44 .05 | 22.89 .25 | 22.55 .99 | ... | ... | 1.50 .08 | 26.97 |
| 101P/Chernykh | | | | | | | | | | | | |
| 1977 Oct 9.4 | −157.9 | 0.458 | 4.73 | ... | ... | 25.28 .08 | ... | 24.95 .17 | ... | ... | 2.99 .03 | ... |
| 1977 Oct 5.2 | −132.1 | 0.445 | 5.20 | ... | ... | 24.94 .03 | 23.77 .07 | 24.54 .07 | ... | ... | 2.64 .02 | ... |
| 1977 Oct 15.2 | −122.1 | 0.440 | 4.69 | ... | ... | 24.98 .05 | 23.32 .22 | 24.23 .36 | ... | ... | 2.84 .02 | ... |
| 1977 Oct 15.3 | −122.0 | 0.440 | 4.69 | ... | ... | 25.12 .05 | ... | 24.66 .17 | ... | ... | 2.83 .02 | ... |
| 1977 Oct 19.3 | −118.0 | 0.438 | 4.50 | ... | ... | 24.78 .07 | ... | 24.48 .20 | ... | ... | 2.82 .02 | ... |
| 1977 Oct 19.3 | −118.0 | 0.438 | 4.50 | ... | ... | 25.00 .05 | ... | 24.51 .19 | ... | ... | 2.82 .02 | ... |
| 1977 Oct 19.3 | −118.0 | 0.438 | 4.50 | ... | ... | 25.04 .05 | ... | 24.56 .17 | ... | ... | 2.83 .02 | ... |
| 1977 Oct 19.3 | −118.0 | 0.438 | 4.50 | ... | ... | 24.91 .06 | ... | 24.24 .31 | ... | ... | 2.82 .02 | ... |
| 1977 Dec 8.1 | −68.2 | 0.420 | 4.44 | ... | ... | 24.53 .08 | ... | *und* | ... | ... | 2.54 .02 | ... |
| 1977 Dec 8.2 | −68.1 | 0.420 | 4.44 | ... | ... | 24.59 .08 | ... | 24.12 .24 | ... | ... | 2.50 .02 | ... |
| 1977 Dec 9.1 | −67.2 | 0.419 | 4.44 | ... | ... | 24.69 .07 | ... | *und* | ... | ... | 2.51 .02 | ... |
| 1977 Dec 9.1 | −67.2 | 0.419 | 4.44 | ... | ... | 24.86 .05 | ... | 23.58 .56 | ... | ... | 2.50 .02 | ... |
| 1977 Dec 9.2 | −67.1 | 0.419 | 4.44 | ... | ... | 24.73 .06 | ... | 24.06 .27 | ... | ... | 2.50 .02 | ... |
| 1977 Dec 9.2 | −67.1 | 0.419 | 4.44 | ... | ... | 24.77 .06 | ... | 23.77 .43 | ... | ... | 2.52 .02 | ... |
| 114P/Wiseman-Skiff | | | | | | | | | | | | |
| 2020 Jan 14.1 | +0.1 | 0.198 | 4.48 | 26.60 .05 | *und* | 24.12 .04 | 23.24 .18 | 23.48 .21 | *und* | 1.02 .14 | 0.96 .16 | 26.64 |
| 2020 Jan 24.2 | +10.1 | 0.199 | 4.52 | 26.75 .04 | 23.85 .33 | 24.17 .02 | 22.68 .42 | 23.57 .20 | 1.06 .25 | 1.00 .16 | 0.89 .21 | 26.79 |
| 2020 Jan 24.2 | +10.1 | 0.199 | 4.52 | 26.69 .06 | 23.11 .85 | 24.16 .03 | 23.23 .20 | 23.56 .19 | *und* | 1.03 .15 | 1.06 .15 | 26.73 |
| 123P/West-Hartley | | | | | | | | | | | | |
| 2019 Jan 31.3 | −4.8 | 0.328 | 4.65 | 27.14 .07 | 24.80 .19 | 24.61 .08 | 24.08 .15 | 23.99 .19 | 1.18 .43 | 1.91 .06 | 2.02 .05 | 27.11 |
| 2019 Jan 31.3 | −4.8 | 0.328 | 4.65 | 27.10 .06 | 23.29 .99 | 24.65 .02 | *und* | *und* | 1.95 .10 | 2.00 .05 | 2.07 .04 | 27.07 |
| 2019 Feb 26.3 | +21.2 | 0.329 | 4.63 | 26.94 .05 | 24.06 .40 | 24.55 .02 | *und* | 23.85 .24 | 2.04 .07 | 2.13 .03 | 2.08 .04 | 26.91 |
| 2019 Feb 26.3 | +21.2 | 0.329 | 4.63 | 27.07 .04 | 24.33 .27 | 24.58 .02 | 23.16 .35 | 22.34 .99 | 1.99 .08 | 2.06 .04 | 2.13 .03 | 27.04 |
| 2019 Mar 25.2 | +48.1 | 0.334 | 4.65 | 26.83 .07 | *und* | 24.54 .02 | 22.99 .50 | 23.74 .34 | 1.96 .10 | 2.04 .04 | 2.07 .05 | 26.80 |
| 2019 May 30.2 | +114.1 | 0.362 | 4.84 | 26.63 .16 | 24.46 .33 | 24.31 .06 | 23.43 .46 | *und* | *und* | 1.97 .08 | 1.94 .09 | 26.59 |
| 2019 May 30.2 | +114.1 | 0.362 | 4.84 | 26.37 .31 | 24.75 .23 | 24.35 .04 | *und* | *und* | 1.82 .26 | 1.80 .13 | 1.97 .09 | 26.32 |
| 2019 May 30.3 | +114.1 | 0.362 | 4.84 | 26.26 .35 | 24.43 .39 | 24.38 .04 | *und* | *und* | 2.11 .17 | 1.67 .17 | 1.86 .11 | 26.21 |
| 126P/IRAS | | | | | | | | | | | | |
| 1983 Sep 10.3 | +17.6 | 0.232 | 3.91 | 27.74 .05 | 23.80 .97 | 25.06 .02 | 23.75 .12 | 24.20 .10 | 2.17 .04 | ... | 2.30 .01 | 27.76 |
| 1983 Oct 5.4 | +42.7 | 0.245 | 4.54 | 27.44 .10 | 25.17 .17 | 24.93 .02 | 23.70 .30 | 23.58 .33 | 1.82 .20 | ... | 2.13 .03 | 27.46 |
| 1983 Nov 29.2 | +97.5 | 0.300 | 4.18 | ... | ... | 24.68 .04 | ... | 23.57 .44 | 1.83 .22 | ... | 1.85 .05 | ... |
| 1983 Nov 29.2 | +97.5 | 0.300 | 4.18 | ... | ... | ... | ... | 24.01 .21 | ... | ... | 1.82 .05 | ... |
| 155P/Shoemaker 3 | | | | | | | | | | | | |
| 2003 Jan 5.4 | +21.5 | 0.262 | 4.41 | 26.57 .11 | *und* | 23.98 .05 | *und* | *und* | 1.38 .21 | 1.48 .08 | 1.62 .05 | 26.57 |
| 2003 Jan 5.4 | +21.5 | 0.262 | 4.41 | 26.34 .17 | 23.92 .47 | 24.04 .04 | *und* | 22.75 .80 | 1.00 .42 | 1.51 .07 | 1.51 .07 | 26.34 |
| 2003 Jan 7.3 | +23.4 | 0.263 | 4.41 | 26.62 .12 | 24.51 .21 | 24.07 .05 | *und* | 22.95 .67 | 1.37 .24 | 1.51 .08 | 1.59 .06 | 26.62 |
| 168P/Hergenrother 1 | | | | | | | | | | | | |
| 2012 Oct 9.2 | +7.2 | 0.151 | 4.20 | 27.85 .01 | 25.03 .06 | 25.32 .00 | 23.81 .07 | 24.52 .04 | 2.43 .02 | 2.51 .01 | 2.55 .01 | 27.91 |
| 2012 Oct 9.2 | +7.2 | 0.151 | 4.52 | 27.86 .01 | 24.96 .04 | 25.34 .00 | 23.67 .09 | 24.53 .02 | 2.27 .03 | 2.31 .01 | 2.35 .01 | 27.91 |

Table 4—Continued

| UT Date | $\Delta$T | log $r_H$ | log $\rho$ | log $Q^{a,b}$ (molecules $s^{-1}$) | | | | | log $A(\theta)f\rho^{a,b}$ (cm) | | | log $Q$ [a] |
|---|---|---|---|---|---|---|---|---|---|---|---|---|
| | (day) | (au) | (km) | OH | NH | CN | $C_3$ | $C_2$ | UV | Blue | Green | $H_2O$ |
| 2012 Oct 9.2 | +7.2 | 0.151 | 4.00 | 27.84 .01 | 25.05 .07 | 25.31 .01 | 23.75 .09 | 24.50 .06 | 2.48 .02 | 2.57 .01 | 2.61 .01 | 27.90 |
| 2012 Oct 9.2 | +7.2 | 0.151 | 4.40 | 27.84 .01 | 24.98 .04 | 25.34 .00 | 23.83 .05 | 24.54 .02 | 2.32 .02 | 2.39 .01 | 2.43 .01 | 27.90 |
| 2012 Oct 9.2 | +7.2 | 0.151 | 4.20 | 27.85 .01 | 25.05 .04 | 25.32 .00 | 23.78 .06 | 24.52 .03 | 2.42 .02 | 2.51 .01 | 2.56 .01 | 27.91 |
| 217P/LINEAR | | | | | | | | | | | | |
| 2009 Oct 15.4 | +36.5 | 0.117 | 4.33 | 27.80 .01 | 25.08 .02 | 25.24 .00 | 23.74 .06 | 24.59 .02 | 2.40 .02 | 2.44 .01 | 2.47 .01 | 27.87 |
| 2009 Oct 15.5 | +36.5 | 0.117 | 4.54 | 27.80 .00 | 25.00 .02 | 25.27 .00 | 23.73 .05 | 24.58 .01 | 2.35 .01 | 2.38 .01 | 2.42 .01 | 27.88 |
| 2009 Oct 15.5 | +36.5 | 0.117 | 4.14 | 27.75 .01 | 24.98 .03 | 25.19 .00 | 23.66 .06 | 24.50 .03 | 2.36 .02 | 2.42 .01 | 2.47 .01 | 27.82 |
| 2009 Nov 21.2 | +73.3 | 0.184 | 4.26 | 27.18 .05 | 24.43 .25 | 24.66 .01 | 20.87 .99 | 24.04 .08 | 1.84 .12 | 1.88 .03 | 1.96 .02 | 27.22 |
| 2009 Nov 21.3 | +73.3 | 0.184 | 4.36 | 27.05 .03 | 24.37 .20 | 24.65 .01 | 23.02 .38 | 23.65 .15 | 1.62 .14 | 1.83 .03 | 1.91 .02 | 27.10 |
| 260P/McNaught | | | | | | | | | | | | |
| 2019 Aug 31.3 | −9.7 | 0.153 | 4.35 | 26.98 .13 | 24.46 .20 | 24.37 .02 | *und* | 23.58 .17 | 1.72 .11 | 1.55 .05 | 1.70 .03 | 27.04 |
| 2019 Aug 31.3 | −9.7 | 0.153 | 4.35 | 27.13 .04 | 24.30 .19 | 24.38 .02 | 22.91 .29 | 23.60 .15 | 1.56 .11 | 1.54 .04 | 1.59 .04 | 27.19 |
| 2019 Oct 1.3 | +21.3 | 0.157 | 4.30 | 27.08 .01 | 24.36 .08 | 24.46 .01 | 23.28 .10 | 23.66 .10 | 1.66 .05 | 1.72 .02 | 1.79 .02 | 27.14 |
| 2019 Oct 1.3 | +21.3 | 0.157 | 4.30 | 27.07 .01 | 23.78 .23 | 24.45 .01 | 22.87 .20 | 23.68 .10 | 1.72 .04 | 1.73 .02 | 1.80 .02 | 27.13 |
| 2019 Oct 3.2 | +23.2 | 0.159 | 4.20 | 27.10 .03 | 24.06 .23 | 24.47 .01 | *und* | 23.88 .09 | 1.80 .06 | 1.83 .02 | 1.85 .02 | 27.16 |
| 2019 Oct 3.2 | +23.2 | 0.159 | 4.30 | 27.11 .02 | 24.15 .16 | 24.44 .01 | 22.78 .29 | 23.47 .17 | 1.81 .05 | 1.78 .02 | 1.83 .02 | 27.16 |
| 2019 Nov 25.2 | +76.2 | 0.217 | 4.41 | 26.73 .02 | 24.01 .17 | 24.12 .02 | 22.70 .37 | 23.60 .19 | 1.72 .05 | 1.87 .02 | 1.89 .02 | 26.76 |
| 2019 Nov 25.3 | +76.3 | 0.217 | 4.32 | 26.70 .03 | *und* | 24.15 .02 | *und* | *und* | 1.76 .05 | 1.77 .03 | 1.85 .03 | 26.73 |
| 2019 Nov 25.3 | +76.3 | 0.217 | 4.41 | 26.68 .03 | *und* | 24.13 .01 | *und* | 23.71 .15 | 1.79 .05 | 1.78 .03 | 1.71 .03 | 26.71 |
| 290P/Jager | | | | | | | | | | | | |
| 1998 Dec 8.2 | −91.9 | 0.362 | 4.79 | 27.67 .06 | 25.19 .14 | 25.21 .02 | 24.33 .15 | *und* | 2.24 .15 | 2.64 .02 | 2.71 .02 | 27.63 |
| 1998 Dec 11.3 | −88.8 | 0.360 | 4.36 | 27.69 .08 | ... | 25.22 .01 | 24.02 .11 | 23.39 .76 | 2.69 .03 | 2.73 .01 | 2.78 .01 | 27.65 |
| 2014 Feb 24.2 | −16.3 | 0.335 | 4.42 | 27.48 .05 | *und* | 24.87 .03 | 23.85 .17 | 24.44 .15 | 2.06 .12 | 2.16 .05 | 2.22 .04 | 27.45 |
| 2014 Feb 24.3 | −16.2 | 0.335 | 4.72 | 27.30 .07 | 24.53 .35 | 24.90 .02 | *und* | 24.33 .14 | 2.20 .11 | 2.06 .06 | 2.08 .06 | 27.26 |
| 2014 Mar 25.2 | +12.7 | 0.334 | 4.80 | 27.30 .06 | *und* | 24.81 .04 | 24.06 .16 | 24.17 .18 | 1.63 .29 | 1.95 .08 | 1.97 .08 | 27.27 |
| 2014 Mar 25.2 | +12.7 | 0.334 | 4.80 | 27.33 .05 | *und* | 24.83 .02 | 24.11 .11 | 23.39 .60 | 2.09 .13 | 1.80 .11 | 2.10 .06 | 27.29 |
| 398P/Boattini | | | | | | | | | | | | |
| 2020 Dec 16.1 | −10.6 | 0.118 | 4.12 | 26.69 .10 | *und* | 24.03 .02 | 22.99 .18 | 22.90 .36 | 1.28 .12 | 1.38 .04 | 1.42 .03 | 26.77 |
| 2020 Dec 16.1 | −10.6 | 0.118 | 4.12 | 26.88 .04 | 24.32 .13 | 24.03 .02 | 22.84 .20 | 23.47 .13 | 1.28 .10 | 1.32 .04 | 1.35 .03 | 26.96 |
| 2021 Jan 6.2 | +10.4 | 0.118 | 4.14 | 26.59 .03 | 23.50 .31 | 24.00 .01 | 22.75 .18 | 23.29 .16 | 1.24 .08 | 1.34 .03 | 1.40 .03 | 26.66 |
| 2021 Jan 6.2 | +10.4 | 0.118 | 4.14 | 26.62 .03 | 23.60 .27 | 23.97 .01 | *und* | 23.34 .14 | 1.35 .06 | 1.31 .03 | 1.37 .03 | 26.69 |
| 2021 Jan 6.2 | +10.4 | 0.118 | 4.14 | 26.61 .03 | 23.72 .22 | 23.98 .02 | 22.38 .37 | 23.05 .25 | 1.22 .08 | 1.40 .03 | 1.42 .03 | 26.68 |
| C/Shoemaker 1984s (1984 U2) | | | | | | | | | | | | |
| 1984 Nov 20.3 | −44.6 | 0.143 | 3.80 | 27.25 .02 | 24.25 .19 | 24.41 .01 | 22.91 .17 | 23.43 .20 | 1.78 .02 | ... | 1.91 .01 | 27.32 |
| 1985 Jan 27.3 | +23.4 | 0.102 | 3.66 | 27.35 .02 | 24.41 .11 | 24.47 .01 | 23.26 .08 | 23.77 .09 | 1.68 .02 | ... | 1.81 .01 | 27.43 |
| 1985 Jan 28.3 | +24.4 | 0.104 | 3.51 | ... | ... | 24.54 .01 | ... | 23.92 .10 | 1.83 .02 | ... | 1.85 .01 | ... |
| 1985 Mar 24.3 | +79.4 | 0.229 | 3.99 | 26.90 .14 | ... | 24.32 .04 | ... | *und* | 1.54 .15 | ... | 1.55 .03 | 26.92 |

[a] *"und"* stands for "undefined". For the gases, this means that the emission flux was measured but was less than zero following sky and continuum removal. For the continuum, this means the continuum flux was measured but following sky subtraction it was less than zero. Data for comets 21P/Giacobini-Zinner and 73P/Schwassmann-Wachmann 3 are found in Schleicher (2022) and Schleicher & Bair (2011), respectively.

[b] Production rates followed by the upper, i.e. the positive, uncertainty. Uncertainties are based on photon statistics and reflect the 1- $\sigma$ values derived from the propagation of the observational uncertainties. The "+" and "−" uncertainties are equal as percentages, but unequal in log-space; the "−" values can be computed.

Table 5. Mean abundance ratios for the strongly carbon-chain depleted comets and ranges in composition for relevant compositional classes.

| Comet | $r_H$ (au) | log Production Rate Ratio[a] NH/OH | CN/OH | $C_3$/OH | $C_2$/OH | NH/CN | $C_3$/CN | $C_2$/CN | $C_3$/$C_2$ | GC/OH[b] |
|---|---|---|---|---|---|---|---|---|---|---|
| 21P/G-Z* | 1.306 | −2.91 .02 | −2.58 .04 | −3.71 .15 | −3.12 .04 | −0.20 .03 | −1.29 .04 | −0.52 .02 | −0.77 .03 | −25.26 .12 |
| 31P/S-W 2* | 2.176 | −2.24 .15 | −2.38 .09 | −3.79 .13 | −3.74 .15 | +0.27 .12 | −1.33 .13 | −1.04 .16 | −0.30 .18 | −24.74 .11 |
| 43P/W-H* | 1.677 | −2.70 .16 | −2.50 .02 | −4.00 .07 | −3.47 .04 | −0.18 .17 | −1.48 .07 | −0.94 .04 | −0.51 .08 | −25.03 .03 |
| 57P/d-N-D | 1.754 | −3.07 .26 | −2.49 .04 | −3.99 .30 | −3.20 .05 | −0.61 .25 | −1.54 .30 | −0.71 .05 | −0.88 .30 | −24.93 .05 |
| 73P/S-W 3* | 1.210 | −2.69 .03 | −2.53 .02 | −4.13 .05 | −3.29 .04 | −0.09 .04 | −1.53 .06 | −0.70 .03 | −0.35 .22 | −25.23 .04 |
| 87P/Bus | 2.234 | *und* | −2.74 .17 | −3.98 .11 | −4.08 .20 | *und* | −1.07 .22 | −1.12 .26 | −0.21 .13 | −25.54 .19 |
| 101P/Chernykh | 2.707 | ... | ... | ... | ... | ... | −1.35 .18 | −0.61 .07 | −0.83 .06 | ... |
| 114P/W-S* | 1.582 | −3.29 .24 | −2.53 .03 | −3.54 .13 | −3.14 .02 | −0.72 .25 | −1.03 .12 | −0.61 .01 | −0.41 .13 | −25.28 .06 |
| 123P/W-H* | 2.199 | −2.18 .16 | −2.24 .09 | −3.66 .19 | −3.53 .17 | −0.03 .12 | −1.21 .20 | −1.12 .17 | +0.30 .25 | −24.39 .11 |
| 126P/IRAS* | 1.864 | −2.56 .29 | −2.59 .08 | −3.85 .11 | −3.67 .13 | −0.05 .29 | −1.27 .04 | −1.06 .12 | −0.08 .20 | −25.09 .05 |
| 155P/Shoemaker 3 | 1.830 | −2.41 .20 | −2.46 .09 | *und* | −3.80 .18 | +0.07 .23 | *und* | −1.37 .18 | *und* | −24.64 .06 |
| 168P/Hergenrother 1 | 1.417 | −2.83 .02 | −2.52 .01 | −4.08 .03 | −3.33 .01 | −0.31 .02 | −1.55 .03 | −0.80 .00 | −0.75 .02 | −25.09 .04 |
| 217P/LINEAR* | 1.398 | −2.74 .02 | −2.51 .03 | −4.16 .10 | −3.24 .04 | −0.23 .02 | −1.64 .10 | −0.71 .05 | −0.88 .11 | −24.85 .02 |
| 260P/McNaught* | 1.505 | −2.92 .11 | −2.62 .02 | −4.33 .14 | −3.34 .09 | −0.28 .10 | −1.68 .14 | −0.72 .08 | −0.86 .14 | −24.77 .06 |
| 290P/Jager* | 2.206 | −3.00 .22 | −2.48 .03 | −3.46 .11 | −3.32 .15 | −0.56 .23 | −0.96 .10 | −0.81 .15 | +0.32 .18 | −24.85 .04 |
| 398P/Boattini* | 1.311 | −2.93 .14 | −2.67 .04 | −4.01 .13 | −3.43 .07 | −0.19 .18 | −1.32 .13 | −0.75 .08 | −0.41 .19 | −24.91 .05 |
| C/Shoemaker 1984s* | 1.405 | −2.97 .03 | −2.75 .09 | −4.20 .11 | −3.86 .19 | −0.11 .05 | −1.33 .12 | −0.87 .14 | −0.51 .00 | −25.04 .04 |
| Ranges of Taxonomic Classes Discussed in Text[c] | | | | | | | | | | |
| Strongly Carbon-chain Depleted | | | | | | | | | | |
| min | | −3.29 | −2.75 | −4.33 | −3.86 | −0.72 | −1.68 | −1.12 | −0.88 | −25.28 |
| max | | −2.18 | −2.24 | −3.46 | −3.12 | +0.27 | −0.96 | −0.52 | +0.32 | −24.39 |
| Moderately Carbon-chain Depleted | | | | | | | | | | |
| min | | −2.54 | −2.83 | −3.88 | −3.35 | −0.03 | −1.08 | −0.72 | −0.90 | −25.13 |
| max | | −1.92 | −2.41 | −3.24 | −2.61 | +0.74 | −0.83 | −0.16 | −0.49 | −24.60 |
| Moderately $C_2$ Depleted | | | | | | | | | | |
| min | | −2.77 | −2.88 | −3.49 | −3.39 | −0.02 | −0.71 | −0.54 | −0.42 | −25.09 |
| max | | −2.12 | −2.26 | −2.84 | −2.52 | +0.55 | −0.52 | −0.18 | +0.33 | −23.97 |
| Typical | | | | | | | | | | |
| min | | −2.79 | −3.00 | −3.44 | −2.84 | −0.52 | −0.81 | −0.13 | −0.83 | −26.58 |
| max | | −1.72 | −2.10 | −2.58 | −2.00 | +0.68 | −0.24 | +0.31 | +0.15 | −24.01 |
| mean | | −2.25 | −2.53 | −3.06 | −2.43 | +0.32 | −0.52 | +0.11 | −0.60 | −25.09 |

[a]The mean log production rate ratios followed by the upper, i.e. the positive, standard error. The "+" and "−" uncertainties are equal as percentages, but unequal in log-space; the "−" values can be computed. *"und"* stands for "undefined"; this means that the emission flux was measured for the numerator but was less than zero following sky and continuum removal for all measurements.

[b]The dust-to-gas ratio, represented by the phase adjusted green continuum (log $A(\theta)f\rho$) ratioed to OH

[c]Minimum and maximum values in log-space for members of the Strongly Carbon-chain Depleted, Moderately Carbon-chain Depleted, Moderately $C_2$ Depleted, and Typical compositional classes discussed in the text. The mean values for each ratio for the typical compositional class, derived from the typical comets in the well-determined restricted subset of our database, are also included.

*Comets denoted with an asterisk are included in the restricted subset of 135 comets used for taxonomic analyses.

Table 6. Active areas and active fractions for the strongly carbon-chain depleted comets.

| Comet | Active area ($km^2$) | $r_n$[a] (km) | Active fraction (%) |
|---|---|---|---|
| 21P/G-Z | 6.5 | $1.82 \pm 0.05$ | 36 |
| 31P/S-W 2 | 3.6 | $1.65 \pm 0.11$ | 10 |
| 43P/W-H | 3.0 | $2.10 \pm 0.19$ | 5 |
| 57P/d-N-D | 12 | $0.96 \pm 0.21$ | 104 |
| 73P/S-W 3 | 2.7 | $0.53 \pm 0.02$ | 76 |
| 87P/Bus | 4.2 | $0.26 \pm 0.01$ | 499 |
| 101P/Chernyk | ... | $0.98 \pm 0.09$ | ... |
| 114P/W-S | 0.4 | $0.78 \pm 0.05$ | 0.6 |
| 123P/W-H | 1.4 | $2.18 \pm 0.23$ | 2 |
| 126P/IRAS | 4.5 | $1.57 \pm 0.14$ | 14 |
| 155P/Shoemaker 3 | 0.5 | ... | ... |
| 168P/Hergenrother 1 | 5.1 | $0.48 \pm 0.04$ | 175 |
| 217P/LINEAR | 3.4 | ... | ... |
| 260P/McNaught | 0.9 | $1.54 \pm 0.09$ | 3 |
| 290P/Jager | 4.8 | ... | ... |
| 398P/Boattini | 0.3 | ... | ... |
| C/Shoemaker 1984s | 1.3 | ... | ... |

[a]The radius of the nucleus; references in the text of Section 3.2.